\documentclass[a4paper,onecolumn,11pt,accepted=2023-03-23]{quantumarticle}
\pdfoutput=1
\usepackage[utf8]{inputenc}
\usepackage[english]{babel}
\usepackage[T1]{fontenc}
\usepackage{amsmath}
\usepackage{amsthm}
\usepackage{amsfonts}
\usepackage{color}
\usepackage{mathtools}
\usepackage{amssymb}

\usepackage{hyperref}

\usepackage{tikz}
\usepackage{lipsum}

\newtheorem{lemma}{Lemma}
\newtheorem{example}[lemma]{Example}
\newtheorem{remark}[lemma]{Remark}
\newtheorem{prop}[lemma]{Proposition}

\newtheorem{definition}[lemma]{Definition}

\newcommand{\eps}{\varepsilon}
\newcommand{\unit}{1\!\!1}

\DeclarePairedDelimiter\ceil{\lceil}{\rceil}
\DeclarePairedDelimiter\floor{\lfloor}{\rfloor}

\newcommand{\ket}[1]{|#1\rangle}

\usepackage[backend=bibtex, citestyle=numeric-comp, sorting=none]{biblatex}
\fancyheadoffset{0cm} 

\newcommand\blfootnote[1]{%
  \begingroup
  \renewcommand\thefootnote{}\footnote{#1}%
  \addtocounter{footnote}{-1}%
  \endgroup
}

\usepackage[none]{hyphenat}
\begin{document}

\title{Practical randomness amplification and privatisation with implementations on quantum computers}

\author{Cameron Foreman}
\email{cameron.foreman@quantinuum.com}
\affiliation{Quantinuum, Partnership House, Carlisle Place, London SW1P 1BX, United Kingdom}
\affiliation{Department of Computer Science, University College London, London, United Kingdom}

\author{Sherilyn Wright}
\affiliation{Quantinuum, Partnership House, Carlisle Place, London SW1P 1BX, United Kingdom}

\author{Alec Edgington}
\affiliation{Quantinuum, Terrington House, 13–15 Hills Road, Cambridge CB2 1NL, United Kingdom}

\author{Mario Berta}
\affiliation{Institute for Quantum Information, RWTH Aachen University, Aachen, Germany$^{\dagger}$}
\affiliation{Department of Computing, Imperial College London, United Kingdom$^{\dagger}$}

\author{Florian J. Curchod}
\email{florian.curchod@quantinuum.com}
\affiliation{Quantinuum, Terrington House, 13–15 Hills Road, Cambridge CB2 1NL, United Kingdom}

\begin{abstract}
We present an end-to-end and practical randomness amplification and privatisation protocol based on Bell tests. This allows the building of device-independent random number generators which output (near\nobreakdash-)perfectly unbiased and private numbers, even if using an uncharacterised quantum device potentially built by an adversary. Our generation rates are linear in the repetition rate of the quantum device and the classical randomness post-processing has quasi-linear complexity\,---\,making it efficient on a standard personal laptop. The statistical analysis is also tailored for real-world quantum devices.

Our protocol is then showcased on several different quantum computers. Although not purposely built for the task, we show that quantum computers can run faithful Bell tests by adding minimal assumptions. In this semi-device-independent manner, our protocol generates (near\nobreakdash-)perfectly unbiased and private random numbers on today's quantum computers.
\blfootnote{$^\dagger$MB’s contribution was made when he was also with Cambridge Quantum (now Quantinuum).}
\end{abstract}

\maketitle

\tableofcontents

\section{Overview}

\subsection{Introduction}

Unpredictable numbers, often termed randomness or entropy, are the cornerstone of numerous applications in computer science. In cryptography for example, so-called keys need to be generated using (near\nobreakdash-)perfectly uniform and private numbers for secrecy not to be compromised. In quantum cryptography too, randomness is essential, e.g. in quantum key distribution in which distant parties consume local randomness to generate a shared secret key. In addition, randomness is a crucial resource for mathematical simulations such as Monte-Carlo techniques, for gambling, or for assuring a fair, unbiased, choice in a political context. Consequently, an equally fundamental and practical question is:
\begin{itemize}
\item[] \textit{How can one be sure that some generated numbers truly are unpredictable and private, i.e. (near\nobreakdash-)perfectly uniformly distributed and independent of an adversary's information?}
\end{itemize}

A possible approach is to aim directly at building a trustful random number generator (RNG). Such a device should then be characterised well enough to always function as promised or the security of the task could be compromised. 
Physical RNGs generate random numbers from a physical process that is either chaotic \cite{stipvcevic2014true} or quantum \cite{stipcevic2012quantum,herrero2017quantum}.\footnote{Pseudo RNG are mathematical functions expanding a short \textit{seed} of (near\nobreakdash-)perfectly random numbers into a larger output. They assume access to (near\nobreakdash-)perfect randomness as a resource and their security relies on computational assumptions. Hence, they are not of interest here.} The idea is then that the outcomes of this process are hard to predict or even, in the case of certain quantum processes, intrinsically random. Unfortunately, there are at least three problems following this approach:
\begin{enumerate}
\item An accurate model of the underlying physical process is necessary yet hard to build. It is also challenging to completely isolate the desired process from undesired noise or its environment. In short, hardware characterisation is difficult and prone to errors. Moreover, the RNG provider should be trusted or the device seriously inspected.
\item Many RNGs require an initial (near\nobreakdash-)perfect random seed as a resource\footnote{The only exceptions to this are RNGs that either directly output (near\nobreakdash-)perfect randomness without post-processing, or output such that a so-called deterministic randomness extractor can be used (see \cite{shaltiel2011introduction} for an introduction to randomness extraction). Such RNGs then need to be very precisely characterised and thus suffer from the drawback of the first point.}, the existence of which is very difficult to justify. 
\item Most RNGs do not offer security against a quantum adversary which might share quantum correlations with the device.
\end{enumerate}

An alternative approach is to accept that the direct building of an RNG outputting (near\nobreakdash-)perfect randomness is challenging, if not practically impossible. The objective is then to build a scheme in which a source of randomness is instead \textit{amplified} in a way that the amplified output is provably (near\nobreakdash-)perfectly random and private -- i.e. relaxing the need to build a trustful RNG directly. 

It is known that an imperfect source of randomness alone can not be amplified using classical processes without making strong assumptions about the source \cite{santha1986generating}.\footnote{More precisely, one can not amplify a Santha-Vazirani source \cite{santha1986generating} without added assumptions, see Sec.\ref{WSR} below.} This changes when one has access to quantum resources \cite{colbeck2012free}. Indeed, with the addition of a quantum device it is possible to perform \textit{device-independent randomness amplification and privatisation} \cite{Friedman17}. That is, an RNG whose output is neither uniform nor private to the user is amplified to generate provably (near\nobreakdash-)perfect uniform and private numbers \cite{gallego2013full,wojewodka2017amplifying,chung2016general,chung2014physical,Brandao16,ramanathan2015randomness,ramanathan2018generic,Friedman17}. The device-independent approach allows to \textit{certify} the random and private nature of the output without the need to model the internal functioning of the quantum device, which can essentially be seen as a black box and therefore requires minimal trust (see \cite{acin2016certified} for a review). This is an important feature especially in the field of quantum technologies since quantum hardware is notoriously noisy.

\subsubsection{The need for device-independence}

As opposed to building directly a trustful RNG, we follow the idea outlined above, namely, building so-called \textit{device-independent} protocols for randomness certification, in which only minimal assumptions are made on the quantum hardware that is used. By seeing the devices essentially as black boxes, it is possible to obtain lower bounds on the entropy generated without relying on a precise description of the internal functioning of the devices. Because of this, we can obtain a higher level of security that is mostly independent of hardware assumptions and therefore solving problem 1 mentioned above. This is possible because of the violations of Bell inequalities, which we explain in more detail in Sec.\ref{Sec:BellIneqs}. To understand the benefits of the device-independent certification approach that we follow, we give several examples of known attacks on existing cryptographic systems that are avoided. A famous example is the vulnerability discovered in dual EC, a pseudo-RNG that was favoured by the U.S. National Security Agency (NSA) and standardised by the National Institute of Standards and Technology (NIST). A weakness in the design allowed to predict future outcomes from a small sample of generated ones \cite{checkoway2014practical,bernstein2016dual,checkoway2016systematic}. More generally, numerous weaknesses and attacks on pseudo-RNG were found and implemented \cite{dodis2013security,heninger2012mining,kelsey1998cryptanalytic}. Attacks on physical RNGs based on non-quantum (e.g. chaotic) processes should also not be underestimated, as for example side-channel attacks \cite{zhou2005side} -- in which leaking information from the device is exploited -- or active implementation attacks, e.g. by injecting undetected errors to compromise the system \cite{boneh1997importance}. Finally, quantum hardware is also known to be vulnerable to different attacks. The popular quantum random number generator (QRNG) based on quadrature measurements on shot-noise limited states are susceptible to attacks if the hardware is not well characterised (or can not be trusted fully) \cite{thewes2019eavesdropping}. Quantum key distribution (QKD) systems have also been successfully attacked, for example by exploiting mismatches between theory and implementation \cite{gerhardt2011full}, but also by active light injection to extract useful information \cite{lydersen2010hacking,garcia2020attacking}. Another generic problem with quantum devices is that badly characterised measurement can lead to wrong claims, e.g. in state tomography or witnessing entanglement \cite{rosset2012imperfect}, opening avenues for systematic errors and potential attacks. Even without active attacks, QRNG requiring trust in their components are known to suffer from defects showing up in advanced statistical tests, see for example \cite{hurley2017quam,hurley2020quantum}.\\

In contrast, the approach that we follow offers the following features:
\begin{itemize}
\item Device-independence: the mismatch between the theory and the physical implementation is reduced to a minimum. In particular, the number of possible active implementation attacks is reduced and unavoidable implementation imperfections are allowed.
\item Continuous self-checking: the device tests itself continuously, certifying that the output is freshly generated and can be used safely. This check accounts for undesired effects, including experimental noise and tampering by an eavesdropper, leading to aborting the protocol if sufficient randomness and privacy can not be guaranteed. Silent failure and false claims are avoided.
\item Privacy: the protocol processes a public source of randomness -- one whose output is not private to the user -- into a provably private output, i.e. it generates and certifies privacy.
\item Composability: the random numbers can be used safely in another cryptographic application\footnote{We come back to the problem of re-using the devices when implementing a device-independent protocol below, when discussing the security definition, see Sec.\ref{Sec:SecurityDef}.}, for example in a public-key algorithm or a QKD protocol.
\item Quantum-proof security: the protocol is secure against an unbounded quantum adversary\footnote{Note that we do not consider a general no-signalling (NS) adversary for several reasons, including that it is not known how to perform randomness extraction in this case. The only technique that we are aware of is a reduction (with a nasty penalty) to using a classical-proof randomness extractor, as done in \cite{Brandao16}, i.e. this would imply sub-linear (in $n$) randomness generation rates, making it less suited to practical implementation.}, who can be entangled with the devices or use a futuristic powerful quantum computer.
\end{itemize}

\subsubsection{The advantages of randomness amplification}\label{sec:expansionvsamplification}

The device-independent (DI) framework that we follow allows for a very high level of security, as explained above. Nevertheless, the framework does not specify  \textit{what the protocol achieves}. Indeed, several DI tasks are possible and it is very important to understand their relevance in a cryptographic scenario. We now discuss two different tasks related to randomness generation: DI randomness amplification and expansion. We argue that, although randomness expansion is a useful task, randomness amplification is both strictly stronger and necessary in practice.

In contrast to an amplification protocol, in a randomness expansion protocol an initial seed of (near\nobreakdash-)perfect randomness is expanded into a longer output. It is therefore \textit{assumed} that initial (near\nobreakdash-)perfect randomness is available as a resource, an assumption that is often difficult to justify (and often not discussed). Moreover, in the task of randomness amplification, correlations between the RNG to amplify and the quantum devices are allowed. This reduces the need for an independence-type assumption between the devices which might be unjustified, especially if they share the same environment. Once generated from an amplification protocol, the (near\nobreakdash-)perfect randomness may then subsequently be used as the (now well justified) seed in an expansion protocol if desired, for example to increase the generation rates\footnote{One could have argued, in the past, that an advantage of randomness expansion was the possibility to generate private randomness from a non-private (aka public) source. This was true when randomness amplification protocol required the weak source to be private, but is not the case since the work of \cite{Friedman17} (and our work in consequence).}.

\subsubsection{The impossibility of cryptography with weak randomness}

An important resource needed in almost all cryptographic applications is the access to a (near\nobreakdash-)perfect source of randomness. This assumption, as discussed before, is generally very hard to justify in practice. Another important question is then ``what is the impact of using weak randomness in cryptography?''. In other words, what happens if the randomness is only somewhat unpredictable and not essentially perfectly random? We will see that it leads to security being compromised in many situations.\\

In \cite{dodis2004possibility}, the authors show that using randomness that is not near\nobreakdash-perfectly unbiased, i.e. for which every single bit is not almost unpredictable\footnote{This means that all min-entropy sources, even Santha-Vazirani sources \cite{santha1986generating}, are insufficient for the listed tasks.}, are insecure for encryption, bit commitment, secret sharing, zero-knowledge proofs (interactive or not) and two party computations. This result holds even against computationally bounded adversaries. In \cite{bosley2007does}, the authors show that in order to encrypt a message with unconditional security, one needs to be able to deterministically extract a (near\nobreakdash-)perfectly random string at least as long as the message from the randomness that is consumed -- which, as discussed before, is impossible with most weak sources of randomness alone \cite{santha1986generating}. The authors then generalise their results to include all privacy primitives that are perfectly or statistically binding, e.g. commitment or computationally secure private- and public-key cryptography. Other works also tackle the question of the impact of weak randomness, as for example \cite{mcinnes1990impossibility,dodis2002non,dodis2012differential}, but their results can be seen as special cases of the ones described before. 

Some positive results also exist, for example in tasks with differential privacy \cite{dodis2012differential} or for authentication \cite{dodis2004possibility} (intuitively requiring no secrecy). Finally, note that many of these results we have mentioned discuss the impact of a weakly random shared secret key -- i.e. shared randomness and not directly local randomness. We use those results to show the consequences of using imperfect RNGs to create (or distribute) those shared keys.


\subsection{Our results}

\paragraph{Theory and software for randomness amplification and privatisation.} The first part of our work amounts to the continuation of \cite{colbeck2012free,Brandao16,ramanathan2015randomness,Friedman17} on device-independent randomness amplification and privatisation (DIRAP) -- in particular we follow the techniques for the statistical analysis as in \cite{Friedman17}. Our main result is to give an end-to-end and practical DIRAP protocol.\\

The technical contributions in the first part of our work are:
\begin{itemize}
\item The Bell inequality and statistical analysis are optimised for real world quantum devices, using three quantum bits in an entangled state.
\item The resulting protocol has a large noise tolerance and, in the noiseless limit, can amplify all non-deterministic Santha-Vazirani (SV) sources (see Sec.\ref{WSR} for the definition of these sources). Note that the good noise resistance of our protocol is impossible to achieve in the simplest set-up for device-independence (with 2 parties, inputs and outputs).
\item The classical post-processing in the form of randomness extractors is designed, implemented and optimised for randomness amplification. In particular, we implemented several seeded and 2-source randomness extractors in near\nobreakdash-linear complexity and used the Number Theoretic Transform (NTT) for efficiency and security\footnote{Contrary to usual efficient implementations using the Fast Fourier Transform (FFT), the NTT does not suffer from potential rounding errors when implemented and allows us to maintain information-theoretic security throughout the entire protocol.}\,---\, a result of independent interest. This allows us to reach rates of several Mbits/sec for large block sizes using a standard personal laptop. We will make several of our randomness extractors software implementations available in a future work \cite{future}.
\end{itemize} \vspace{0.3cm}


\paragraph{Implementations of our protocol on different quantum computers.} The main objective of the second part of our work is to provide implementations of our protocol on real-world devices available today. Indeed, although they do not allow for a loophole-free Bell test, today's quantum computers are now widely accessible and awaiting real-world applications. Our protocol makes today's quantum computers useful to generate (near\nobreakdash-)perfect randomness for cryptography in a semi-device-independent manner in which the hardware is only partially trusted.\\

The contributions in the second part of our work are:
\begin{itemize}
\item We show that one can use today's quantum computers in order to run faithful Bell tests under minimal added assumptions -- making the implementation semi-device-independent only. For this, we develop methods to account for undesired signalling effect (e.g. cross-talk) in devices which do not close the locality loophole. At a high level, our method amounts to trusting that the quantum computer has not been purposely built to trick the user, but otherwise allows for the device to remain mostly uncharacterised.
\item We showcase our software with implementations on different quantum computers and different types of physical qubits: those from the IBM Quantum Services (superconducting), from Quantinuum (ion traps) and from AQT/UIBK (Univ. of Innsbruck based on AQT system; ion traps). By tailoring the Bell inequality, statistical analysis and circuit implementation, we obtain high Bell inequality values allowing our protocol to generate random numbers for cryptography. Our protocol can also be understood as a way of benchmarking and comparing the performances of the different devices.
\item We illustrate the quantum advantage of our protocol by showing that several pseudo-RNG (PRNG), a classical RNG based on a chaotic process and a commercially available QRNG are successfully amplified through our protocol. More precisely, we show that the numbers generated by the PRNG and QRNG fail at certain statistical tests, but pass them successfully once amplified by our protocol implemented on quantum computers. This suggests that, from a statistical perspective, our protocol was successful. To strengthen our results, we also show an example that the (weaker) classical alternative for randomness amplification, i.e. 2-source extraction on two PRNGs, is unable to generate numbers passing the statistical tests.
\end{itemize}


\section{Relation to previous work}

\subsection{Other works on device-independent randomness amplification}

The first ones to consider the task of randomness amplification were Colbeck and Renner, providing the proof-of-concept work \cite{colbeck2012free}. Later work focused on obtaining some noise resistance and the possibility to amplify imperfect sources with arbitrary bias \cite{gallego2013full}, but has the caveat of requiring an unrealistic number of devices, and having vanishing generation rates, making it unsuitable for implementations. In some other works \cite{wojewodka2017amplifying,chung2016general,chung2014physical}, more general correlations between the imperfect RNG and the quantum device are allowed\footnote{In our setup, we require that a Markov-type independence condition holds between the imperfect RNG and the quantum devices that are used for the Bell test during the execution of the protocol, see assumption 6 in Sec.\ref{assumptions} for the formal statement.} 
, although this comes at a high cost: amplification is possible for very small bias only ($\delta < 0.0144$ \cite{wojewodka2017amplifying} in \eqref{SVsource} below), large number of devices (polynomial in $1/\eps_{sec}$ \cite{chung2014physical} and exponential in $1/\eps_{sec}$ \cite{chung2016general}, where $\eps_{sec}$ is the protocol error, see \eqref{sec_def} below), low or no noise tolerance \cite{wojewodka2017amplifying,chung2016general,chung2014physical} and/or extremely computationally expensive processing steps (in \cite{chung2016general, chung2014physical} a quantum-proof randomness extractor with $2^d$ steps needs to be applied at the start of the protocol, to perform extraction on the imperfect RNG and every possible seed, where $d$ is the seed length). These issues make these works unsuited for implementation. The only works that could allow for a potential implementation are \cite{Brandao16,ramanathan2015randomness,ramanathan2018generic,Friedman17}. However, our work is the only one to offer all the following features:
\begin{itemize}
\item Our protocol is efficient: the randomness generation rates go linearly with the runtime of the quantum device. The only other work with this property is \cite{Friedman17}, although it is unclear if this protocol is practical to implement. The protocols in \cite{Brandao16,ramanathan2015randomness,ramanathan2018generic} would give at best an output that is sublinear in the runtime of the quantum device\footnote{This is because, in these works, the adversary is allowed to be post-quantum in the sense that it only respects the so-called no-signalling principle. Although this is in principle a potentially stronger security criterion than in our work, it implies only a sub-linear rate of randomness generation because of the absence of randomness extractors secure against such adversaries.}.
\item Our randomness post-processing has near\nobreakdash-linear complexity and was implemented using the Number Theoretic Transform, guaranteeing information-theoretic security and making it fast in practice on a standard personal laptop. This is not the case in all other works, which have generic polynomial complexity and/or generally use the Fast Fourier Transform (FFT) for efficiency, therefore having rounding issues opening up potential attacks.
\item We perform both randomness amplification and privatisation, as otherwise only done in \cite{Friedman17}. Although our statistical analysis relies on the results of \cite{Friedman17}, our work has the advantage of offering a much larger noise resistance, which is impossible to {be achieved} in the simpler setup that they consider. This noise resistance difference was clearly observed when we implemented the two different protocols on quantum computers{, allowing ours to be implemented in practice}. {A detailed explanation of our claim is given in Sec.\ref{Sec:BellIneqs}.}
\item As a consequence of all the other points, we could optimise and implement our protocol on real-world quantum computers.
\end{itemize}

As said before, our statistical analysis mostly consists of applying the latest techniques developed in \cite{Brandao16,Friedman17,arnon2018practical,Friedman16} to a set-up allowing for a practical implementation.

\subsection{Other quantum randomness ``generators'': one name, many meanings}\label{sec:otherQRNGs}

As discussed before, there exist other protocols that fall into the generic name of randomness \textit{generation}, in particular device-independent randomness expansion as discussed in Sec. \ref{sec:expansionvsamplification}. For randomness expansion, the seminal works of \cite{colbeck06,pironio2010random} were later followed by loophole-free Bell tests implementations \cite{bierhorst2018experimentally,liu2018device,shalm2021device}, with rates of about 180 bits/sec reported in \cite{liu2018device} and 3.6k bits/sec in \cite{shalm2021device} (albeit against adversaries with classical side-information only). This was made possible by building on the great efforts in closing all loopholes in Bell experiments \cite{hensen2015loophole,rosenfeld2017event,giustina2015significant,shalm2015strong}. As discussed previously, the difference between our work (randomness amplification) and the ones mentioned (randomness expansion) is that we relax the assumption that a source of (near\nobreakdash-)perfectly unbiased random numbers {is required to be used as a resource}, but also that no correlations exist between that source and the quantum devices' behaviour, i.e. that they are independent. Remark that both protocols for randomness expansion and amplification require the use of another RNG -- either as the weak source to be amplified or to provide the seed to be expanded -- and therefore, perhaps better understood as ``physical'' randomness extractors.\\

Another line of research is the work on semi-device-independence (SDI), in which the device-independent framework is followed but some additional assumptions are made about the devices. Some of these SDI protocols have the advantage that they do not require the generation of entanglement, greatly increasing the repetition rate of the device. Without entanglement, the set-up is the one in which a preparation device sends different quantum states to a measurement device -- there is therefore no space-like separation constraint and an additional assumption is needed. Examples of such added assumptions are a bound on the Hilbert space dimension \cite{brunner2008testing,gallego2010device,bowles2014certifying,lunghi2015self}, an overlap assumption \cite{brask2017megahertz} or a photon number type bound \cite{van2017semi,van2019correlations,rusca2020fast} on the prepared states. Those protocols, although interesting in terms of generation rates, require additional assumptions on certain components of the devices, making them less secure. Finally, different protocols have appeared in which some parts of the devices are treated in a device-independent manner, but the other parts still need to be fully trusted. An example is \cite{drahi2020certified}, in which the state of the source may remain uncharacterised but the rest of the device needs to be fully trusted (in particular the measurement device). In addition, all of these works focus on the task of randomness expansion. Although our protocol allows for a device-independent implementation, our implementation using quantum computers falls into the category of SDI, in the sense that additional assumptions are made (a detailed analysis of which is the subject of Sec.\ref{sec:QC_validity}). Comparing different SDI protocols is usually complicated and amounts to choosing the additional assumption(s) that are believed to be most valid.\\

Finally, the last category are ``standard'' quantum random number generators\footnote{As previously said, we are not interested in classical RNG, be it either pseudo-RNG or even physical RNG based on non-quantum physical processes, because those are ultimately deterministic (which does not mean that they are useless for certain applications, even in cryptography).} (QRNG), in which a quantum process is measured to generate random outcomes, see \cite{herrero2017quantum} for a review. Such QRNG require a high level of trust in the components and, as said before, are more prone to errors and implementation attacks.

\begin{figure}[!ht]\begin{center}
\scalebox{0.2}{\includegraphics{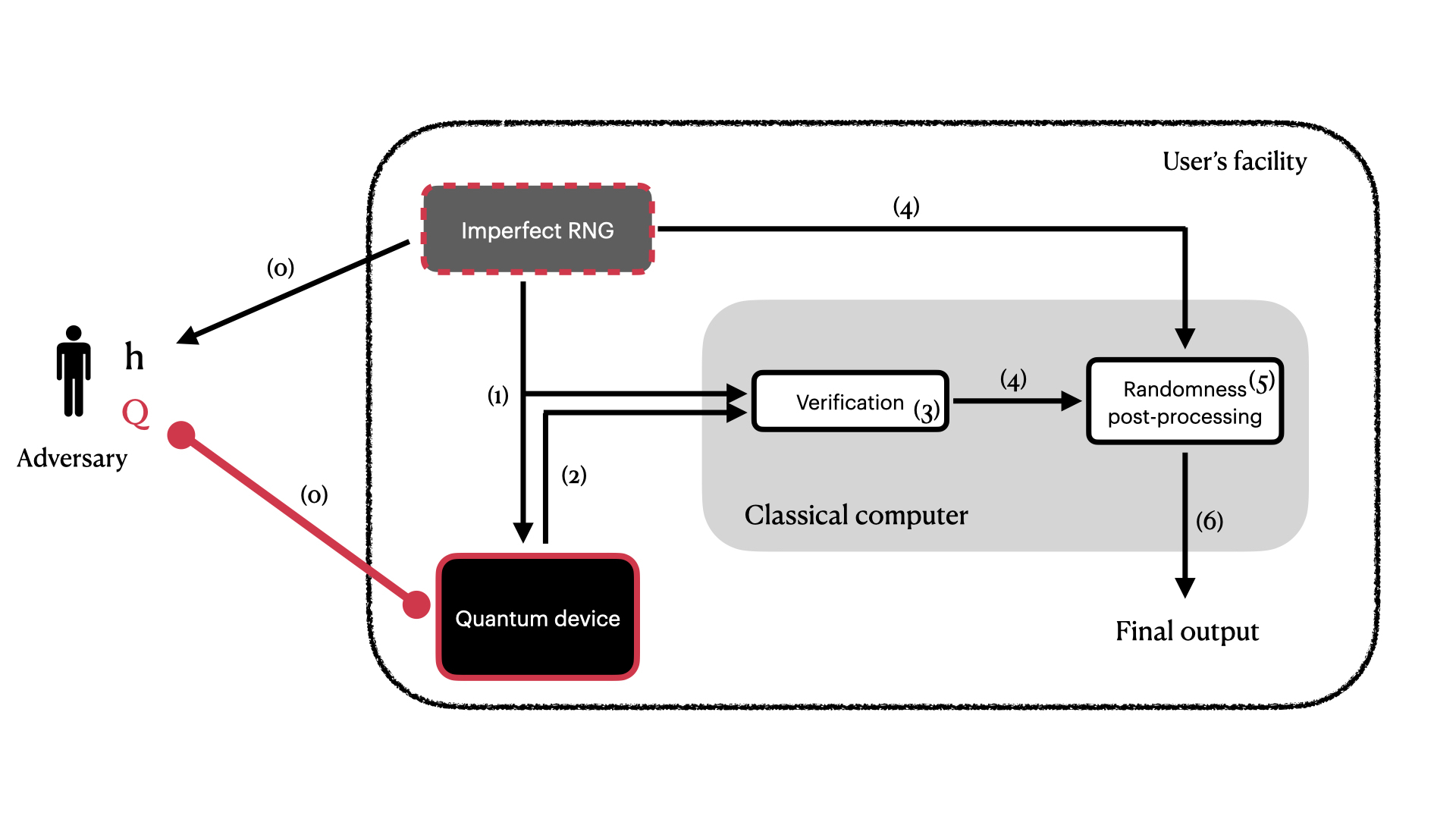}}
\end{center}
\caption{\label{Fig:DIRAsetup} 
The setup for device-independent randomness amplification is of the same type as in previous work \cite{colbeck2012free,Brandao16,ramanathan2015randomness,Friedman17}. The parts that require quantum hardware have been highlighted in red, i.e.~the quantum device and optionally the imperfect RNG. The user's facility is assumed to be in a safe environment shielded from the outside once the protocol starts. The steps are as follows: (0) Before the beginning of the protocol, the adversary may have received numbers generated by the imperfect RNG and generally a description of it, this is the history or side-information $h$. The adversary may also have built the quantum device, using the information $h$, with which it might still be correlated by storing quantum systems $Q$ that are entangled with it. (1) The imperfect RNG serves to challenge the quantum device by repeatedly sending it inputs. (2) The quantum device generates outputs each time that inputs are given to it. (3) After numerous interaction rounds, a verification is performed on the input-output statistics, which serves to \textit{certify} the unpredictability  of the device's outcomes. (4) Upon successful verification, the outcomes of the quantum device together with a fresh string of numbers from the imperfect RNG, are sent to the randomness post-processing step. (5) Classical algorithms process the two strings of numbers and output a near\nobreakdash-perfect random and private string of numbers\,---\,the final output of the protocol (6).}
\end{figure}


\section{Idea of the protocol}

The idea and main ingredients of the protocol should be understandable for non-experts in quantum cryptography. The technical material with all the details and proofs is deferred to Appendices \ref{APP:Bell} -- \ref{APP:Signaling}.

\subsection{Setup}

Our setting is depicted in Fig.~\ref{Fig:DIRAsetup}. In order to run a device-independent randomness amplification protocol, one needs an initial imperfect RNG, a quantum device capable of running a Bell test and a classical computer for storing data, performing the randomness verification step and post-processing.


\subsection{Interaction with the quantum device\,---\,data collection}

The first part of the protocol consists of collecting data which will serve to analyse the behaviour of the quantum device. It is the only step requiring quantum hardware. The quantum device is being driven in different settings, called inputs, and its response, called outputs{, are recorded}. Both inputs and outputs are saved for later analysis. After sufficiently many rounds of such interactions with the quantum device, one can build a faithful joint input-output probability distribution for the device\,---\,this is its \textit{observed behaviour} {that} will serve to certify that it truly generates unpredictable outputs.


\subsection{Randomness certification}

In the second step, the collected data is analysed in order to certify private randomness in the output of the quantum device. There exist certain input-output statistics that can only be generated by devices relying on specific quantum processes. Observing such distinctive statistics therefore serves as a certificate that the underlying process in the device truly is quantum. Furthermore, this opens up the possibility to show that the output of the device has some private randomness\footnote{Depending on the assumptions made, non-locality does not necessarily imply private randomness, but this can be the case in certain circumstances.}. Note that, with this approach, the user does not \textit{assume} that a specific implementation generates randomness from a quantum process, but instead \textit{verifies} it from the behaviour of the device. In the device-independent approach that we follow, this verification additionally only requires a minimal modelling of the internal machinery of the quantum hardware, which is essentially seen as a black box. The security of the protocol is then mostly independent of the implementation of the quantum hardware.


\subsection{Randomness post-processing}

The third and final step consists of \textit{extracting} the private randomness that has been certified in the outcomes of the quantum device. This step is performed by a classical computer. The outcomes of the quantum device, which are only partially private and random, are processed by algorithms on the classical computer together with a fresh string from the imperfect RNG. The function of these algorithms, or \textit{extractors}, is to transform the partially random and private strings into a shorter output that is near\nobreakdash-perfect.


\section{Main tools and ingredients}

\subsection{What is cryptographic randomness?}\label{Sec:SecurityDef}

The concept of randomness is present in numerous disciplines and its definition varies for different applications. Here we ask for the most stringent definition as given by randomness for cryptography. In particular, randomness in the cryptographic setting that we follow means that the generated output is unpredictable to any adversary that is only assumed to respect the laws of quantum physics. We do not, for example, rely on computational assumptions on the adversary (is it really random otherwise?). This unpredictability requires two concepts: uniformity and privacy. Indeed, even if used in a safe environment protected from the outside, a device generating a pre-determined sequence of numbers would not make a good RNG. The same applies to random numbers that are truly unpredictable when generated but immediately known to an adversary afterwards\footnote{As an example, imagine using a QRNG that is in reality only one half of a quantum key distribution (QKD) device, with the other half in possession of an adversary. The numbers, although truly unpredictable in advance, are then correlated to the ones generated in the other invisible half of the QKD device.}. In both cases, the numbers are not suited for cryptographic use.\\

As the security criterion, we ask that \cite{portmann2014cryptographic,renner2008security} the joint state of the user (describing the random numbers that are generated at the output of the protocol) and of the adversary is essentially indistinguishable from a state in which the user's state is uniform and uncorrelated to the adversary's:
\begin{align}\label{sec_def}
 \frac{1}{2} \left\|\rho_{\cal{UE}}-\bar{\unit}_{\cal{U}} \otimes \rho_{\cal{E}} \right \|_1 (1-p_{abort}) \leq \eps_{sec}{,}
\end{align}
in which $\cal{U}$ denotes the system of the user, $\cal{E}$ the one of the adversary, $\bar{\unit}_{\cal{U}}$ the (normalised) identity state on the user's side, and $||.||_1$ is the trace distance. The security of the protocol is conditioned on the probability of not aborting, $1-p_{abort}$. As an example, when the protocol does not abort and in the trivial case $\eps_{sec}=1$, there is no constraint on the joint quantum state $\rho_{\cal{UE}}$ of the user and adversary, which may therefore be correlated. Condition \eqref{sec_def} reflects the requirement that the adversary's system $\cal{E}$ be uncorrelated to the system $\cal{U}$ held by the user and that the state of the user is the uniform one, i.e. privacy and uniformity as discussed above. The security parameter $\eps_{sec} \in [0,1]$ quantifies the joint probability of not aborting and the probability of distinguishing the joint state $\rho_{\cal{UE}}$ from the ideal one $\bar{\unit}_{\cal{U}} \otimes \rho_{\cal{E}}$\,---\,even to an extremely powerful adversary possessing information $h$ and $Q$ about the quantum device (see Fig. \ref{Fig:DIRAsetup}). Note that the adversary is only assumed to respect the laws of quantum physics and is otherwise unbounded. It may for example possess a powerful futuristic quantum computer.\\

Importantly, this security definition is \textit{composable} \cite{portmann2014cryptographic}, which means that the generated random numbers can safely be used in another protocol. Note that composability for device-independent protocols only holds, strictly speaking, when the physical devices used to run the protocol are not used again (and that the adversary is never given access to the devices afterwards). Indeed, as noted in  \cite{barrett2013memory}, if the devices were to be re-used, for example to run the same protocol again, then there is in principle nothing forbidding a malicious device from storing information from the first execution and leaking it during the second one\footnote{This has nothing to do with using the devices repeatedly during the Bell test, which is not a problem.}. This obliges one to make the assumption that the devices do not use such memory effects between two executions of protocols using the same physical devices -- which is implicit when fully trusting the devices, but not in the device-independent framework that we follow.

Finally, because of this stringent definition, random numbers that are useful for cryptographic applications can also be used in all other applications such as mathematical simulations, computations, gambling, etc.


\subsection{Imperfect random number generators}\label{WSR}

The starting point of the protocol is an imperfect RNG that needs to be amplified into cryptographic randomness satisfying the definition in \eqref{sec_def}. We consider RNGs that output sequentially, i.e.~output bits $r_i \in \{0,1\}$ with $t(r_i)<t(r_{i+1})$ the time at which each bit is generated. Contrary to other approaches for randomness generation, as for example in randomness expansion, in randomness amplification those bits are not assumed to be completely unpredictable, private to the user nor distributed in an identical and independent way (the IID assumption). The starting assumption is that each bit is only \textit{somewhat} unpredictable, conditioned on the previously generated bits and on any additional information $h$ an external observer has about this source (see Fig.~\ref{Fig:DIRAsetup}). Following the literature, we say that such imperfect RNGs generate \textit{weak} randomness only\footnote{Throughout this manuscript, we use \textit{imperfect RNG} as synonymous of a device outputting as a \textit{weak source} satisfying the SV condition below, in \eqref{SVsource}.}. Such a weak source of randomness is called a Santha-Vazirani (SV) source and was first studied in \cite{santha1986generating}. The quality of {an} SV source is quantified by the parameter $\delta \in [0,\frac{1}{2}]$ such that:
\begin{equation}\label{SVsource}
\frac{1}{2}-\delta \leq p(r_i|\vec{r}_{i-1},h) \leq \frac{1}{2} + \delta \hspace{1cm} \forall i{,}
\end{equation}
where $\vec{r}_{i-1} = (r_{i-1},r_{i-2},...,r_{1})$ are all the bits that were previously generated and $p(r_i|\vec{r}_{i-1},h)$ denotes the probability of guessing outcome bit $r_i$ given the history $h$ and the previous bits generated during the protocol. It is known that it is impossible to amplify {an} SV source using classical processes without additional assumptions \cite{santha1986generating}. More precisely, it is impossible to process the outcomes of the SV source with $\delta > 0$ into an outcome string with $\delta' < \delta$. Additionally, in our work the output of the SV source is not assumed to be private. Such a \textit{public} source of randomness is one that is not perfectly predictable before the numbers are generated, but once generated these numbers are possibly known to anyone. An example of such a public source is a randomness beacon available on the internet. Such numbers are obviously not (directly) usable in cryptographic applications requiring privacy.\\

With additional quantum resources, it becomes possible to amplify SV sources \cite{colbeck2012free}. The objective of a protocol for randomness amplification and privatisation is to process the outcomes of a public SV source with parameter $\delta \in [0,\frac{1}{2}]$ into a final output that is provably near\nobreakdash-perfectly random and private, i.e. $\delta \rightarrow 0$ (in particular, this is the case if \eqref{sec_def} is satisfied for some $\eps_{sec}$). For this, one needs an additional quantum device.

\begin{figure}[!ht]\begin{center}
\scalebox{0.2}{\includegraphics{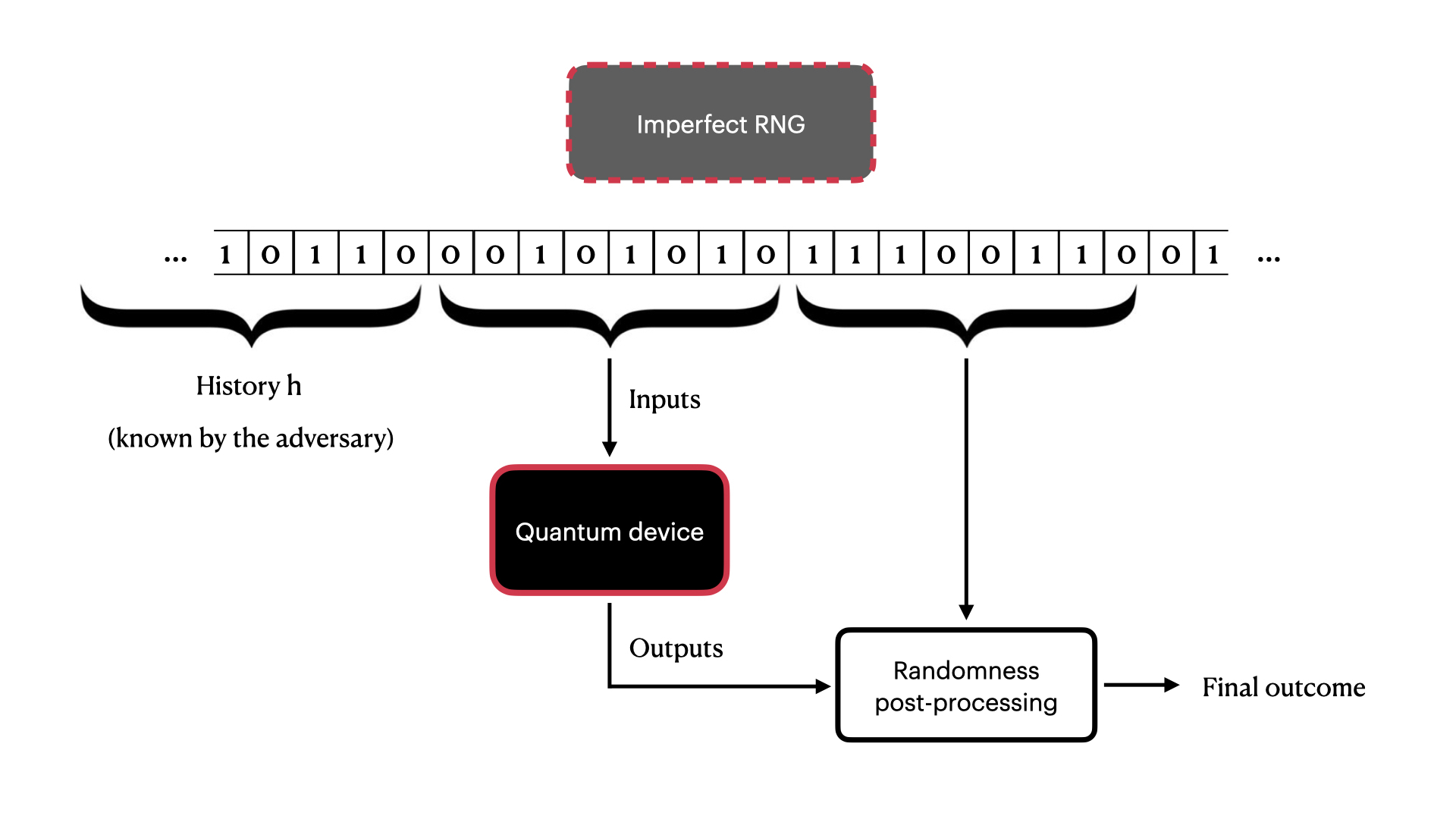}}
\end{center}
\caption{\label{Fig:ImperfectRNG} In a randomness amplification and privatisation protocol, the imperfect RNG is used twice: first to generate the inputs to drive the quantum device and then as an input to the randomness extractors. We assume that the external adversary had access to the imperfect RNG prior to the beginning of the protocol and hence holds information $h$ about it (see Fig.~\ref{Fig:DIRAsetup}). The quantum device might have been built using information $h$, for example to {partially} correlate its behaviour to the one of the imperfect RNG.}
\end{figure}


\subsection{Quantum devices, Bell tests, and guessing probabilities}\label{Sec:BellIneqs}

\begin{figure}[!ht]\begin{center}
\scalebox{0.17}{\includegraphics{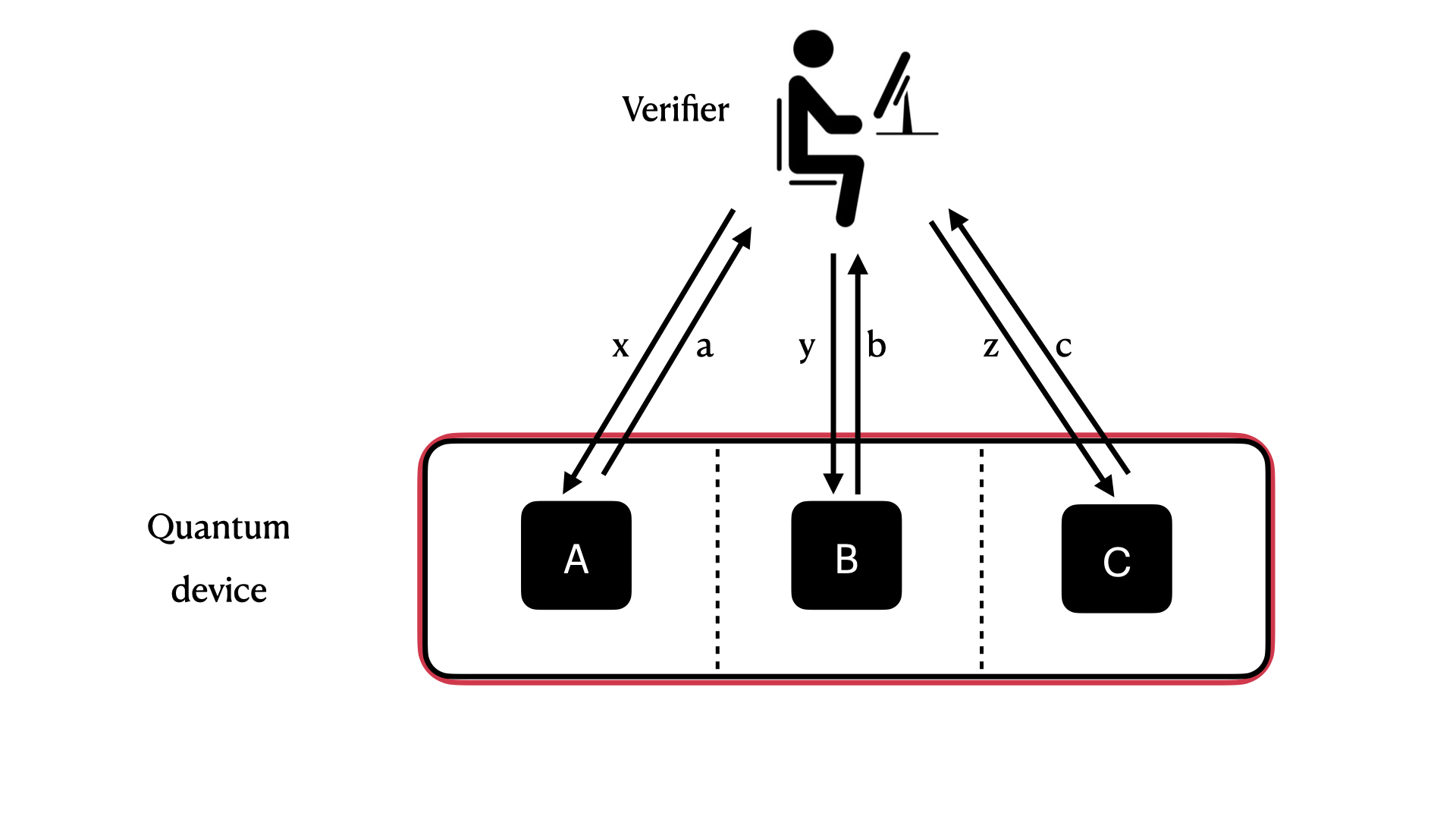}}

\end{center}
\caption{\label{Fig:QuantumDevice} The verifier makes rounds of interactions with the quantum device in order to analyse its behaviour. The quantum device is itself made of three separate parts $\cal{A,B,C}$ that are kept from communicating with each other during each interaction round (indicated by dashed lines), {for example by imposing spacial separation or shielding}. For every round, each of the three parts of the quantum device is being driven with fresh inputs $x,y,z$ and generates outputs $a,b,c$ which are recorded. After sufficiently many rounds, one can build a faithful estimate of the input-output distribution $\vec{P}_{obs} \equiv \{p(abc|xyz)\}^{a,b,c}_{x,y,z}$ of the three parts\,---\,its \textit{observed behaviour}. This behaviour is then later analysed in order to certify randomness in the outcomes of the quantum device.}
\end{figure}

The central building block of any device-independent protocol is the quantum device that is used together with the certification process associated to it. {In this work, the quantum device is composed of three parts that are shielded\footnote{{As opposed to imposing space-like separation, a form of physical shielding may be used to block communication, as is common practice in many types of cryptographic device.}} from each other or separated so that communication is impossible between them during each interaction round (see Fig.~\ref{Fig:QuantumDevice})}. The three parts are labelled, respectively, $\cal{A,B,C}$ and are seen as black boxes of which we do not model the internal functioning. The objective is to interact with these three boxes in order to \textit{verify} that they indeed make measurements on quantum states with certain properties, i.e. to discard any alternative classical (and hence deterministic) explanation for their behaviour. To do so, the verifier (the user) repeatedly chooses different inputs to the black boxes, which then generate outputs at each round. The inputs to the three boxes $\cal{A,B}$ and $\cal{C}$ are labelled, respectively, $x,y,z$ and the generated outputs of each box $a,b,c$. In our set-up, all variables are bits $x,y,z,a,b,c \in \{0,1\}$. After many rounds of such inputs-outputs interactions with the three boxes, one can estimate the joint conditional probability distribution
\begin{align}
\vec{P}_{obs} \equiv \{p(abc|xyz)\}^{a,b,c}_{x,y,z}
\end{align}
called the observed \textit{behaviour} of the device. In the device-independent approach that we follow, one is not allowed to rely on a description of the internal functioning of the boxes. Instead, everything needs to be done working with the observed behaviour $\vec{P}_{obs}$ alone. The objective is to build a quantum device which outputs in such a way that it proves to the verifier that it indeed relies on quantum processes. This verification by the user is done with a Bell test, i.e.~by evaluating a so-called Bell inequality. An ideal Bell test will be implemented in order to avoid the possibility of tricking the verification process, called a loophole (see \cite{brunner2014bell} for a review on Bell tests and in particular Sec.VII B about loopholes).

In our work, we evaluate the Mermin inequality \cite{mermin1990extreme}, which reads
\begin{equation}\label{Mermin_ineq}
M_{obs} \equiv {f}(\vec{P}_{obs}) = \langle A_0B_1C_1 \rangle + \langle A_1B_0C_1 \rangle + \langle A_1B_1C_0 \rangle - \langle A_0B_0C_0 \rangle \leq 2{,}
\end{equation}
where $\langle A_xB_yC_z \rangle \equiv \sum\limits_{a,b,c = 0,1} \big( p(a \oplus b \oplus c = 0|xyz) - p(a \oplus b \oplus c = 1|xyz) \big)$ and $\oplus$ {denotes} the sum modulo $2$.

The violation of the Mermin inequality $M_{obs} > 2$ is only possible when the three boxes share quantum systems in an entangled state on which they perform quantum measurements. Its violation therefore \textit{certifies} their true quantum nature from the observed statistics only. The advantage of using the Mermin inequality for randomness amplification is that, in the noiseless limit, a quantum device can reach the algebraic maximum {$M_{obs} = 4$}. This property is what allows our protocol to generate cryptographic randomness from any SV source that is not completely deterministic, i.e. $\forall \delta<\frac{1}{2}$ {in \eqref{SVsource}} (amplifying it to $\delta \rightarrow 0$).\\

{These properties of the Mermin inequality are what provided us with a practical advantage over the setup (and Bell inequality) used in  \cite{Friedman17}, in that we were able to amplify much weaker SV sources (i.e. larger $\delta$ in \eqref{SVsource}) on quantum computers. Indeed, the quantum correlations that need to be generated in \cite{Friedman17}, although requiring a simpler setup (with only two qubits), are closer to the set of classical correlations and therefore tolerate very low noise. As opposed to this, the correlations generated in our work lie in a region for which the distance to the set of classical correlations is greater -- in turn allowing for greater noise resistance. For example, if adding white noise, i.e. mixing the ideal (noise free) correlations with the uniform distribution, we get a maximal amount of $50\%$ fraction of tolerated noise in our work versus the maximal amount of $1-\frac{1}{\sqrt{2}}\approx 30\%$ in the setup with two players, inputs and outputs considered in  \cite{Friedman17} (see, e.g. \cite{cabello2005bell} for a discussion on this). {Note} that, in general, the tolerated noise might not compensate for the added complexity of the set up (e.g. adding qubits), but in our case it allowed for an implementation.}\\

In addition, using this inequality requires the evaluation of 4 input settings which can be described by just two input bits, $x,y$. Each setting can be constructed using the result that the last input bit is the sum modulo $2$ of the first two input bits, namely, $x,y,z=x \oplus y$. \\

In turn, from the violation of a Bell inequality it is also possible to bound the predictive power that the external adversary (modelled in Fig. \ref{Fig:DIRAsetup}) has on the outcomes of the boxes. This predictive power is formalised by the maximum guessing probability $P_g(ABC|x^{\ast},y^{\ast},z^{\ast},Q, M)$\footnote{We use the standard notation of using capital letters for random variables, e.g. output $A$, and minuscules to denote a particular realisation of that random variable, e.g. $a$ is short for  $A=a$.} {, which is the maximum probability that an external observer manages to guess the outcomes given some inputs $x^{\ast},y^{\ast},z^{\ast}$, a quantum system $Q$ which may be entangled with the device (see Fig.~\ref{Fig:DIRAsetup})  and for some $M$ which is calculated from the observed behaviour. Specifically $M$ is a function of $M_{obs}$ based on the protocol security analysis, such that $M \leq M_{obs}$, see the details in \eqref{adjustedM}.} Note that this guessing probability only concerns the outcomes of the quantum device and is different from the security of the \textit{final} outcomes of the protocol (after extraction) in \eqref{sec_def}. In our protocol we upper bound the adversarial guessing probability of the three outcomes from a known analytical bound \cite{woodhead2018randomness} on any two of the three outcomes since $P_g(ABC|x^{\ast},y^{\ast},z^{\ast},Q,M) \leq P_g(AB|x^{\ast},y^{\ast},z^{\ast},Q,M)$,
\begin{equation}\label{Pg_M}
\begin{array}{ll}
P_g(M) \equiv P_g(ABC|x^{\ast},y^{\ast},z^{\ast},Q, M)\leq P_g(AB|x^{\ast},y^{\ast},z^{\ast},Q, M) \\

\indent \leq \begin{cases} \frac{3}{4}-\frac{M}{8} + \sqrt{3}\sqrt{\frac{M}{8}(\frac{1}{2}-\frac{M}{8})} \hspace{1cm} \textrm{if } M \geq 3 \\
\frac{3}{2} - \frac{M}{4} \hspace{4cm} \textrm{if } 2 < M \leq 3\end{cases}{.}
\end{array}
\end{equation}
Note that $P_g(M)$ in \eqref{Pg_M} holds for all input triplets $(x^{\ast},y^{\ast},z^{\ast})$, so we took the freedom to get rid of them in the notation. The min-entropy of the outputs $A,B,C$ is then $H_{min}(ABC|x^{\ast},y^{\ast},z^{\ast},Q,M) = -\log_2(P_g(M))$, $\forall x^{\ast},y^{\ast},z^{\ast}$. In our set-up, it is important to note that the inputs are chosen from a public source of randomness, i.e. the inputs are known to the adversary in order to guess the outputs.


\subsection{Bell tests with inputs drawn from a weak source of randomness}

In Sec.\ref{Sec:BellIneqs}, we have implicitly assumed that the inputs $x,y,z$ were chosen independently of the device, i.e. we considered no possible correlations between the source of inputs and the device's behaviour. This is not the case in our scenario, in which we only have access to an imperfect RNG generating weak randomness and which moreover might be {partially} correlated to the quantum device through the adversary information $h$ and quantum systems $Q$ (see Fig \ref{Fig:DIRAsetup})\footnote{{Note that this is not possible if allowing the imperfect RNG and quantum device to be perfectly correlated, as illustrated by the related example in Sec.2 of \cite{wojewodka2017amplifying}.}}. In such a set-up, standard Bell inequalities, such as {$M_{obs}$} in \eqref{Mermin_ineq}, can not be used directly since it can not be assumed that the measurement inputs are chosen independently of the device's behaviour.\\

In order to circumvent this problem, we consider instead another type of inequality which is valid in our set-up with correlations between the inputs and the device. {This type of inequality was termed a \textit{measurement dependent locality} (MDL) inequality in \cite{putz2016measurement} and can serve to bound the possible values of $M$ which are compatible with the observed MDL inequality value (through the observation of $M_{obs}$). This $M$ then allows us to bound the adversarial guessing probability using \eqref{Pg_M}.} In other words, we bound the power given to the adversary when allowed to correlate the underlying state and measurements of the device with the source generating the inputs (or measurement choices) through the classical side-information $h$. We have detailed the derivation in App.\ref{app:mermin}.\\

As stated earlier, the Mermin inequality \eqref{Mermin_ineq} only requires input-output statistics from 4 (of 8 possible) input settings. This means we can use two weakly random input bits to select one of the four input settings, by generating the inputs as $x, y, z = x \oplus y$. This gives an improvement to the bound. Intuitively, this improvement in the bound can be seen to come from the adversary only getting additional guessing power of the input setting from 2 weak inputs versus 3 weak inputs.\\


The result is that one can use the following bound on the value $M$ for the {guessing probability} in \eqref{Pg_M}. This bound accounts both for the effect of choosing the inputs with an SV source  of quality $\delta$ and finite statistical effects,
\begin{equation}\label{adjustedM}
M \geq 4-\frac{4-M_{obs}+\Delta_f}{4(\frac{1}{2}-\delta)^2}\,.
\end{equation}
Here, $\Delta f$ denotes the width of the statistical confidence interval for the estimation test and we refer to App.\ref{app:mermin} for the technical details and computation.  Note that this bound is valid when $p_{obs}(x,y,z)=\frac{1}{4} \hspace{0.1cm} \forall x,y,z$ appearing in the Mermin inequality \eqref{Mermin_ineq}, i.e. the observed frequencies of each relevant input triplet is the same. This is easily generalisable to different observed inputs statistics (following App.\ref{SingleRound_APP}).


\subsection{Statistical analysis}\label{sec:stats}

The previous section explained how the outputs of the quantum device could be certified to contain some randomness and privacy. In this subsection, we evaluate how such randomness \textit{accumulates} through multiple rounds of the data collection process. We discuss everything in terms of the guessing probability $P_g(M)$ of the adversary, but this can also equivalently be understood in terms of min-entropy $H_{min}(M) = -\log_2 \big(P_g(M) \big)$, which is (roughly) the quantity that can be extracted during post-processing.\\

\paragraph{Identically and independently distributed rounds in the limit of large $n$.}\label{IID} In the case that the different rounds of interaction with the quantum device are assumed to be independent and identical (I.I.D.), then the probability $P_g(A^nB^nC^n|Q,h,M)$ of guessing the outcomes $(A^nB^nC^n)$ generated by $n$ uses of the quantum device is simply the product of the guessing probabilities $P_g(ABC|Q,h,M)$ of the outcomes generated at each round
\begin{align}\label{IIDminE}
p^{IID}_Q[n] \equiv P_g(A^nB^nC^n|Q,h,M) = \bigg(P_g(ABC|Q,h,M)\bigg)^n\,.
\end{align}
For details we refer to App.\ref{app:identical}. {Assuming that the quantum device behaves identically and independently may be a reasonable assumption in certain cases, for example if the device provider can be trusted and functions at very slow speed (possibly avoiding memory effects).}


\paragraph{Accounting for memory based quantum attacks.}\label{par:EAT} In the most general case, the adversary is allowed to perform memory based quantum attacks (MBQA). Indeed, assuming that a device built by an adversary behaves identically and independently each round might be {a too restrictive assumption}. {To generalise the results to account for the most general MBQA, we apply the entropy accumulation theorem as developed in \cite{dupuis2016entropy,arnon2018practical} to the Mermin inequality and the guessing probability described in Sec.\ref{Sec:BellIneqs}.} The result is that the guessing probability $P_g(A^nB^nC^n|Q,h,M)$ in $n$ uses of the quantum device is upper bounded as
\begin{equation}\label{EAT}
p_Q[n]\equiv P_g(A^nB^nC^n|Q,h,M)\,\leq\,2^{-nt+v\sqrt{n}}{,}
\end{equation}
where $v$ and $t$ are related to the single round guessing probability $P_g(ABC|Q,h,M)$ \eqref{Pg_M} as well as some other parameters\footnote{For example a smoothing parameter and a (small) error probability of the entropy accumulation routine.}, details on how $v$ and $t$ relate to the single round guessing probability are deferred to App.\ref{app:memory} (\ref{MBQA_entropy}), which is written in terms of min-entropy rather than guessing probability, using the fact that $H_{min}=-\log_2(P_g)$. This guessing probability can be loosely understood as the one that would be obtained assuming I.I.D. rounds as in \eqref{IIDminE}, giving the term $2^{-nt}$, but with a penalty multiplicative term $2^{v \sqrt{n}}$ to account for the most general attacks by the adversary and memory effects in the device. We refer to App.\ref{app:memory} for the details and computations of all parameters.


\subsection{Post-processing randomness}\label{Sec:RandProc}

\paragraph{Overview.} Whenever the verification was successful, the last step of the protocol is to turn the raw string of numbers that are hard to guess into bits that are (near\nobreakdash-)indistinguishable from perfectly random numbers in the sense of \eqref{sec_def}. This is achieved by post-processing the outcomes of the quantum device with so-called randomness extractors, which are classical algorithms from the theory of pseudo-randomness in theoretical computer science \cite{Vadhan07}. In this work, we use two different types of extractors: multi-source and seeded extractors. Multi-source randomness extractors take multiple sources of weakly random numbers and turn them into a shorter string of information-theoretically secure random bits defined in \eqref{sec_def} (see \cite{Li15,Li16} for the latest developments). Seeded extractors use a seed of (near\nobreakdash-)perfect randomness and another input from a weak source, outputting a secure bit string that is longer than the seed size. No quantum hardware is needed for the implementation of this last step.

{In order for our protocol to be secure against unbounded quantum adversaries, it is crucial to employ randomness extractors that are themselves secure against potential attacks from quantum adversaries. Such adversaries are malicious third parties that have quantum technologies at hand, for example allowing them to store information in a quantum memory \cite{Julsgaard04}. It is well-known that not all randomness extractors fulfil this stringent security criterion \cite{gavinsky2007exponential,kasher2012two} and for that reason we work in the quantum-secure Markov chain framework developed in \cite{Friedman16}. This allows us to build secure randomness extractors even in the presence of quantum adversaries.}

In Section \ref{assumptions} we collect the precise security assumptions of our model. For full technical details about the randomness post-processing discussed in this section, we refer to Appendices \ref{APP:math} -- \ref{app:Raz}.\\

\begin{figure}[!ht]
\begin{center}
\scalebox{0.2}{\includegraphics{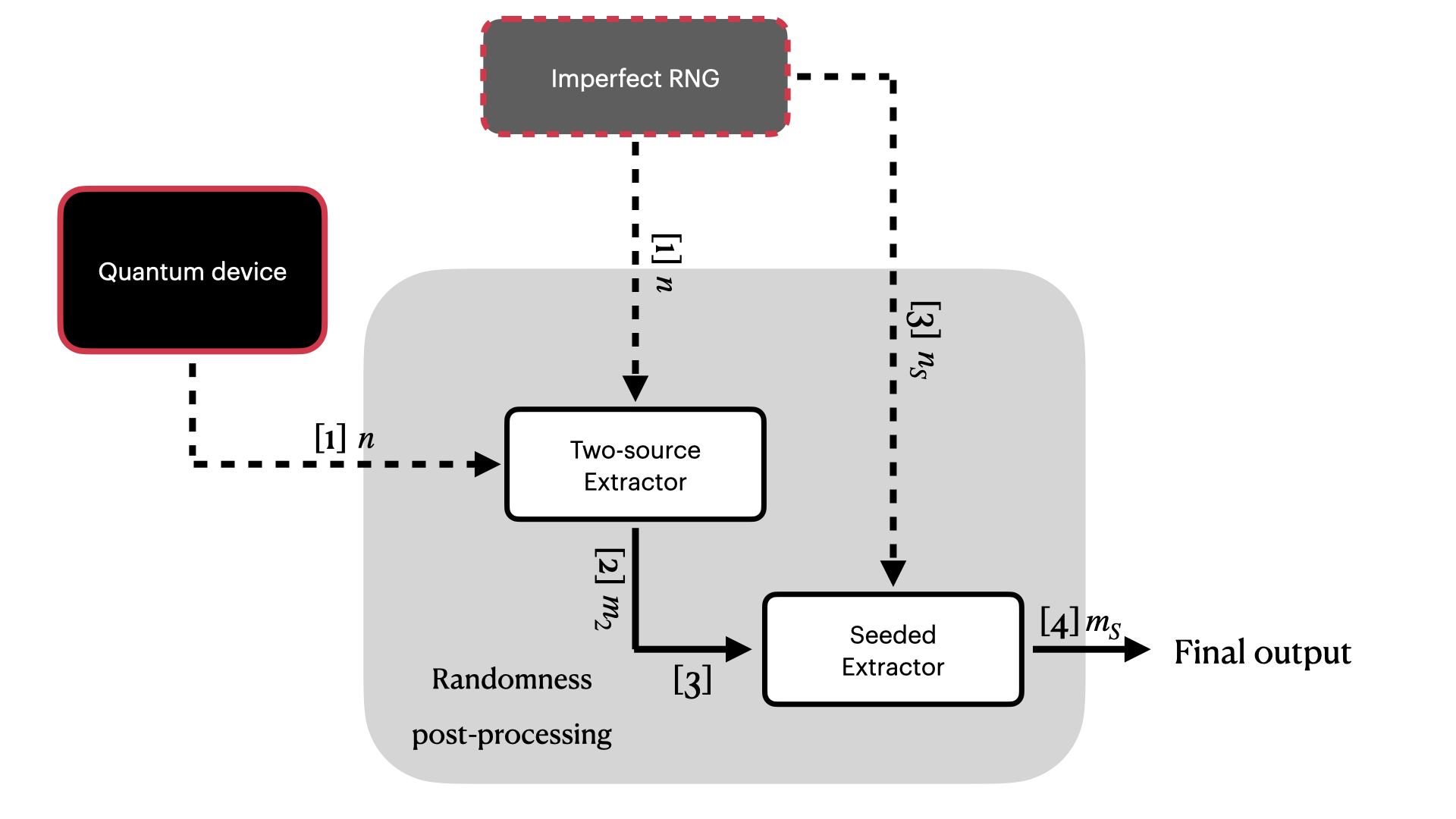}}
\end{center}
\caption{\label{Fig:RandProc_rev} The randomness post-processing flow (Box 5 in Fig.~\ref{Fig:DIRAsetup}) for {\it randomness amplification but not privatisation}. All steps are performed by mathematical functions on a classical computer: [1] The outcomes of the quantum device, together with a string of numbers from the imperfect RNG, are processed by a two-source randomness extractor. The two incoming bit strings are only somewhat hard to guess but not perfectly random in an information-theoretic sense\,---\,indicated by the dashed lines. [2] The two-source randomness extractor transforms the two input strings into a string of physically secure random numbers\,---\,indicated by the solid line. [3] The generated string of physically secure random numbers together with a string of numbers from the imperfect RNG, are processed by a seeded randomness extractor. [4] The seeded randomness extractor outputs an extended, final string of physically secure random numbers.}
\end{figure}

\begin{figure}[!ht]
\begin{center}
\scalebox{0.2}{\includegraphics{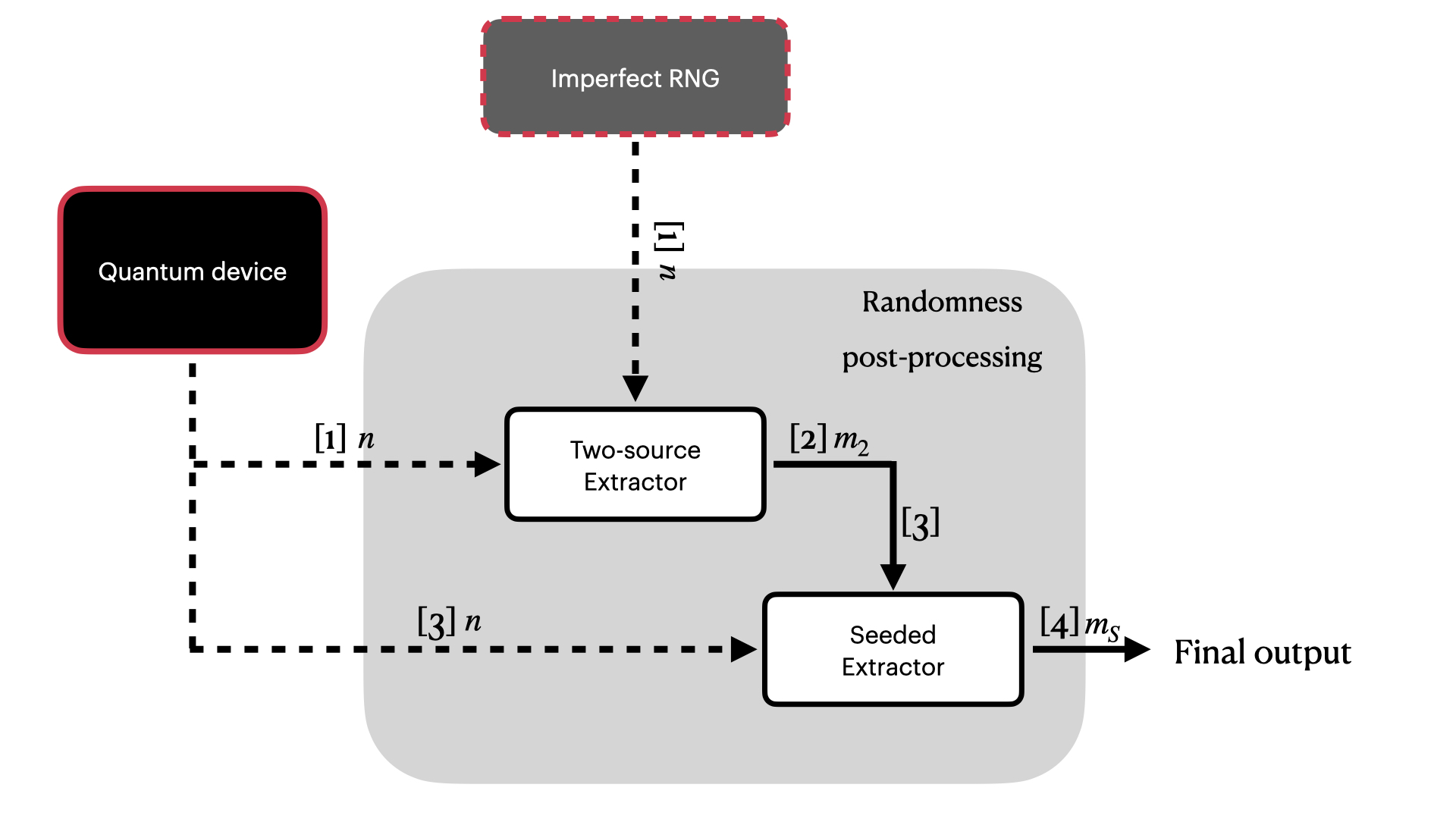}}
\end{center}
\caption{\label{Fig:RandProc} The randomness post-processing flow (Box 5 in Fig.~\ref{Fig:DIRAsetup}) for {\it randomness amplification and privatisation}. All steps are performed by mathematical functions on a classical computer: [1] and [2] are the same as for randomness amplification in Fig.~\ref{Fig:RandProc_rev}. [3] The outcomes of the quantum device, together with the generated string of physically secure random numbers, are processed by a seeded randomness extractor. [4] is the same as for randomness amplification in Fig.~\ref{Fig:RandProc_rev}.}
\end{figure}


\paragraph{Contribution.} We distinguish two slightly different tasks:
\begin{itemize}
\item Randomness amplification from private, imperfect RNGs as depicted in Fig.~\ref{Fig:RandProc_rev}.
\item Randomness amplification and privatisation from public (non-private), imperfect RNGs as depicted in Fig.~\ref{Fig:RandProc}.
\end{itemize}
For both tasks we describe the setting and the randomness extractors we have implemented. We follow the theoretical approach laid out in \cite{Brandao16}\,---\,together with the statistical analysis from \cite{Friedman17}.

For randomness amplification as in Fig.~\ref{Fig:RandProc_rev}, the output of the imperfect RNG is assumed to be private. We first feed the outcomes of the quantum device together with an additional string of bits from the imperfect RNG into a {\it two-source randomness extractor}. Second, the resulting short string of near\nobreakdash-perfect private and random bits is extended by means of a {\it seeded randomness extractor} using the bits from the imperfect RNG. For randomness and privacy amplification as in Fig.~\ref{Fig:RandProc}, the RNG is no longer assumed to be private. The first step of the protocol is identical, but for the second step we extend the resulting string of near\nobreakdash-perfect private and random bits by employing a seeded randomness extractor that uses the outcomes of the quantum device.\\

For the software implementation of these steps, it is crucial that the randomness extractors used do not only have a polynomial runtime in principle, but that they can be efficiently implemented in practice. In particular, sensible security parameters for realistic quantum hardware dictate the need for input blocks larger than approximatively $n\approx10^7$ bits in order to achieve non-zero output size when using the MBQA analysis. Furthermore, asking the post-processing to be done on a standard laptop machine only really leaves algorithms of {\it linear runtime} to be practically feasible. As such, the main contribution of our work on randomness extractors is twofold:
\begin{itemize}
\item We improve the complexity of some theoretically available randomness extractor schemes from a generic polynomial dependence to quasi-linear time $O(n\log(n))$ in the input size $n$.
\item We give explicit implementations of these algorithms based on the Number Theoretic Transform (NTT) \cite{Assche06}. In contrast to alternative schemes based on the Fast Fourier Transform (FFT) \cite[App.C]{Hayashi16}, the NTT has the advantage of being information-theoretically secure and therefore preventing potential attacks stemming from rounding issues related to the finite implementation of the FFT.\footnote{Concerning this issue, we also refer to the discussion in \cite[App.A]{Hayashi16}.}
\end{itemize}

Importantly, the software implementation of our randomness extractors reaches rates of the order of several Mbits/sec using a standard laptop machine with input blocks of $n\approx10^7$ bits. In fact, with our current code they can even be run with input sizes up to $n\approx10^9$ bits. In the following, we give a more detailed description of this implementation.\\


\paragraph{Santha-Vazirani source.} As mentioned in Sec.\ref{WSR}, we model the imperfect RNG as a Santha-Vazirani source \eqref{SVsource} with parameter $\delta>0$. Hence, any $n$ raw bits generated by the imperfect RNG can be guessed by the adversary with probability $p_{\text{SV}}[n]$ at most
\begin{align}
p_{\text{SV}}[n]\leq 2^{n\cdot \log_2 (1/2+\delta)}\,.
\end{align}
Thus, the probability of guessing an $n$-bit string generated by a Santha-Vazirani source decreases exponentially with $n$. All the logarithms are always taken in base 2 in this work. \\


\paragraph{Two-source extractor.}\label{Dodis} Our first extractor is based on the work of Dodis {\it et al.}, employing the near optimal\footnote{Namely, we lose a single output bit compared to the optimal theoretical construction.} cyclic-shift matrices approach for the construction \cite[Sec.3.2]{Dodis04}{, outlined in App.\ref{sec:two-source}}. For two $n$-bit input sources with a guessing probability of $p_{\text{SV}}[n]$ and $p_Q[n]$, respectively, the constructed two-source extractor secure against quantum adversaries\footnote{Note that if asking for security against a classical adversary, one can multiply the output in~\eqref{DodisOut} by roughly 5.} has output size
\begin{align}\label{DodisOut}
m_2[n]=\frac{1}{5}\left(-\log\Big(p_{\text{SV}}[n]\cdot p_Q[n]\Big)-8\log\frac{1}{\eps_{\text{sec}}}+9-4\log3-n\right)\,,
\end{align}
where $\eps_{\text{sec}}>0$ denotes the security parameter of the output string. That is, for sufficiently large block sizes $n$, this extractor allows to distil near perfect randomness roughly as soon as the sources have the quality
\begin{align}\label{eq:Dodis}
\text{$p_{\text{SV}}[n]\cdot p_Q[n]\leq2^{-n \cdot c}$ for {some} constant $c>1$.}
\end{align}
{For the details around the theory of the construction we refer to App.\ref{sec:two-source}.}

To put in some numbers, for our statistical analysis we have the guessing probabilities
\begin{align}
\text{$p_{SV}[n]\leq 2^{-n\cdot c_{SV}}$ and $p_Q[n]\leq 2^{-n\cdot c_Q}$ with constants $c_{SV},c_Q>0$}
\end{align}
and we then get an output string of perfectly random numbers of size roughly
\begin{align}\label{tworates}
m_2[n] \approx \frac{c_{SV}+c_Q-1-\xi}{5} \cdot n\,
\end{align}
with $\xi>0$ a free parameter relating the output size $m_2[n]$ to the security parameter of the extractor $\eps_{sec}[n]\approx2^{-\xi\cdot n/8}$. For example, if we obtain $c_Q = 0.35$ from an experiment, when combining this with an imperfect RNG of quality $\delta = 0.036$ ($c_{SV} = 0.9$), we find for the linear output rate
\begin{align}
m_2[n] \approx \frac{0.9+0.35-1}{5} \cdot n = 0.05 \cdot n\,.
\end{align}
The crucial technical step for the implementation of the Dodis {\it et al.}~extractor is efficient finite field multiplication in the binary Galois field $\text{GF}[2^n]$. For that, we employ the scheme proposed in \cite[App.D]{Hayashi16} that is based on the efficient algebra of circulant matrices via the NTT\,---\, resulting in quasi-linear complexity $O(n\log(n))$ for certain input sizes $n$. Even though this comes at the cost of some polynomial time pre-processing based on prime testing, we emphasise that this additional one-time step runs immediate in practice for the relevant range of parameters. For more we refer to App.\ref{sec:code-dodis}.\\


\paragraph{Seeded extractor.}\label{Hayashi} Our second extractor is based on an explicit implementation of the work of Hayashi and Tsurumaru \cite{Hayashi16}, that is known to be secure against quantum adversaries \cite[Sec.III.D]{Hayashi16}. These concepts were originally developed for quantum key distribution networks, but some adaptations make the work applicable to our settings. In particular, for an $n_S$-bit input source with a guessing probability quality $p[n_S]$ and a seed of $m_2=n_S-m_S$ bits of perfect randomness, the output size is\footnote{We notice that as opposed to, e.g., Trevisan based constructions \cite{trevisan01}, the seed size $m_2=n_S-m_S$ is not logarithmic in $n_S$. However, $m_2$ still gets small for applications with $m_S\approx n_S$.}
\begin{align}
m_S[n_S]=-\log p[n_S]-2\log\frac{1}{\eps_{sec}}-\log\left\lceil\frac{n_S-m_2}{m_2}\right\rceil\,,
\end{align}
where $\eps_{sec}>0$ denotes the security parameter of the output string. This leads to linear output rates as long as the guessing probability is
\begin{align}\label{eq:Hayashi-source}
\text{$p_S[n]\leq 2^{-\alpha\cdot n}$ for some $\alpha\in(0,1]$.}
\end{align}
Here, the input source of quality $p_S[n]$ may come from either the imperfect RNG or the quantum device, i.e.~depending on the application $p_S[n]\in\{p_{\textrm{SV}}[n],p_Q[n]\}$. For the details around the theory of the construction we refer to App.\ref{app:seeded}.

To put in some numbers, for a source as in \eqref{eq:Hayashi-source} we choose\footnote{The condition \eqref{seededrates} imposes that the source has guessing probability $p_S[n]=2^{-\alpha\cdot n}$ with $\alpha>1/2$ and hence only works with sources that are already sufficiently strong.}
\begin{equation}\label{seededrates}
\begin{aligned}
\text{$m_S=(c-1)\cdot m_2$ for some multiple $c\in \mathbb{N}$ with $c\leq \left\lfloor\frac{1}{1-\alpha}\right\rfloor$} \\
\text{and error $\eps_{sec} \leq \sqrt{c-1} \cdot 2^{-m_2(1+c(\alpha-1))/2}$.}
\end{aligned}
\end{equation}
For example, having $\alpha=9/10$ leads to $c\leq 10$ and for $c=9$ we get an output size $m_S = 8\cdot m_2$ with error $\eps_{sec} \leq10^{-150}$ for the seed size $m_2=10^4$.

Strongly building on the work of Hayashi and Tsurumaru \cite{Hayashi16}, our implementation is again based on the efficient algebra of circulant matrices via the NTT leading to quasi-linear complexity $O(n_S\log(n_S))$ for certain input sizes $n_S$ and seed sizes $m_2$. For details, we refer to App.\ref{sec:code-ht}.\\


\paragraph{Output rates.} We emphasise that for both our randomness extractors, we get linear output rates $m[n]\propto n$\,---\, see \eqref{tworates} and \eqref{seededrates}. As discussed, this comes from our statistical bounds on the guessing probability decreasing exponentially with the input block size $n$. We note that in the previous works \cite{Brandao16,ramanathan2015randomness,ramanathan2018generic}\,---\,secure against not only quantum but so-called non-signalling adversaries\,---\,the output is only of sublinear size.\\


\paragraph{Extensions.}\label{sec:Raz} Whereas our implementation thus far is fully explicit and efficient, it can not amplify two arbitrarily weak sources of randomness. Consequently, we consider the following extensions:
\begin{itemize}
\item For the two-source extractor, the construction of Raz \cite[Theorem 1]{Raz05} works for sources with lower quality than needed for the Dodis type construction in \eqref{eq:Dodis}. On paper, this would translate to a higher noise tolerance of the quantum hardware used and for that reason we improved the constants in the Raz construction for our specific applications. We find that for two $n$-bit input sources with a guessing probability quality of $p_{\text{SV}}[n]$ and $p_Q[n]$, respectively, the constructed two-source extractor secure against quantum adversaries has for any $\delta>0$ with 
\begin{align}
\text{$p_{\text{SV}}[n]\leq 2^{n\cdot \log_2 (1/2+\delta)}$ roughly an output size $m_2[n]=\frac{\delta}{18.5}\cdot\Big(-\log p_Q[n]\Big)$}{,}
\end{align}
for a security parameter $\eps_{\text{sec}}\leq\sqrt{3}\cdot2^{-1.375}\cdot2^{-m_2[n]/8}$ of the output string. Notice that in principle this allows for an arbitrarily low value in the guessing probability $p_Q[n]$ of the quantum source. For the details around the theory of the construction we refer to App.\ref{app:Raz}.

\item For the seeded extractor, Trevisan based constructions \cite{trevisan01} are known to be quantum-proof \cite{portmann09} and work with exponentially shorter seed size $m_2\approx\log(n_S)$ compared to the Hayashi-Tsurumaru construction with $m_2=n_S-m_S$. For some of our settings, this allows in principle the extraction of higher rates of randomness. Unfortunately, Trevisan based constructions come with the downside of a cubic runtime $O(n^3)$ in the input size $n_S$. Nevertheless, implementations of Trevisan based constructions have been optimised in \cite{Mauerer12}.

In particular, in the setting of randomness and privacy amplification (Fig.~\ref{Fig:RandProc}) employing a noisy quantum device generating outcomes with $p_Q \leq 2^{-\alpha_Q\cdot n}$ for $\alpha_Q<1/2$, a seeded randomness extractor capable of extracting from such a weak source is required. This is not the case for the implemented Hayashi-Tsurumaru construction, but can indeed be achieved with the off-the-shelf Trevisan based constructions from \cite{Mauerer12}.
\end{itemize}


\paragraph{Outlook.} It is important to further improve on the parameters of the implemented randomness extractors:
\vspace{-2mm}
\begin{itemize}
\item For the two-source extractor, Raz' construction is on paper again outperformed by Li's two-source extractor \cite{Li16}. It would be interesting to work out the practical efficiency of this construction. Importantly, this extension would allow the use of arbitrarily low quality SV sources.
\item For the seeded extractor, for further improvements one would need to show that the state-of-the-art constructions are secure against quantum adversaries. We refer to \cite{berta2016quantum} for an overview.
\item A follow-up work \cite{Mario2Source} implemented another efficient 2-source extractor based on the Toeplitz construction and made efficient using the FFT. An advantage of this implementation over ours is that the output size for quantum-proof security comes without a penalty (our output is roughly divided by $5$).
\end{itemize}


\subsection{List of assumptions}\label{assumptions}

For clarity, we here collect a list of all the assumptions needed to run our device-independent randomness amplification and privatisation protocol. One can find such a list in other works such as \cite{Friedman17}, to which we have added some additional assumptions necessary for a realistic implementation:
\begin{enumerate}
\item Quantum mechanics is correct and any potential adversary respects its laws. 
\item The classical computer that is used (see Fig. \ref{Fig:DIRAsetup}) functions properly.
\item The user's facility in which the protocol is run is shielded from the outside\,---\,in particular there is no back-door.
\item The quantum device is made of three separated parts which do not exchange information during a round of the experiment (see Fig.~\ref{Fig:QuantumDevice}){.}\footnote{{In the case of a loophole-free Bell tests, this assumption is enforced by physically separating the devices. Alternatively, this could be enforced by shielding the 3 parts from one-another. In our case this is natural since a shielding assumption is already required (assumption 3), however this shielding solution indeed needs to be enforced.}}
\item The adversary only holds classical information $h$ about the imperfect RNG{, that is, a (public) SV source}. Whenever explicitly stated, the output of the imperfect RNG is additionally assumed to be private -- in which case the protocol only performs randomness amplification (and not privatisation).
\item Given the classical and quantum information of the adversary{, denoted by $h$ and $Q$ respectively,} the imperfect RNG and the quantum device are independent, see App.\ref{APP:math} for the formal statement.
\item If the devices running the Bell test are later re-used, they do not leak any relevant information about previous protocols that were run on them\footnote{This has nothing to do with the fact that the devices are used multiple times to run the protocol.}, see the discussion in Sec. \ref{Sec:SecurityDef}.
\end{enumerate}
We note that without Assumptions 2 and 3 no cryptography would be possible. Assumption 1 has been generalised in some works \cite{Brandao16,ramanathan2015randomness,ramanathan2018generic} to adversaries who are not necessarily constrained by the laws of quantum mechanics, but only by the more general constraints of no-signalling\footnote{The no-signalling principle holds in quantum mechanics and states that information cannot be transmitted at infinite speed.}. We remark that, firstly, there is today no evidence that quantum mechanics is not correct and, secondly, the no-signalling generalisation obliges to reduce the output size to sublinear rates (and therefore severely reduces the efficiency of the protocol). Assumptions 4, 5 and 6 are related to our specific setting and are necessary to obtain security. Finally, note that Assumption 4 can {and should} be verified by minimal inspection of the device.

\begin{table}[h!]
\begin{center}
\begin{tabular}{ |p{15cm}| }
 \hline
 \multicolumn{1}{|c|}{\large Box 1: Randomness amplification and privatisation protocol } \\
 \hline \vspace{0.15cm}
 
\textbf{1. Data collection}\\
During $n$ rounds, do:
\begin{itemize}
\item[a.] Generate 2 bits $x,y$ with the imperfect RNG of quality $\delta$ defined in \eqref{SVsource}.
\item[b.] Drive the quantum device with settings $x,y,z = x \oplus y $ and collect the 3 output bits $a,b,c$. Save the 6 bits of that round.
\end{itemize}\\
\textbf{2. Verification}
\begin{itemize}
\item[a.] Compute the observed behaviour $P_{obs} \equiv \{p(abc|xyz)\}^{a,b,c}_{x,y,z}$ and observed Bell value $M_{obs} = M(P_{obs})$ using~\eqref{Mermin_ineq}.
\item[b.] If $M_{obs}$ is sufficiently high to verify the device's behaviour, continue to randomness post-processing, otherwise abort. 
\end{itemize}\\

\textbf{3. Randomness post-processing}
\begin{itemize}
\item[a.] Collect the three outputs, $a,b,c$, for each of the $n$ rounds. (Note: if device is I.I.D, only collect $a,b$ each round) 
\item[b.] This bit string, of size $3n$ ($2n$ if device is I.I.D), is sent to a two-source extractor together with a fresh string of $3n$ ($2n$ if device is I.I.D) bits from the imperfect RNG.
\item[c.] The two-source extractor outputs an  $m_2$-bit string of physically secure random numbers in the sense of the security definition in \eqref{sec_def}.
\item[d.] (Optional) The $m_2$-bit string is further expanded by a seeded extractor either re-using the string of outcomes from the quantum device (randomness amplification and privatisation) or another fresh string from the imperfect RNG (amplification only). The output $m_S$ of the seeded extractor is a larger bit string $m_S > m_2$ of physically secure random numbers.
\end{itemize}\\
 \hline
\end{tabular}
\end{center}
\label{ListProtocol}
\end{table}

\section{Protocol and concrete numerical examples}

We use this section to illustrate the results that can be obtained with our protocol. All results are first given directly at the output of the two-source extractor that we implemented and then other theoretical constructions for comparison. We then also give results when appending a seeded extractor to increase the output, and therefore randomness generation rates. Remark that our protocol can be used in two sensibly different manners: $a$- together with a \textit{public} imperfect source of randomness, it generates a near\nobreakdash-perfectly private and uniform output -- i.e. randomness amplification and privatisation; $b$- together with a \textit{private} imperfect source of randomness -- i.e. randomness amplification only. In case $b$, although relying on the stronger assumption that the imperfect source of randomness output is private to the user, one can then repeatedly feed a seeded extractor with fresh outputs from the imperfect source in order to obtain, after a latency, randomness generation rates linear in the output rate of the imperfect source. If we do not state explicitly that the results are for randomness amplification only, the results are given for the task of randomness amplification and privatisation.\\

At the end of this section, we give some results from a showcase of our protocol ran on today's available quantum computers. These results show that, under some added assumptions, these devices are capable of performing randomness amplification (and privatisation) today (the validity and implementation is discussed in Sec.\ref{sec:QC_implementation}). Then, in Sec.\ref{sec:stattests}, we show how we took several imperfect RNG consistently failing statistical tests and used our showcase protocol to obtain an output successfully passing statistical tests -- showing, from a statistical perspective, our protocol was successful in performing randomness amplification. The imperfect RNGs that we used were pseudo-RNGs, a classical hardware RNG built in house and a QRNG available commercially. We also show how an example classical alternative to our protocol fails, even though our showcase protocol is successful, illustrating the advantage of our protocol (and thus, the use of quantum resources) from this statistical perspective.

\subsection{Steps of the protocol}

For clarity, we have summarised the steps of our protocol in Box. 1. These are for the task $a-$ of randomness amplification and privatisation. The only difference when performing task $b$ (i.e. when the {imperfect RNG} is private) is the last step, where the seeded randomness extractor is instead fed with a freshly generated output from the imperfect source instead of the quantum device's output.


\subsection{Efficiency of the protocol}

\begin{figure}[!ht]\begin{center}
\scalebox{0.45}{\includegraphics{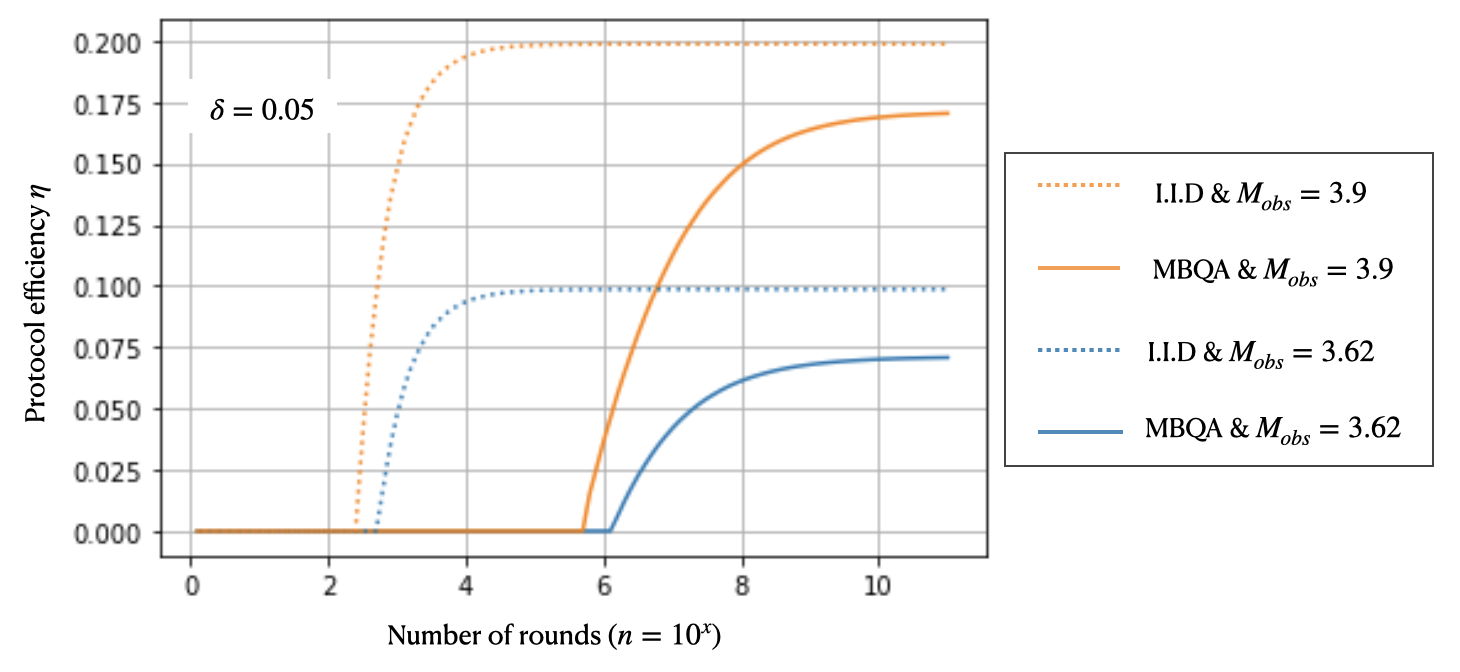}}
\end{center}
\caption{\label{Plot:IIDvsEAT} The protocol efficiency $\eta=\frac{m_2}{n}$ at the output of the 2-source extractor (i.e. without a seeded extractor), which is the number of output bits $m_2$ from the protocol per use of the quantum device, as a function of the total number of quantum device uses $n$ (where $n=10^x$). We chose $\delta=0.05$ -- corresponding to an imperfect RNG that is roughly 86\% random $M_{obs} = 3.62$ is the best Mermin value that we obtained from superconducting quantum computers and $M_{obs}=3.9$ is the best we attained from Ion-trap devices, see Table. ~\ref{ResultsQC}. MBQA: most general memory based quantum attacks and I.I.D: assumption that the rounds are identical and independent, see Sec.~\ref{sec:stats}.}
\end{figure}

\begin{figure}[!ht]\begin{center}
\scalebox{0.5}{\includegraphics{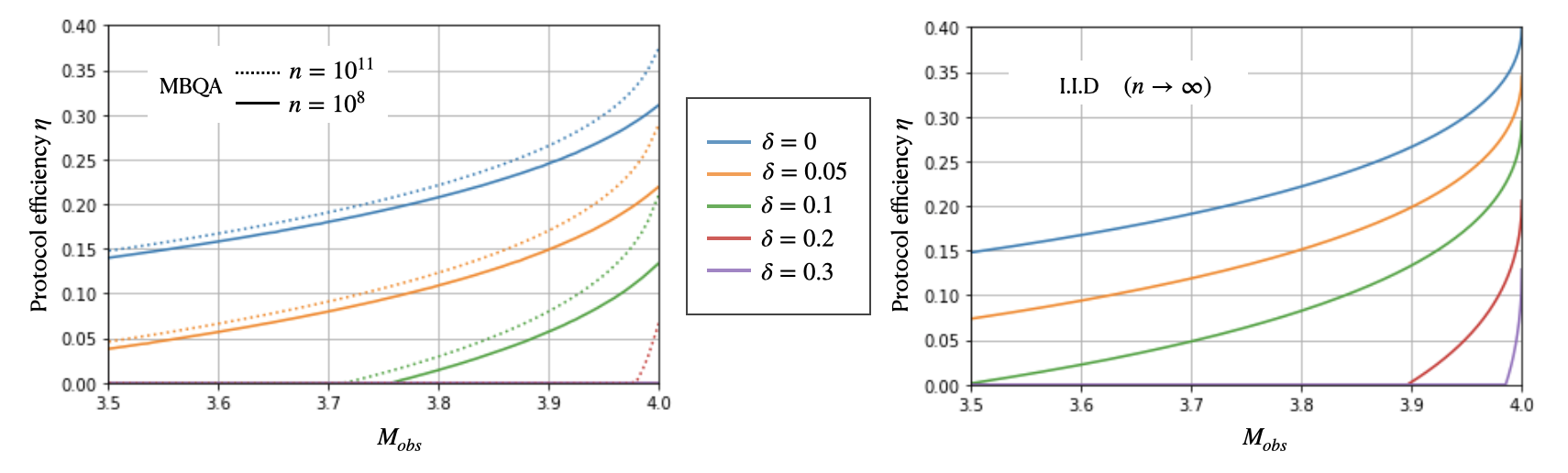}}
\end{center}
\caption{\label{Plot:OutvsEps} The protocol efficiency $\eta = \frac{m_2}{n}$ at the output of the 2-source extractor (i.e. without a seeded extractor) as a function of the observed Mermin value $M_{obs}$ for differing qualities of imperfect RNG ($\delta$). (On the left:) MBQA with $n=10^8$ when the lines are full and $n=10^{11}$ when the lines are dashed. (On the right:) I.I.D. in the asymptotic limit ($n\rightarrow \infty$). A similar plot of the protocol efficiency when using an appended (Trevisan's) seeded extractor is given in Fig.~\ref{Plot:seeded_efficiency}.}
\end{figure}

An important measure of the quality of our protocol is its overall \textit{efficiency} $\eta = \frac{m_2}{n}$ (or $\eta_S = \frac{m_S}{n}$ when appending a seeded extractor), i.e.~the total output size of the protocol $m_2$ (or $m_S$) divided by the total number of uses of the quantum device $n$. The derivation and exact formula for $m_2$ are given in \eqref{FinalEntropy_2} in App.\ref{Altogether_security}. The randomness generation rate of a particular implementation will then be the product of the repetition rate of the quantum device with the efficiency of the protocol. For what follows, we set $\eps_{sec} \leq 2^{-32}$ and $\Delta_f = 0$, where $\eps_{sec}$ is the total protocol security parameter and $\Delta_f$ is the penalty to account for finite statistics. These choices are made to for simplicity, since $\Delta_f$ decreases exponentially with $n$, and note that $\eps_{sec}\leq2^{-32}\approx 10^{-10}$.The efficiency $\eta = \frac{m_2}{n}$ as a function of the total number of uses $n$ of the quantum device is plotted in Fig.~\ref{Plot:IIDvsEAT}, and of $M_{obs}$ in Fig.~\ref{Plot:OutvsEps}. The results for the greater efficiency $\eta_S = \frac{m_S}{n} > \frac{m_2}{n}$ obtained by appending a seeded extractor is discussed in Sec.\ref{Sec:Improving_with_seeded} and plotted in Fig.~\ref{Plot:seeded_efficiency}. \\
As a reminder, $\delta=0$ corresponds to a source that is already perfectly random. $\delta=0.05$ (each bit can be predicted correctly with $0.45 \leq p(r_i|\vec{r}_{i-1},h) \leq 0.55$, see \eqref{SVsource}) means that the imperfect RNG is about $86\%$ random ($-\log_2(\frac{1}{2} + \delta)$), $\delta=0.1$ about $74\%$ random, $\delta=0.2$ about $51\%$ random and $\delta=0.3$ about $32\%$ random only.\\

\begin{figure}[!ht]\begin{center}
\scalebox{0.5}{\includegraphics{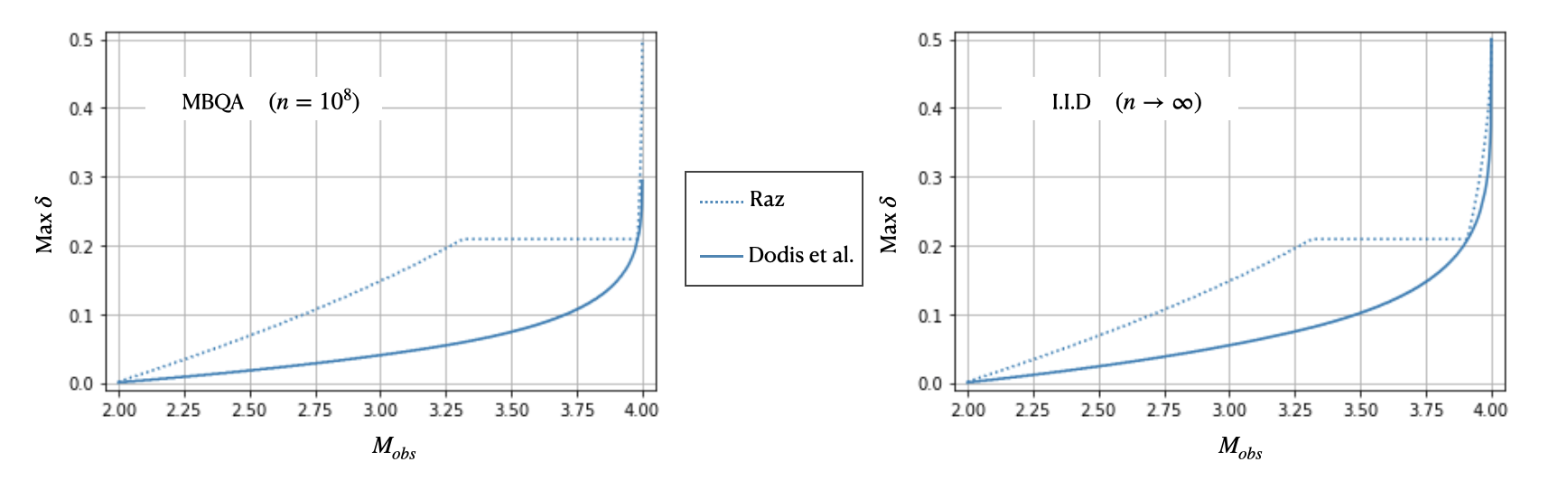}}
\end{center}
\caption{\label{Plot:RazVSDodis} The maximum $\delta$ that can be amplified as a function of the observed Mermin value $M_{obs}$ when different two-source randomness extractors are used: Dodis et al.~\cite{Dodis04} (solid lines) is our implemented extractor with near\nobreakdash-linear complexity and Raz' construction (dashed lines) is based on \cite{Raz05}. (On the left:) Memory based quantum attacks (MBQA) with $n=10^8$. (On the right:)  Identical and independent rounds (I.I.D.) in the asymptotic limit $n \rightarrow \infty$.} 
\end{figure}


\subsection{Maximum $\delta$ that can be amplified}

Fig.~\ref{Plot:RazVSDodis} shows a plot of the maximum $\delta$ that can be amplified as a function of the observed Mermin value. In this plot, it shows the case when using our implemented Dodis \textit{et al.} two-source extractor and compares it to what would be obtained if implementing the 2-source extractor from \cite{Raz05} (see Sec.~\ref{sec:Raz}). 
Remark that, although all parameters are explicit, we did not find an implementation of the extractor from \cite{Raz05}. Moreover, it does not have quasi-linear complexity but it can amplify larger $\delta$ in general.  \\
\\
As shown in the figure, full amplification and privatisation is possible in the I.I.D case using the Dodis two-source extractor, but not in the MBQA case, where the maximum is $\sim0.3$. Note the non-trivial behaviour of Raz's extractor in the region $M_{obs} {\gtrsim} 3.3 $, this is because the min entropy rate of one source must always be above $0.5$, which the imperfect RNG does not satisfy for $\delta {\lesssim} 0.21$. This means that, in order for any imperfect RNG's to be amplified with $\delta {\lesssim} 0.21$, the quantum output must have min entropy rate above $0.5$, which only happens close to $M_{obs} {\gtrsim} 3.97$ in the MBQA case for $n = 10^8$ and $M_{obs} {\gtrsim} 3.9$ in the I.I.D case.


\subsection{Improving the efficiency by appending a seeded extractor}\label{Sec:Improving_with_seeded}
In order to increase the protocol efficiency and therefore the generation rates too, one can append a seeded extractor as explained in Sec.\ref{Sec:RandProc} and depicted in Fig. \ref{Fig:RandProc_rev} and \ref{Fig:RandProc}. We present the results with both our implementation based on \cite{Hayashi16} and using a Trevisan-based construction for randomness amplification and, optionally, privatisation.\\

\begin{figure}[!ht]\begin{center}
    \scalebox{0.25}{\includegraphics{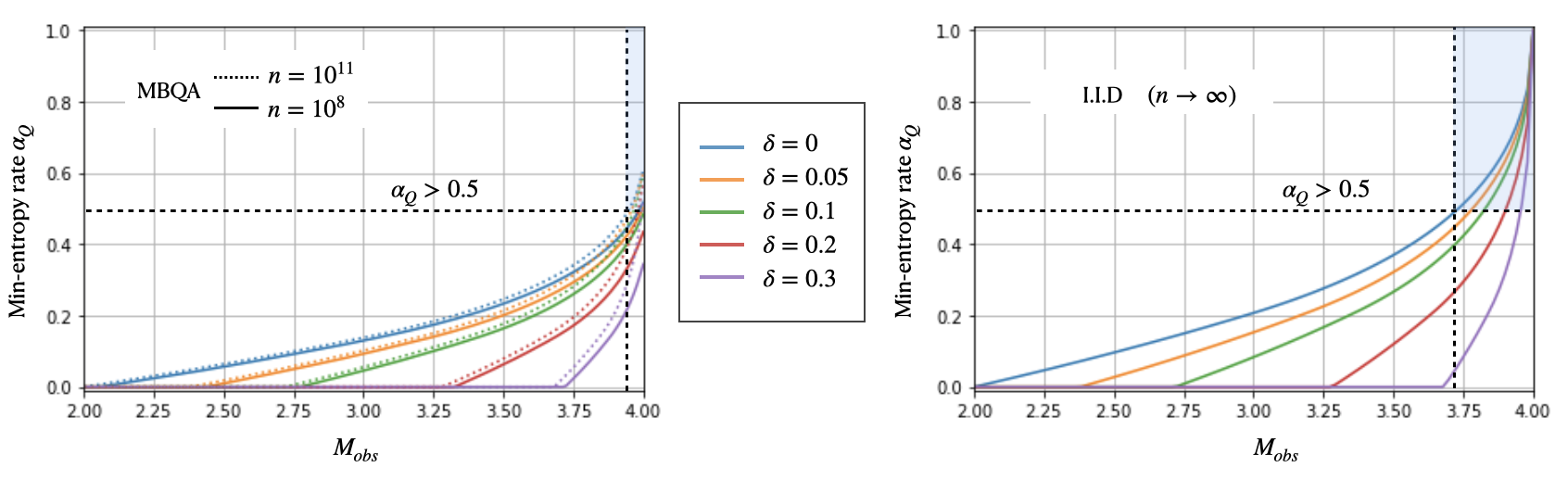}}
    \end{center}
    \caption{\label{Plot:GoodRNG} The min-entropy rate ($\alpha_Q \in [0,1]$) of the outputs of the quantum device as a function of the observed Mermin value $M_{obs}$, for different qualities of imperfect RNG, $\delta$. Note that, for MBQA we obtain 3 outputs per round which must go to post-processing and for I.I.D. we obtain two outcomes per round which go to post-processing. (On the left:) Memory based quantum attacks (MBQA) with $n=10^8$ when the lines are full and $n=10^{11}$ when the lines are dashed. Here $\alpha_Q = - 1/3n \cdot\log_2(p_Q[n])$, where $p_q[n]$ is defined in \eqref{EAT}. (On the right:)  Identical and independent rounds (I.I.D) in the asymptotic limit $n \rightarrow \infty$. Here $\alpha_Q = - 1/2n \cdot \log_2(p_Q[n])$, where $p_q[n]$ is defined in \eqref{IIDminE}. The blue highlighted region illustrates when $\alpha_Q > 0.5$.}
\end{figure}

In Fig.~\ref{Plot:GoodRNG}, we have plotted the amount of randomness per use of the quantum device, as a function of the observed Mermin value $M_{obs}$, for different values of $\delta$. The region in which the quantum device is good enough to apply our implementation of an efficient seeded extractor is highlighted in blue\,---\,in this region the post-processing has quasi-linear complexity only.

\paragraph{Our implementation based on \cite{Hayashi16}.}
This construction has the advantage of having quasi-linear complexity, allowing it to run on a standard laptop with relevant block sizes. The drawback of our construction is that it requires that one of the sources has a high min-entropy rate -- namely, a guessing probability of $p_S[n] \leq 2^{-\alpha \cdot n}$ with $\alpha\geq 0.5$. The entropy gain $m_S-m_2$, where $m_S$ is the output of the seeded extractor and $m_2$ of the 2-source extractor, when appending this seeded extractor, can then be computed roughly as $m_S-m_2 = (c-1) \cdot m_2$ with $c = \floor{\frac{1}{1-\alpha}}$. For example, if $\alpha= 0.9 + \delta > 0.9$ with any $\delta>0$, then the output of the seeded extractor is $m_S = 9 \cdot m_2$, i.e. at least 9 time larger than the one of the 2-source extractor. The error added is then basically negligible for our typical choice of parameters and block size, see \eqref{seededrates}. This is particularly advantageous in the setup for randomness amplification without privatisation. Note that in our case, to re-use the quantum device's output as input to the seeded extractor in order to perform randomness amplification and privatisation, requires $\alpha_Q>0.5$, which occurs when $M_{obs}{\gtrsim} 3.9$ in the MBQA case and $M_{obs}{\gtrsim} 3.72$ in the I.I.D case, see the blue highlighted region in Fig.\ref{Plot:GoodRNG}.\\

\begin{figure}[!ht]\begin{center}
\scalebox{0.45}{\includegraphics{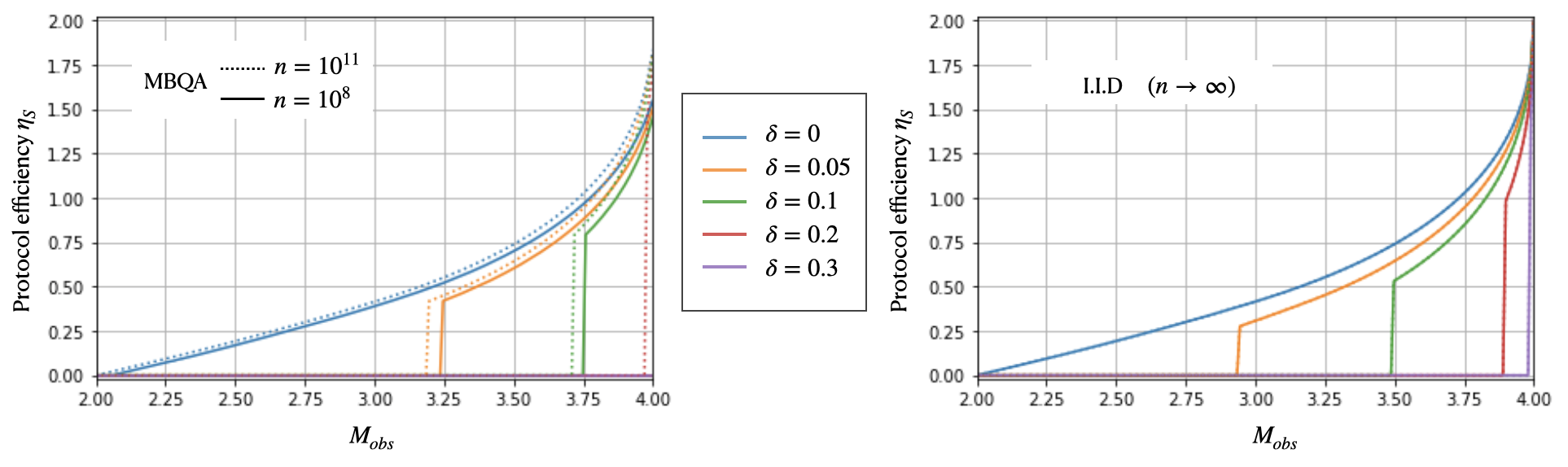}}
\end{center}
\caption{\label{Plot:seeded_efficiency} The protocol efficiency $\eta_S = \frac{m_S}{n}$ -- i.e. the total number of generated random bits per use of the quantum device -- at the output of the protocol with Trevisan's extractor as a function of the observed Mermin value $M_{obs}$ for different $\delta$ and $n$. (On the left:) Memory based quantum attacks (MBQA). (On the right:) Identical and independent rounds (I.I.D) in the asymptotic limit $n \rightarrow \infty$. Note that $\eta_S \in [0,2]$.}
\end{figure}

\paragraph{Trevisan-based construction.}
This construction of seeded extractors has higher complexity $O(n^3)$ but has the benefit of extracting roughly all entropy from the source using a short seed only. More precisely, when using Trevisan's extractor one can extract $\alpha_Q \cdot n-4\log\frac{1}{\eps} -6$ bits from a quantum output with total min-entropy $\alpha_Q \cdot n$, requiring a seed (the output $m_2$ from the 2-source extractor) of size logarithmic in $\alpha_Q \cdot n$. For example, when using the modular implementation in \cite{Mauerer12}, given a source with min-entropy $k$, one obtains an output of size $m_S = k-4\log\frac{1}{\eps} -6$ bits where $\eps$ is the error per bit. For randomness amplification and privatisation, the protocol efficiency $\eta_S = \frac{m_S}{n}$ then becomes roughly $\eta_S \approx \alpha_Q$. The protocol efficiency $\eta_S$ with Trevisan's extractor as a function of the observed Mermin value $M_{obs}$ and different $\delta$ is plotted in Fig.\ref{Plot:seeded_efficiency}, for both the MBQA and I.I.D cases.

\newpage 

\subsection{Protocol efficiency and randomness generation rates in concrete examples}

We discuss three concrete examples:
\begin{itemize}
\item \textbf{The previous highest violation of the Mermin inequality from \cite{maxMermin}.} 

The value we found is $M_{obs}=3.57$, dating back to 2006. This value, when using our protocol already allows the amplification of an imperfect RNG of parameter $\delta {\leq} 0.11$ (i.e.~roughly $71\%$ random) and an overall protocol efficiency between $\eta = 5\%$ and $\eta=9\%$ with $\delta=0.05$ depending on the number of rounds and whether the I.I.D. assumption is made (and without a seeded extractor). With  Raz' extractor \cite{Raz05} as discussed in Sec.~\ref{Sec:RandProc}, one can amplify a WSR with $\delta\leq 0.2$ (i.e.~roughly $50\%$ random).

Appending a seeded extractor, one can then obtain a final protocol efficiency, $\eta_S$ of between around $60\%$ ($\delta=0$) and $75\%$ ($\delta=0.05$) efficiency in the MBQA case. In the I.I.D case, one can obtain around $72\%$ ($\delta=0$) and around $80\%$ ($\delta=0.05$) efficiency.

\item \textbf{Our semi-device-independent implementations on quantum computers.} 

\begin{table}
\begin{center}
\begin{tabular}{ |p{2.5cm}||p{2.7cm}|p{1cm}|p{1.3cm}|p{1.3cm}|p{1cm}|p{1cm}|}
 \hline
 \multicolumn{7}{|c|}{Results from our quantum computer implementations} \\
 \hline
Device & Type & $M_{obs}$ & $\max \delta$ (Raz) & $\max \delta$ (Dodis) & $\eta$ & $\eta_S$\\ 
\hline \hline
 \textbf{AQT/UIBK}  & Ion-trap & 3.9 & 0.209 & 0.132 & 0.15 & 1.16 \\ \hline
 \textbf{Quantinuum H1}  & Ion-trap & 3.88 & 0.209 & 0.127 & 0.14 & 1.11 \\ \hline
 Quantinuum H0 & Ion-trap  & 3.83 & 0.209 & 0.114 & 0.12 & 1.01\\ \hline
 \textbf{IBM \textit{ibmq\textunderscore toronto}} & Superconducting  & 3.62 & 0.209 & 0.082 & 0.06 & 0.72\\ \hline
 IBM \textit{ibmq\textunderscore ourense} & Superconducting & 3.35 & 0.203 & 0.058 & 0.015 & 0.49\\ \hline
 IBM \textit{ibmq\textunderscore valencia} & Superconducting & 3.11 & 0.149 & 0.043 & 0 & 0\\ \hline
\end{tabular}
\caption{Table of results containing, for different quantum computers, the observed Mermin value, the maximum $\delta$ that can be amplified and the protocol efficiencies with or without a seeded extractor. The samples collected from ion-trap devices were small compared to the ones we could collect from superconducting ones: $n=6 \cdot 10^4$ (AQT/UIBK), $n=4 \cdot 10^4$ (Quantinuum H1), $n=3 \cdot 10^4$ (Quantinuum H0) and 50 experiments of $n=10^7$ on each IBM device. The maximum $\delta$, the protocol efficiency $\eta$ (the number of random output bits per use of the quantum device, without appending a seeded extractor) and $\eta_S$ (appending Trevisan's seeded extractor) are given for $n=10^8$, $\eps_{sec}=2^{-32}$, $\Delta_f=0$ and $\delta=0.05$ and the most general MBQA attacks. Specifically, this means we did the statistical analysis as if we had attained $M_{obs}$ from $10^8$ rounds for each device. These parameters were picked based on practicality: they are useful for applications and attainable on quantum computers today.}\label{ResultsQC}
\end{center}
\end{table}

The results of our implementations on quantum computers are summarised in Table. \ref{ResultsQC}. We give the details of these implementations, discuss the validity and added assumptions required when running Bell tests on quantum computers in Sec.\ref{sec:QC_implementation} below. These implementations, because of those added assumptions, are semi-device-independent only.\\

On the superconducting quantum computer \textit{ibmq\textunderscore toronto} of IBM we obtained the value of $M=3.62$. In the MBQA case, this allows the amplification of an imperfect RNG of quality $\delta\leq 0.082$ and a protocol efficiency $\eta= 6\%$ when $\delta = 0.05$, without a seeded extractor. The implementations on Quantinuum's ion-trap device H1 gave the Mermin value $M = 3.88$ and the ion-trap device of AQT/UIBK gave $M = 3.9$ -- the highest reported in the literature, allowing randomness amplification with our 2-source extractor implementation up to $\delta=0.132$ and an efficiency of $15\%$ for $\delta=0.05$, without a seeded extractor (again, depending on the assumptions and number of rounds, see Fig.\ref{Plot:IIDvsEAT}). These are lower bounds on $\delta$ and efficiencies $\eta, \eta_S$, which can be improved by increasing the number of rounds or/and adding the I.I.D assumption.

Appending Trevisan's seeded extractor, we obtain a protocol efficiency between $61\%$ for $\delta=0.1$ and $72\%$ for $\delta=0.05$ on IBM's \textit{ibmq\textunderscore toronto} ($M_{obs}=3.62$) with $n=10^8$. On Quantinuum and AQT/UIBK ion-trap devices with $M_{obs}=3.9$, the protocol efficiency is $108\%$ for $\delta=0.1$ and $116\%$ $\delta=0.05$. Note that an efficiency of $116\%$ means that one obtains $1.16$ final random bits per use of the quantum device.

Another important quantity is the speed at which the quantum computer can perform different circuits. Indeed, the protocol's randomness generation rate per second is the amount of circuits implemented per second multiplied by the efficiency of the protocol (which is a function of the performance of the device, i.e. $M_{obs}$). At the time of the experiments, the quantum computers of the IBM Quantum Services had a fixed repetition rate of $r=2\cdot 10^3$ circuits per second, which severely limits the generation rates of the protocol. In this case, with an efficiency of the protocol of about $\eta = 6\%$ for $M_{obs}=3.62$ and $\delta=0.05$, this gives an output rate of about $\eta \cdot r = 120$ bits per second. Note, however, that this is not a fundamental limitation. Our protocol roughly amounts to performing 2 CNOT gates, which should take roughly $10^3$ nanoseconds on the device. This could in principle take the rates up to about $60$ kilobits per second. One can then append Trevisan's seeded extractor in order to increase the rates as described before. In this case, we obtain generation rates of 1440 bits per second for $\delta=0.05$ with the fixed circuit repetition rate of $r=2\cdot 10^3$ circuits per second and an in principle $720$ kilobits per second if circuits are implemented in $10^3$ nanoseconds. Note that in the latest case, the bottleneck of the protocol becomes the complexity of Trevisan's extractor, which is well below $720$ kilobits per second in our setup with large block sizes (about a few kilobits per second only). Unfortunately, in this case the observed Mermin value of $M_{obs}=3.62$ is insufficient to exploit our efficient seeded extractor to avoid the bottleneck. One could explore the different implementations of the Trevisan extractor that are fast in-practise, for example, by parrellelising the one-bit-extractor step. 

With the H1 device of Quantinuum, working at about $13$ circuits/second for our implementation, we obtain roughly $1.8$ final random bits per second for $M_{obs}=3.88$ and $\delta=0.05$ (without a seeded extractor). Appending Trevisan's seeded extractor we obtain randomness generation rates of about $14$ bits per second ($\delta=0.05$) and $13$ bits/sec ($\delta=0.1$). The AQT/UIBK device ran at $40$ circuits per second, giving randomness generation rates of about $6$ bits per second (without a seeded extractor) and about 46 bits/sec ($\delta=0.05$) or 43 bits/sec ($\delta=0.1$) when appending Trevisan's extractor. With $M_{obs}=3.9$, we are in a setup in which we can use our efficient implementation of a seeded extractor, so we can further increase the efficiency.

\item \textbf{On an ideal quantum device.} This would imply achieving $M_{obs}=4$, which is impossible in practise but is interesting to understand the limits of the protocol and what will happen when the devices get better. In this case, one would be able to amplify an imperfect RNG with $\delta\rightarrow 0.5$ (i.e. almost deterministic) and get a protocol efficiency up to $\eta=40\%$ depending on the number of rounds $n$ and $\delta$ without a seeded extractor.

Appending Trevisan's extractor one can increase the efficiency up to $\eta_S=200\%$ in which 2 bits are generated per run of the quantum device. Remember that without further improvements, the current maximal outputting rate of existing Trevisan extractor implementations do not allow going above a few kilobits per second in our setup. On the other hand, with $M_{obs}=4$ the outputs of the quantum device have a sufficiently high min-entropy rate to make our efficient implementation of a seeded extractor useful. In this ideal case, the final efficiency of the protocol becomes the same as with Trevisan's extractor and gives the maximal $\eta_S=200\%$.
\end{itemize}


\textbf{Statistical tests.}\label{sec:stat_tests} As a sanity check, we have also run the statistical tests of the NIST \cite{NIST}, DieHard\footnote{We refer to DieHarder's \href{https://webhome.phy.duke.edu/~rgb/General/dieharder.php}{webpage}.} and ENT\footnote{See \href{http://www.fourmilab.ch/random/}{http://www.fourmilab.ch/random/}.} test suites on  sets of 5 x 10Gb generated using our showcase quantum computer implementation with different imperfect RNGs as inputs. As expected, all tests were passed.

The different imperfect RNGs used as the input to the protocol were: different pseudo-RNGs, a classical chaotic process from the avalanche effect in a reverse-biased diode and a commercially available QRNG - Legacy Quantis-USB. All imperfect RNGs consistently failed, or gave 'suspicious'/'weak' results, in some statistical tests before being processed by our protocol. Once amplified using our showcase protocol implemented on quantum computers, the final output randomness passed all tests. We detail those results in the next section, together with a comparison against classical alternatives to our protocol to illustrate the quantum advantage of the protocol from a statistical perspective.

The amplification protocol was implemented using $\delta = 0.05$ for the imperfect RNG and the \textit{ibmq\textunderscore ourense} quantum computer from the IBM Quantum Services. 

\subsection{Our protocol versus classical alternatives}\label{sec:stattests}

From a statistical perspective we show that several sources of randomness, both hardware quantum and classical RNGs and software pseudo-RNGs are successfully amplified by our protocol. By this we mean that the results of statistical tests are consistently improved with our protocol. These results will be explained in more detail in \cite{future}. \\

\textbf{Versus a classical RNG built in-house.} As mentioned above, this RNG was based on a classical chaotic process from the avalanche effect in a reverse-biased diode built in house, which when tested gave good results but several 'weak' tests. Interestingly, when testing the output generated at the end of our amplification protocol we observed that there were no longer any 'weak' or 'suspicious' test results, so, from a statistical perspective it seemed there had been an improvement in the quality of the random numbers being generated.\\

\textbf{Versus a quantum RNG available commercially.} We reproduced the results of \cite{hurley2017quam,hurley2020quantum} showing that the output the Legacy Quantis-USB QRNG exhibits consistent patterns as witnessed by the ENT statistical test suite, failing the Chi-squared test. After using this QRNG as the input to our showcase protocol, we obtained a final output which passed all statistical tests. Again, this means that from a statistical perspective, it seemed the randomness quality had been improved. \\

\textbf{Versus pseudo-RNG.} Here, the goal was to test whether classical alternatives could produce similar results to our showcase protocol. We started by finding a PRNG that failed some NIST statistical tests: a class of 64-bit linear congruential generator called MMIX \footnote{See \href{http://mmix.cs.hm.edu/index.html}{http://mmix.cs.hm.edu/index.html}.}. On average, this PRNG passed only $6/15$ of the NIST statistical tests. 

\begin{itemize}
    \item (Quantum resources:) Using the showcase protocol on quantum computers with MMIX as the imperfect RNG, the output randomness passed 15/15 NIST tests (on average).
    \item (Classical resources:) Using our implemented Dodis \textit{et al.} extractor on randomness generated from MMIX and a maximal period LFSR, the output randomness passed 6/15 NIST tests (on average).
\end{itemize}

This is a simple example (from a statistical perspective) where we were unable to amplify the quality of the MMIX PRNG with classical resources only, whilst it was successful using our showcase protocol with quantum resources. 


\section{Implementations on quantum computers}\label{sec:QC_implementation}

\subsection{Overview}

The second part of this work serves as a real-world example of the usefulness and accessibility of quantum technologies, which is one of our main objectives. Although the ideal implementation of our protocol would be to use a quantum device running a loophole-free Bell test, today these are notoriously hard to build and would achieve randomness generation rates that are very low or lead to an almost trivial randomness amplification protocol\footnote{Indeed, to achieve a useful amplification protocol, i.e. one having $\delta << \frac{1}{2}$, one needs relatively high Bell inequality violations. Moreover, by ``high'' we mean that the observed violation of the Bell inequality should be high compared to the algebraic maximum of the inequality (and not only the maximum value achievable with quantum resources).}. In contrast, available today are a wide range of usable quantum technologies and in particular promising quantum computers which are waiting for real-world applications. Such quantum computers moreover have several features making them attractive to run Bell tests, as for example superconducting \cite{clarke2008superconducting} and ion-traps devices \cite{bruzewicz2019trapped} do not suffer from the so-called detection loophole \cite{ansmann2009violation} and ion-trap devices have extremely low cross-talk \cite{pino2021demonstration}. 

In this context, our results are:
\begin{itemize}
\item Under minimal added assumptions which we describe in detail, today's quantum computers can be trusted to run faithful Bell tests and therefore run our protocol securely. For this, we develop a method to account for some signalling effects (e.g. cross-talk) in the statistical analysis, {if the signaling can be shown to occur in a fraction of the rounds only. In complement to other techniques \cite{bacon2003bell,silman2013device} which allow to account for other forms of signalling, for example considering contributions of join measurements instead of the desired separable ones \cite{silman2013device}, this allows to consider different signalling effects on the same device}. At a high level, this amounts to trusting that the quantum computer has not been purposely built to trick the user, but allows for unavoidable imperfections in its implementation.
\item By optimising the circuit implementation as well as the parameters of our protocol to the specific hardware, all the quantum computers we used achieved sufficiently high Bell inequality values to run our protocol, for certain $\delta > 0$, and generate random numbers in a semi-device-independent manner. In particular, although from a small sample only, the results obtained from the devices of Quantinuum and AQT/UIBK gave the highest Mermin inequality value $M_{obs} \approx 3.9$. This is also the case on IBM's device \textit{ibmq\textunderscore toronto} giving $M_{obs}=3.62$. The previous highest Mermin inequality value that we could find was $M_{obs}=3.57$  \cite{maxMermin} (albeit in 2006).
\item We showed in the previous section how our results compare to classical alternatives from a statistical perspective. By running extensive statistical tests, we showed that the output generated by (other commercially available) quantum and classical hardware RNGs and also existing software pseudo-RNGs, is successfully amplified through our protocol run on quantum computers by using them as the imperfect source of randomness. This means that the random numbers from those sources fail, or perform badly, at some commonly used statistical tests without being processed by our protocol, but succeed after being amplified by our protocol.
\end{itemize}

The numerical results that we obtained are summarised in Table. \ref{ResultsQC} and the remainder of this section is organised as follows. Sec.\ref{sec:QC_validity} serves to discuss and describe the necessary added assumptions for running a faithful Bell test on quantum computers but also how to account for and quantify some signalling (e.g. cross-talk) in the implementations. We then describe the different quantum computer implementations, in particular how we optimised the circuits for each device, in Sec.\ref{sec:QC_implementations}.


\subsection{Validity of quantum computers for Bell experiments and added assumptions}\label{sec:QC_validity}

We want to address the questions:
\begin{itemize}
\item[] \textit{How valid is the use of quantum computers to perform Bell tests? Which added assumptions are required?}
\end{itemize}

Quantum computers are not built purposely for the task of running Bell tests and in particular open the so-called locality loophole. Indeed, in a quantum computer the qubits are {not strictly shielded from} each other and cross-talk can occur. In a loophole-free implementation, the qubits are isolated from each other and the experiment synchronised such that there is no time for possible communications between the different parts of the quantum device during a round (see Sec.~\ref{Sec:BellIneqs}). We term such possible undesired communication between the sub-parts \textit{signalling} and cross-talk is a particular type of it. In the presence of signalling, it is known that quantum correlations become possible to generate without quantum resources, given that the amount of such signalling is sufficient \cite{bacon2003bell,brask2017bell}.

In order to account for signalling in Bell tests, we developed methods that include these undesired effects in the statistical analysis, reducing the amount of certified randomness accordingly. Each of these methods implies a different additional assumption about the quantum device's functioning and therefore reduces the implementation to semi-device-independent. Note that the user only needs to make one of the below listed assumptions, not all of them. These assumptions are abstract and only consider the effect of signalling at the level of the observed statistics. One could also consider taking a different approach, {for example} by allowing for a weak form of signalling each round \cite{silman2013device} (still requiring additional assumptions about the devices, as we do here), {or in fact combining the different techniques (for example using ours in conjunction with \cite{silman2013device} or following the techniques of \cite{bacon2003bell}) to account for a wider range of signalling effects. Because these other techniques have been described in other works, we here focus on ours only.}\\

To use the Bell inequality values obtained from a quantum computer implementation, the user needs to make one of the following assumptions:
\begin{itemize}
\item \textit{Assumption $A$: The effect of signalling (e.g., cross-talk) is random, in the sense that it is not tailored to the specific Bell inequality that is used.}

\underline{or}

\item \textit{Assumption $B$: The effect of signalling (e.g., cross-talk) is not random in the sense of $A$, but is fixed in the sense that its effect is the same each time it occurs.}

\underline{or}

\item \textit{Assumption $C$: The effect of signalling (e.g., cross-talk) in the quantum computer is a mixture of the effects described in assumptions $A$ and $B$.}
\end{itemize}

Assumption $A$ is probably generic if the device was not purposely built in a malicious way and is similar to random noise. Indeed, accidental signalling or other classical effects are complex and changing, making it unlikely to be exactly such that they contribute positively to the Bell inequality that is used. Moreover, for each Bell test there exist several equivalent Bell inequalities that can be used, each requiring a different tailored signalling effect\,---\, which was not observed when we tested the devices.

Assumption $B$ considers the opposite situation in which, for some reason, the signalling occurs in a way that positively contributes to the Bell inequality that is used. In this case, the assumption is that this positive contribution occurs the same way every time. This could be thought of happening, for example, if there was a systematic imperfection in the device leading to a fixed signalling effect. The opposite situation, in which this effect is not systematic but random, is captured in Assumption $A$.

Assumption $C$ allows to consider a mixture of the effects in Assumptions $A$ and $B$, which could in principle occur side by side. Indeed, one could imagine that there is a systematic signalling effect in the device but, for some reason, in some rounds this effect gets randomised because of other phenomena.\\

For the sake of clarity, note that it is important that the Bell test is run on a device that is \textit{trusted} to be a quantum device. Although the device might be noisy or mostly uncharacterised, if the Bell test is run, for example, on a classical computer simulating a quantum device there is no way for the user to distinguish it from a fair Bell experiment without inspecting the device. Such a simulator would violate the assumptions we made above, but there may be no way to witness it for the user. The user therefore needs to make sure that the Bell test is indeed run by what has been built, in good faith, as a quantum device. This can be insured, for example, by inspecting the device or by trusting the provider.\\ 

In order to account for signalling effects in our statistical analysis, we follow a worst-case approach and apply the largest hit on the generated randomness these could imply. We show that the signalling effect in assumption $A$ actually increases the amount of generated randomness that can be certified and therefore the worst-case is to ignore its contribution\footnote{This is not too surprising, as the random form of signalling does not contribute positively to the Bell value\,---\,which in turn inflates the one from the no-signalling rounds from which we certify randomness (see App.\ref{APP:Signaling}).}. This is a very positive sign that when random forms of cross-talk (in the sense of assumption $A$) diminish in quantum computers, the efficiency of our protocol will get even higher. The effect of signalling as in Assumption $B$ is negative on the amount of randomness that can be certified, but because of its fixed assumed effect it can be quantified, and therefore bounded, from the observed statistical behaviour of the device. This contribution is then accounted for in a worst-case manner: the Bell value and the number of rounds that can be used for certifying randomness diminish. Assumption $C$, in the worst-case scenario, amounts to taking the hit from Assumption $B$ alone. The details are given in App.\ref{APP:Signaling}.\\
\begin{figure}[!ht]\begin{center}
\scalebox{0.53}{\includegraphics{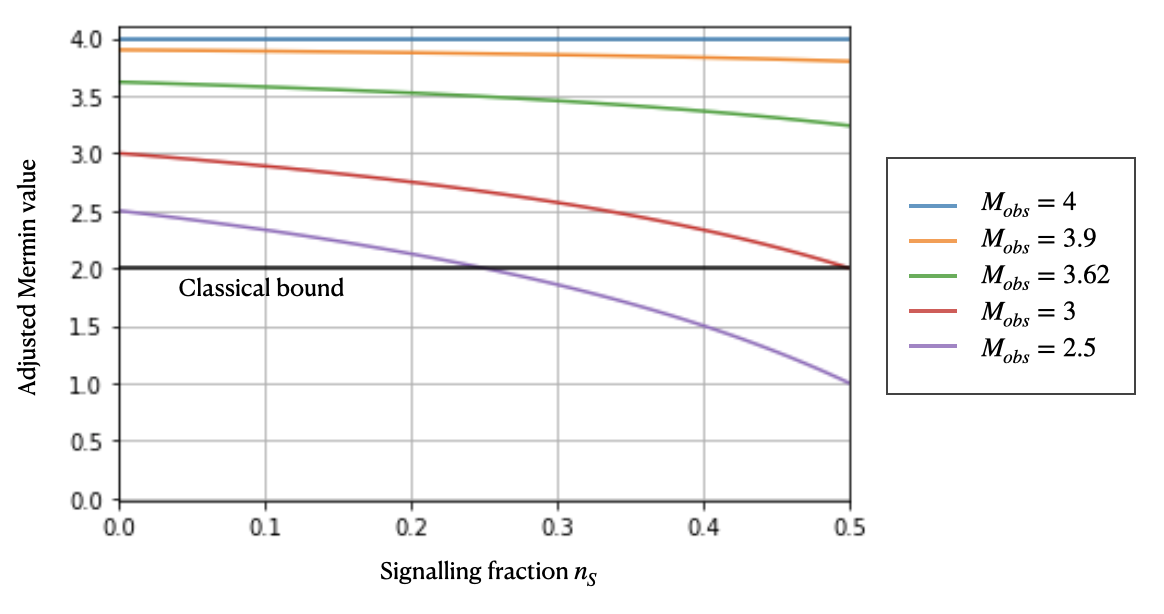}}
\end{center}
\caption{\label{Fig:Signaling} The adjusted Mermin value as a function of the signalling amount $n_s$ measured from the observed behaviour $P_{obs}$ for different observed Mermin values $M_{obs}$. The green curve corresponds to the observed Bell inequality value obtained from the quantum computer \textit{ibmq\textunderscore toronto} and the typical maximal amount of signalling fraction was $n_S\in [0.0014,0.0140]$ (from 10 experiments with $n=10^7$). The red curve corresponds to the observed Bell inequality value obtained from the AQT/UIBK quantum computer. Randomness can then be certified in a fraction $(1-n_s)$ of the rounds only, because randomness can only be certified in rounds when no-signalling occurs. We refer to App.\ref{APP:Signaling} for details.}
\end{figure}

From the experimental results that we obtained, the penalty for accounting for signalling in the quantum computers that we used is quite low and does not impact on the capacity of quantum computers to run our protocol. The impact of the signalling effect of Assumption $B$ and the typical effect we observed in the superconducting quantum computer \textit{ibmq\textunderscore toronto} of the IBM Quantum Services are plotted in Fig.~\ref{Fig:Signaling} (green). {We note here that we believe that on superconducting devices we do not have a strong reason to consider that the cross-talk only occurs in a fraction of the rounds (as opposed to, for example, weak cross-talk occurring every round \cite{silman2013device}). In this case, it would then be desirable to combine our technique with the techniques in \cite{silman2013device} in order to account for signalling in a more general way.} When using ion-trap devices, the effect of potential cross-talk is so low \cite{pino2021demonstration} that it could be ignored, but one can also apply the same techniques if desired. {As opposed to superconducting devices, in the case of ion-traps (Quantinuum's and AQT's) we additionally believe that using our technique is well justified because of the particular device structure and functioning. More precisely, in those devices ions (i.e. qubits) are located in traps and cross-talk can occur because of a failure in focusing the laser \cite{parrado2021crosstalk,pino2021demonstration}, therefore affecting another ion, or from light scattered from one ion to another during measurement, which can also happen with some probability \cite{parrado2021crosstalk,gaebler2021suppression}. Because of this probability to fail to shield one ion from another, we believe that our model of signalling is particularly well justified for this specific case.} For completeness, we also plot the impact on the observed Mermin value from the AQT/UIBK device in Fig.~\ref{Fig:Signaling} (orange). Again, we believe that our assumptions are sensible if the quantum device was not built in a malicious way. This is reasonable to expect, for example, from devices that are readily available to other users running quantum algorithms. Indeed, we find it hard to believe that quantum computers were built in order to trick the specific users that will be running our protocol. The advantage of our method using quantum computers is that it allows the use of a non-malicious yet mostly uncharacterised quantum device. This is in contrast to the standard physical methods for generating randomness. In this sense, it at least partially solves the problem in question 1 (as well as 2 and 3) stated in the introduction.\\


\subsection{Implementations of Mermin inequality violations on quantum computers}\label{sec:QC_implementations}

In order to use quantum computers to perform the Bell test in~\eqref{Mermin_ineq}, the implementations used circuits first preparing the so-called Greenberger-Horne-Zeilinger (GHZ) state of three qubits \cite{greenberger1989going}
\begin{align}
\ket{\textrm{GHZ}} = \frac{1}{\sqrt{2}}\Big(\ket{000}+i\ket{111}\Big)\,.
\end{align}
The prepared state is then measured with the Pauli $X$ or $Y$ measurement on each qubit depending on the circuit that is chosen. Remark that these (graph) states and measurements allow for a very simple circuit implementation, which is what in turn leads to high Bell inequality violations. Also note that in all our implementations we fix the state preparation in all circuits corresponding to different measurements, i.e. only the measurement bases vary from one circuit to another. We believe that our results also serve as a nice way to benchmark the performance of several quantum computers across two types (superconducting and ion-traps), which might be of independent interest.

\begin{figure}[!ht]\begin{center}
\scalebox{0.45}{\includegraphics{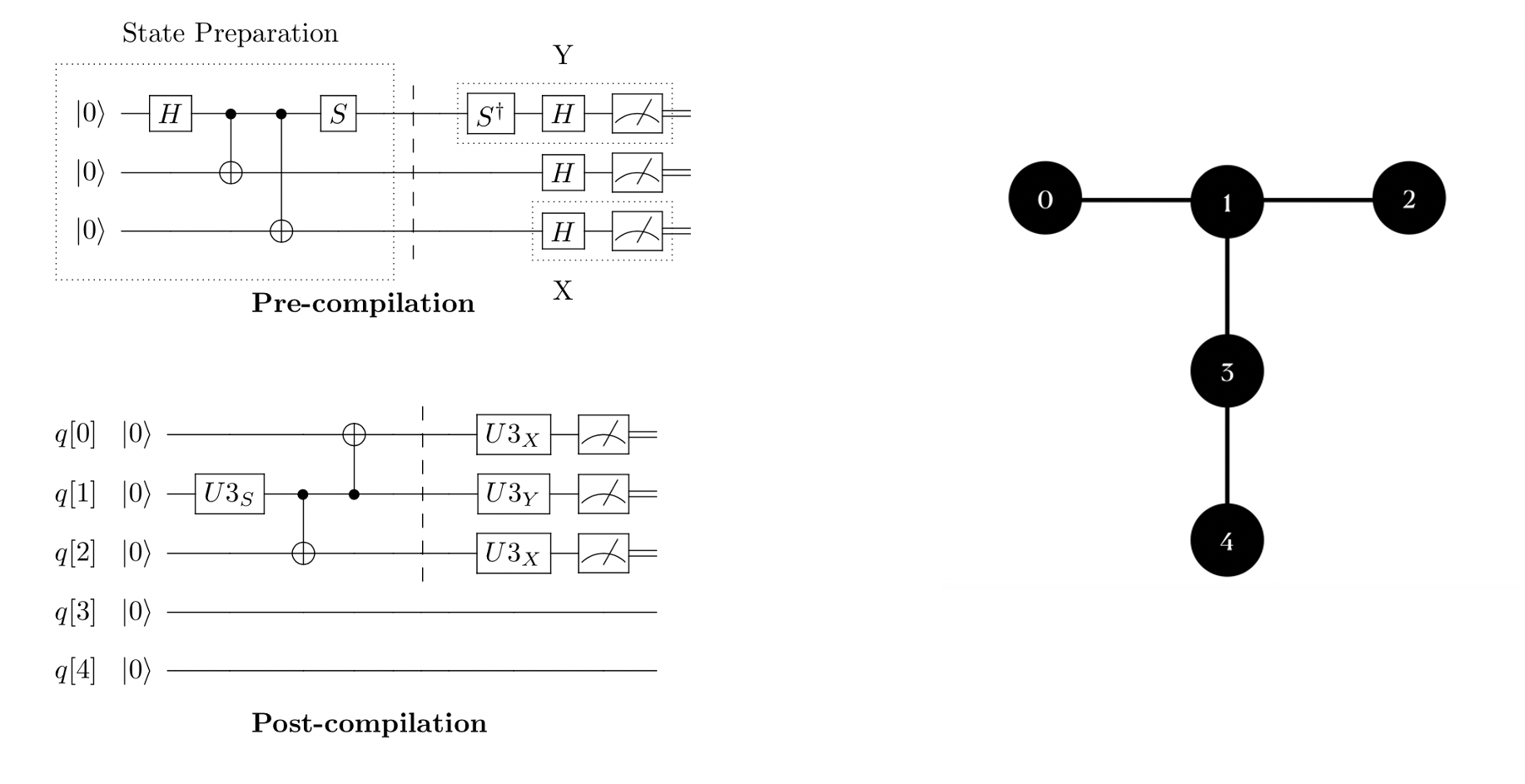}}
\caption{(left) One of the eight circuits that are implemented on IBM quantum computers, through the IBM Quantum Services, before and after compilation with t$\ket{ket}$ \cite{sivarajah2020t}. The state preparation (inside the dashed box) was fixed to be the same on all circuits, as indicated by the vertical dashed line (a barrier). The three inputs bits $x,y,z$ serve to encode the circuit implemented and the measurements on this circuit return the three output bits $a,b,c$. The input-output statistics is tested by evaluating the Mermin inequality \eqref{Mermin_ineq} as explained in Sec. \ref{Sec:BellIneqs}. (right) The physical layout of the qubits on the quantum computers \textit{ibmq\textunderscore ourense} and \textit{ibmq\textunderscore valencia} from the IBM Quantum Services. In both machines the qubits chosen via optimisation were qubits $0,1,2$.}
\label{Fig:QC_circuits}
\end{center}
\end{figure}

\subsubsection{Superconducting quantum computers from the IBM Quantum Services}

We optimised for physical qubits and gate implementation of the circuits on every available quantum computer available on IBM Quantum Services using the compiler t$\ket{ket}$ \cite{sivarajah2020t}. This is what allowed us to achieve high Mermin inequality values. All implemented circuits (after optimisation) have minimal depth $6$, prepare the same quantum state (the measurements only are different) and are run on the optimal physical qubits for each machine. Examples of circuits, for the devices \textit{ibmq\textunderscore ourense} and \textit{ibmq\textunderscore valencia} before and after compilation are given in Fig. \ref{Fig:QC_circuits} (left). The physical qubit layout of certain machines can be found in Fig. \ref{Fig:QC_circuits} (right) and it is the same for both machines \textit{ibmq\textunderscore ourense} and \textit{ibmq\textunderscore valencia}. The physical qubits that were chosen for the implementation by t$\ket{ket}$ were qubits $0,1,2$ on both machines. Implementations on different machines follow the same steps, but have different layouts and therefore optimised circuits.

The highest Mermin value was obtained on the quantum computer \textit{ibmq\textunderscore toronto} ($M_{obs}=3.62$) with 26 qubits although we use 3 of them only. Other good machines were \textit{ibmq\textunderscore ourense} and \textit{ibmq\textunderscore valencia} (respectively $M_{obs}=3.35$ and $M_{obs}=3.11$), both of which are $5$-qubit machines. Numerous other machines from the IBM Quantum Services give sufficiently good Bell values for our protocol. Our results for IBM Quantum Services devices were computed as the average obtained from 50 tests with $n=10^7$ number of circuits, i.e. the number of rounds. Interestingly, some machines performing well have low number of qubits, which is good in order to minimise the required resources. $M_{obs}=3.62$ was the highest Bell inequality value we could find in the literature before the ones that we obtained on ion-trap devices (see below). We summarised all results for the quantum computer implementations in Table. \ref{ResultsQC}. As mentioned in the previous section, the typical maximum signalling fraction we observed from \textit{ibmq\textunderscore toronto} was $n_s \in [0.0014,0.014]$, i.e. a minimal impact on the performance of the device for our protocol (see App.\ref{APP:Signaling} and Fig. \ref{Fig:Signaling}).\\

\subsubsection{Ion-trap quantum computers from Quantinuum and AQT/UIBK}

Our implementation on these devices lead to the very high Mermin value $M_{obs}=3.9$ on AQT/UIBK's device and $M_{obs}=3.88$ on Quantinuum's H1 device, which are the highest reported in the literature. We remind the reader that those values were not obtained in a loophole-free manner. We also obtained the values $M_{obs}=3.835$ from Quantinuum's H0. Note that we optimised the circuits (either by hand or using the compiler t$\ket{\textrm{ket}}$) for all implementations othen than for AQT/UIBK, where they performed the circuit optimisation themselves. Although they gave high violations, the small amount of time and slow repetition rate meant we were only able to collect data from $n \approx 10^4$ (AQT/UIBK) and $n \approx 5 \cdot 10^4$ (Quantinuum), rounds in which we tested the inputs appearing in the Bell inequality only.\\

As said in the previous section, cross-talk on ion-trap devices is notoriously low and we therefore ignored signalling effects with such devices. One can easily include them in the analysis as described before.


\section{Conclusion}

In the first part of our work, we have presented an end-to-end protocol for device-independent randomness and privacy amplification.  The setup, parameters, randomness post-processing, and statistical analysis were all optimised for real-world quantum devices. Our protocol has linear rates in the runtime of the quantum device and maximal noise tolerance. The randomness post-processing was also tailored to the task of randomness and privacy amplification. In particular, it was implemented keeping information-theoretic security whilst its complexity was taken down to near linear \,---\,allowing it to run efficiently on a standard personal laptop with large block sizes.

In the second part of our work, we have implemented our protocol on several quantum computers from the IBM Quantum Services, Quantinuum and AQT/UIBK. This can be understood either as a concrete example of the results that can be obtained with our protocol or, under minimal added hardware assumptions, as a semi-device-independent implementation. In the second case, one can run our protocol on today's quantum computers in order to generate private random numbers. The Mermin inequality violations we observed are the highest reported in the literature, a testimony to the work achieved by the different teams building quantum computers.

\vspace{0.3cm}

\textit{\textbf{Future work --}} Some important further developments to our results are already being worked on, among which:
\begin{itemize}
\item After running a significantly large experiment on the Quantinuum H1 ion-trap quantum computer, we have managed to collect enough output to test our protocol on a number of different imperfect RNGs. With this, we can continue to test the limits and demonstrate the practicality of our protocol in real world applications. 
\item It would be interesting to explore the implication of the recent generalisation \cite{GeneralisedEAT} of the entropy accumulation theorem that we use \cite{dupuis2016entropy,arnon2018practical} to the particular set-up for randomness amplification.
\item It is important to further improve on the randomness post-processing. This could lead to even higher $\delta$ that can be amplified, but also higher efficiencies. The challenge is to manage to keep the complexity low in the actual software implementations, see Sec.\ref{Sec:RandProc} for more details about this. In \cite{future}, we will expose further results on the topic of randomness extractors and their implementations.
\end{itemize}


\paragraph*{Acknowledgements.} We acknowledge discussions with Mafalda Almeida, Fernando Brand\~ao, Silas Dilkes, Christopher Portmann, Volkher B.~Scholz, Kimberley Worrall and Richie Yeung. We also acknowledge access to all the quantum computers, either through IBM Quantum Services or directly through Quantinuum and AQT/UIBK (Christian Marciniak, Ivan Pogorelov, and Thomas Feldker at the University of Innsbruck, who operated the device in the context of the AQTION project, EU H2020-FETFLAG-2018-03 under Grant Agreement No. 820495, based on the AQT System). Note that the views expressed are those of the authors, and do not reflect the official policy or position of the quantum computing providers listed. 


\appendix

\section{Statistical analysis}\label{APP:Bell}

\subsection{Preliminaries}\label{preliminaries}

We start the appendices by formalising the set-up we consider, see Fig.\ref{Fig:DIRAsetup}. An adversary, E, holds side information $h$ about the weak source of randomness which will be used to choose the inputs to the quantum device. This initial source of randomness is assumed to be a Santha-Vasirani source, i.e. it outputs sequentially such that its outputs $r_i$ satisfy \eqref{SVsource}. E is allowed to build the quantum device which is later given to the user, with which she might be entangled by keeping a quantum system entangled with the device in a quantum memory. After it has been handed to the user, the adversary is not allowed to access the device any more.


\subsection{Generalising the Mermin inequality to account for weakly random inputs}\label{app:mermin}


Inequality \eqref{Mermin_ineq} can be written in the equivalent form
\begin{equation}\label{Mermin_probs}
L \equiv L(\vec{P})= \frac{1}{8} \sum\limits_{a,b,c \in \{ 0,1\}} \bigg( \delta_{a \oplus b \oplus c=1} \big( p(abc|011) + p(abc|101) + p(abc|110) \big) + \delta_{a \oplus b \oplus c=0} \cdot p(abc|000) \bigg) \geq \frac{1}{8}{,}
\end{equation}
with $\oplus$ denoting the sum modulo $2$ and where $\frac{1}{8}$ is the local (or classical) bound. We then have the following relation between $L$ and $M$: $L=\frac{4-M}{16}$. The inequality written in this form can be understood as a losing probability at the Mermin-Bell game -- and requires quantum resources to obtain losing probability $L < \frac{1}{8}$. Indeed, under this form one sees that minimising $L$ implies trying to minimise the terms $p(a \oplus b \oplus c=1|011)$, $p(a \oplus b \oplus c=1|101)$, $p(a \oplus b \oplus c=1|110)$ and $p(a \oplus b \oplus c=0|000)$ -- i.e. the combination of outputs the quantum device should avoid outputting, depending on the inputs that were chosen. An important property of this inequality is that there exist quantum states and local measurements such that all these terms vanish (in a noiseless implementation), i.e. $p(a \oplus b \oplus c=1|011)=0$ and similarly with the other terms. \\

As discussed in the main text, standard Bell inequalities can not be used in our set-up in which there are correlations between the input choices and the device's behaviour. To account for such correlations, we first build a new inequality from \eqref{Mermin_probs}, which can be understood as a measurement dependent locality (MDL) type inequality \cite{putz2016measurement}. From an initial Bell inequality, one essentially performs the mapping $p(abc|xyz) \rightarrow p(abcxyz)$, i.e. considers joint distributions of both inputs and outputs instead of conditional ones. We then obtain
\begin{equation}\label{MDL}
L \rightarrow L_{\textrm{MDL}}  \equiv L_{\textrm{MDL}}(\vec{P}) = \sum\limits_{a,b,c \in \{ 0,1\}} \bigg( \delta_{a \oplus b \oplus c=1} \big( p(abc011) + p(abc101) + p(abc110) \big) + \delta_{a \oplus b \oplus c=0} \cdot p(abc000) \bigg){.}
\end{equation}

One can further estimate inequality $L_{\textrm{MDL}}$ by evaluating a \textit{losing condition $l_i$}, which is computed at each round of interaction with the quantum device:
\begin{equation}\label{loosing_cond}
l_i(a_i,b_i,c_i,x_i,y_i,z_i) = \begin{cases}
1 \hspace{1cm} \textrm{ if } a_i \oplus b_i \oplus c_i=1 \textrm{ and } (x_i,y_i,z_i) \in \{(011),(101),(110)\}\\
1 \hspace{1cm} \textrm{ if } a_i \oplus b_i \oplus c_i=0 \textrm{ and } (x_i,y_i,z_i) = (000)\\
0 \hspace{1cm} \textrm{ otherwise }{,}
\end{cases}
\end{equation}
where $a_i,b_i,c_i$ and $x_i,y_i,z_i$ are the outputs and inputs in round $i \in \{1,2,...,n\}$. One then estimates $L_{\textrm{MDL}}$ using $L_{\textrm{MDL}} = \frac{1}{n}\sum\limits_{i=1}^{n} l_i$.\\


\subsection{Single-round min-entropy from the losing probability}\label{SingleRound_APP}

The goal of this subsection is to relate the value of the losing probability (\ref{Mermin_probs}) to the single round min-entropy. The next step will be to see how this single-round quantity can be related to the total accumulated entropy in the $n$ rounds of the protocol (this is why we compute them), which depends on the set-up and assumptions considered.\\ 

Without loss of generality, we assume that the three boxes $\cal{A,B,C}$ and the adversary share a 4-partite quantum state $\rho^{\cal{ABCE}}_{h}$ on which the three parts in the quantum device make measurements $M^{\cal{A}}_{a|x,h},M^{\cal{B}}_{b|y,h},M^{\cal{C}}_{c|z,h}$. Note that, contrary to a standard Bell experiment, the measurements and state might depend on the side-information $h$ of the adversary. The adversary performs a measurement $O^{\cal{E}}_{e|w,h}$ on their share Q of the system, obtaining outcome $e$. All measurements are assumed to act locally on the state, i.e. the user's outcomes statistics is given by
\begin{equation}\label{Born}
p^{\cal{ABC}}(abc|xyz,Q_h)=\textrm{Tr}\big( M^{\cal{A}}_{a|x,h}\otimes M^{\cal{B}}_{b|y,h}\otimes M^{\cal{C}}_{c|z,h}\otimes \unit^{\cal{E}} \cdot \rho^{\cal{ABCE}}_{h} \big){,}
\end{equation}
for some unknown state and measurements (in particular, the Hilbert space dimension is not bounded) and for some $h$ which is not accessible to the user. For simplicity of the notation, we took the freedom to label a specific quantum realisation with quantum state $\rho^{\cal{ABCE}}_{h}$ and measurements $M^{\cal{A}}_{a|x,h}$,$M^{\cal{B}}_{b|y,h}$,$M^{\cal{C}}_{c|z,h}$, $O^{\cal{E}}_{e|w,h}$ by $Q_h$.\\

The guessing power that the adversary $\cal{E}$ has over the outcomes of the experiment is quantified by a \textit{guessing probability} $P_g(ABC|(xyz)^{\ast}, Q, h)$, given their quantum system $Q$ and side-information $h$. Moreover, as in our set-up the SV source is not assumed to be private, we want to quantify the guessing probability for \textit{given} inputs, which we label by $(xyz)^{\ast}$. Remember that $z = x \oplus y$, so $z$ is not generated directly from the imperfect RNG. The objective is to upper bound this quantity, i.e. certify some unpredictability. In our case, we upper bound the guessing probability of the adversary over the outcomes $a,b,c$ by using a bound on the guessing probability over two outcomes, say $a,b$ (without loss of generality in our case). The interested reader is referred to \cite{nieto2014using} for a detailed discussion about guessing probabilities. The result takes the form of the following optimisation problem
\begin{equation} \label{PgOPT}
\begin{aligned}
P_g(ABC|(xyz)^{\ast}, Q, h) \leq P_g(AB|(xyz)^{\ast}, Q, h) \\
= \max\limits_{Q_h} \sum\limits_{ab} p^{\cal{AB}}(ab|(xyz)^{\ast},Q_h) \cdot p^{\cal{E}}(e=(ab)|(xyz)^{\ast},(ab),w,Q_h)\\
\textrm{s.t.} \hspace{0.5cm} L_{\textrm{MDL}}(p(abcxyz,Q_h))=L_{\textrm{MDL}}^{obs}{.}
\end{aligned}
\end{equation}
In words, this optimisation problem corresponds to allowing for the most advantageous quantum realisation (through the maximisation over $Q_h$) for the adversary to correlate their outcome with the outcomes she is trying to guess, i.e. maximise $p^{\cal{E}}(e=(ab)|(xyz)^{\ast},(ab),w,Q_h)$ for each combination of $a,b$. Clearly, this corresponds to a worst-case bound. In addition to requiring that the realisation $Q_h$ is of the form \eqref{Born} (that it is quantum), we also require that the realisation reproduces the observed MDL inequality value \eqref{MDL}. Because we only have a constraint on the MDL inequality, and not on a Bell inequality, it is unclear how to solve this optimisation directly. Instead, we constrain the possible values of a Bell inequality which are compatible with the observed violation of the MDL inequality as follows.\\

For all realisations $Q_h$ (i.e. for any $h$), we note that
\begin{equation} \begin{split} \label{amplifProof1}
L_{\textrm{MDL}}^{obs} = 2 L^{obs} & = \sum\limits_{\substack{abc \\ xyz}} l(a,b,c,x,y,z) \cdot p^{\cal{ABC}}(abcxyz|Q_h) \\
& = \sum\limits_{\substack{abc \\ xyz}} l(a,b,c,x,y,z) \cdot p^{\cal{ABC}}(abc|xyz,Q_h) \cdot p(xyz|h) \\
& \geq (\frac{1}{2}-\delta)^2 \sum\limits_{\substack{abc \\ xyz}} l(a,b,c,x,y,z) \cdot p^{\cal{ABC}}(abc|xyz,Q_h) \\
& = {8}(\frac{1}{2}-\delta)^2 ( L(p^{\cal{ABC}}(abc|xyz,Q_h)){,}
\end{split}\end{equation}
where $l(a,b,c,x,y,z)$ is given in \eqref{loosing_cond}. In the first step, we have used the relation between observable quantities $L_{\textrm{MDL}}^{obs}(p_{obs}^{\cal{ABC}}(abcxyz)) = 2 L^{obs}(p_{obs}^{\cal{ABC}}(abc|xyz))$ from equ. \eqref{MDL} and \eqref{Mermin_probs} respectively, where we assume that $p_{obs}(xyz)=\frac{1}{4}$ $\forall x,y,z$\footnote{We here label the observed frequencies of the inputs as $p_{obs}(x,y,z)$ in order to distinguish it from $p(x,y,z|h)$.} appearing in the Mermin inequality. {In the second line, we have related joint probabilities with conditional probabilities. In the third line, we have used that $(\frac{1}{2}-\delta)^2 \leq p(xyz|h) = p(xy|h) \leq (\frac{1}{2}+\delta)^2$ (the SV assumption \eqref{SVsource} over the source, where we only require 2 bits from the imperfect RNG as $z = x \oplus y$) and in the last step we used the definition of the losing probability of the Mermin game as given in \eqref{Mermin_probs}}. In the end, we can use 
\begin{equation} \begin{split} \label{amplifProof2}
L(p^{\cal{ABC}}(abc|xyz,Q_h)) \leq \frac{L^{obs}}{4(\frac{1}{2}-\delta)^2} \equiv L_b.
\end{split}\end{equation}
In words, and in some sense, we can bound the value that \textit{would have been obtained} for the Bell inequality from the observed MDL inequality value. The constraint $L_{\textrm{MDL}}(p(abcxyz,Q_h))=L_{\textrm{MDL}}^{obs}$ in \eqref{PgOPT} can then be replaced by condition \eqref{amplifProof2}, i.e. a guessing probability as in a standard Bell test (with a worst-case bound on the Bell inequality violation instead). The resulting guessing probability was derived in \cite{woodhead2018randomness} and, when written as a function of $L_b$, reads
\begin{equation}\label{PgLb}
P_g(L_b) \leq \begin{cases} \frac{1}{4} + 2L_b + \sqrt{3}\sqrt{L_b(1-4L_b)} \hspace{1cm} & \textrm{if } L_b \leq \frac{1}{16} \\
\frac{1}{2} + 4L_b \hspace{4cm} & \textrm{if } \frac{1}{16} \leq L_b < \frac{1}{8}.
\end{cases}
\end{equation}

Once again, note that because of the symmetries in our set-up this guessing probability holds for all input triplets $(xyz)^{\ast}$. Finally, the min-entropy of the outcomes is then
\begin{equation}\label{Hmin}
H_{min}(L_b) \equiv H_{min}(ABC|(xyz)^{\ast}, Q, h) = -\log_2 \big( P_g(L_b) \big).
\end{equation}

In the main text, in order to simplify the notation, we have instead used the object $M_{obs}=M(\vec{P}_{obs})$, see \eqref{Mermin_ineq}. Although, as discussed, this can not directly be used as a Bell inequality as in the standard set-up it still represents an intuitive quantity to evaluate. For example, the result in \eqref{adjustedM} can be obtained by using $M_{obs}=4-16L^{obs}$ and that $L_{\textrm{MDL}}^{obs} = 2 L(\vec{P}_{obs})$ if $p_{obs}(x,y,z)=\frac{1}{4} \hspace{0.1cm} \forall x,y,z = x \oplus y$. By doing this, we restrict the analysis to set-ups in which the inputs frequencies are equal. 


\subsection{Identical and independent rounds for large $n$}\label{app:identical}

The first situation that we consider is the one in which one can assume that the different rounds of interactions with the quantum device are identical and independent of each other. More precisely, the global quantum state describing the joint system of the adversary $\cal{E}$ and the one of the quantum device $\cal{ABC}$ (see Fig.\ref{Fig:QuantumDevice}) over the entire run of $n$ interactions with the quantum device is assumed to have the structure\footnote{Remark that one can include the system held by $\cal{E}$ because of the equivalence of all purifications from the systems of $\cal{AB}$ to the ones with $\cal{E}$.}
\begin{equation}\label{IIDstateI}
\rho^n_{\cal{\textbf{ABCE}}} = (\sigma_{\cal{ABCE}})^{\otimes n},
\end{equation}
where $\otimes$ denotes the tensor product and we have labelled the systems over $n$ rounds in bold. Note that we do not assume knowledge of the state $\sigma_{\cal{ABCE}}$, only that such state exists and that the decomposition holds. Similarly, the measurements made in the device $\cal{A}$ are assumed to take the form $(M^a_x)^{\otimes n}$ (and the same holds for the measurements of $\cal{B}$ and $\cal{C}$). These conditions imply that the variables $l_1,l_2,...$ used to evaluate L are independently and identically distributed (I.I.D) random variables, and the same holds for the outcomes $a_1,a_2,...$ (and similarly for outcomes $b_i$ and $c_i$). In the limit of large $n$, we can evaluate the total accumulated entropy
\begin{equation}\label{minEPG}
H_{\min}(A^nB^n|Q, h) = n \cdot H_{\min}(L_b).
\end{equation}

Assuming that the quantum device behaves identically and independently may be a reasonable assumption in certain cases, for example when the device provider can be trusted and functions at slow speed (possibly avoiding memory effects). Importantly, the I.I.D assumption in the asymptotic limit $n\rightarrow\infty$ also serves to test the ultimate limits of the protocol since the results in the most general set-up (MBQA, non I.I.D.) tend to the one in this simpler I.I.D. set-up \cite{arnon2018practical} for large $n$.


\subsection{Memory based quantum attacks}\label{app:memory}

Although making the assumption of identical and independent rounds of interactions can be interesting in some cases, this assumption clearly goes against the mindset of device-independence. Indeed, in general it is very limiting to assume that there is no correlation between the different interaction rounds -- such as memory effects for example. To generalise the security proof to the most general case, in which the I.I.D. assumption is dropped and where the adversary can also make general operations on its share of the quantum state, one can use the framework developed in \cite{arnon2019simple,arnon2018practical, Friedman17}. There, the authors explain how to use the \textit{entropy accumulation theorem} (EAT) \cite{dupuis2016entropy} in the set-up that we are considering.

Even if the results are more general, the idea is still to go through a \textit{reduction} to the single-round quantities introduced in Sec.\ref{SingleRound_APP}. That is, the total of entropy accumulated in $n$ interaction rounds can be computed from single round quantities (which is why we considered them) and some penalty terms.\\

Remark, however, that all structure is not lost. Indeed, the interaction rounds with the quantum device are made in a sequential manner: if the inputs ($x_i,y_i,z_i$) and outputs ($a_i,b_i,c_i$) of one round can {depend} on the ones {generated} in previous rounds, the opposite is not allowed. Namely, the inputs and outputs generated during an interaction round can not depend on the ones generated during a \textit{later} interaction round -- the future does not influence the past. This sequential structure is one of the ingredients that allows to obtain non-trivial results.\\

In order to use the EAT, we first need to show that our experiment and protocol satisfies certain conditions (see \cite{arnon2019simple,arnon2018practical}), namely that
\begin{itemize}
\item $\{A_iB_iC_i\}_{i \in [n]}$, the outputs, are all finite dimensional {measurement outcomes (classical random variables)}. The same holds for $\{L_i\}_{i \in [n]}$ (the losing condition evaluated at each round, see \eqref{loosing_cond}) and $\{X_iY_iZ_i\}_{i \in [n]}$, the inputs, which represent the information that ``leaks'' to the adversary (the SV source is public, i.e. its outputs can not be assumed to be private once generated). Finally, $\{R_i\}_{i \in [n]}$ (the quantum register holding the information about the state at round $i$) is an arbitrary quantum system.
\item At every round of the protocol, the losing condition $l_i$ \eqref{loosing_cond} can be evaluated directly from the classical {systems} $a_i,b_i,c_i,x_i,y_i,z_i$ at that round, i.e. without changing the underlying state.
\item For all $i$, we have the Markov chain condition
\begin{equation} 
A_1,B_1,C_1,A_2,...,A_{i-1},B_{i-1},C_{i-1} \leftrightarrow X_1,Y_1,Z_1,X_2,...,X_{i-1},Y_{i-1},Z_{i-1},E \leftrightarrow X_{i},Y_{i},Z_{i}{,}
\end{equation}
i.e. in our set-up at each round $i$ the inputs $X_{i},Y_{i},Z_{i}$ do not reveal ``new'' information about the previous outcomes $A_1,B_1,C_1,A_2,...,A_{i-1},B_{i-1},C_{i-1}$ (information that was not already obtained through the previous inputs made public and the side-information E of the adversary). Remark the importance of assumption number 6 in Sec.\ref{assumptions}, which excludes the situation in which the weak source of randomness would update its state depending on the outcomes of the Bell test. 
\end{itemize}

We are now in a position where we can apply the EAT and need to compute the necessary quantities. For the details (and why these are the relevant quantities), {we refer the reader to \cite{arnon2019simple} (lemma 10) as adapted for the setup for randomness amplification in \cite{Friedman17} (theorem 33 and claim 1, p.27), but also to \cite{arnon2018practical}.}\\

The derivative of the min-entropy  $H_{\min}(L_b)\geq-\log_2 \big( P_g(L_b) \big)$ {(\ref{minEPG})} is
\begin{equation}
\frac{\textrm{d} H_{\min}(L_{\textrm{MDL}}^{obs})}{\textrm{d} L_{\textrm{MDL}}^{obs}} = \frac{\textrm{d} H_{\min}(L)}{\textrm{d} L_b} \cdot \frac{\textrm{d} L_b}{\textrm{d} L_{\textrm{MDL}}^{obs}} = \begin{cases}
\frac{-2-\frac{\sqrt{3}}{2}\frac{1-8L_b}{\sqrt{L_b(1-4L_b)}}}{\ln(2)(\frac{1}{4}+2L_b+\sqrt{3}\sqrt{L_b(1-4L_b)})} \cdot \frac{1}{4(\frac{1}{2}-\delta)^2} & \textrm{if } L_b \leq \frac{1}{16}\\
\frac{-4}{\ln(2)(\frac{1}{2}+4L_b)}    \cdot \frac{1}{4(\frac{1}{2}-\delta)^2}  & \textrm{if } \frac{1}{16} \leq L_b < \frac{1}{8}{,}
\end{cases}
\end{equation}
with, again, $L_b(L_{\textrm{MDL}}^{obs},\delta)$ given in \eqref{amplifProof2}.\\

We now define the function $f_{\min}$ (our min-tradeoff function, see \cite{arnon2018practical}) as a function of a new variable $L_{cut}$ (a degree of freedom which we will optimise on)
\begin{equation}\label{fminEQU}
f_{\min}(L_b,L_{cut}) = \begin{cases}
H_{\min}(L_b) & \textrm{if } L_b \geq L_{cut}\\
H_{\min}(L_{cut})+(L_b-L_{cut}) \cdot \frac{\textrm{d} H_{\min}}{\textrm{d} L_{\textrm{MDL}}^{obs}} \bigg|_{L_b=L_{cut}} & \textrm{if }  L_b < L_{cut}{.}
\end{cases}
\end{equation}
We can now evaluate the total conditional smooth min-entropy accumulated in the outputs $A^nB^nC^n$ over successive $n$ rounds, which is the main quantity of interest (as it is roughly the randomness that can be extracted). {This is computed similarly to (33) in \cite{Friedman17} (p.28), in the proofs and explanations following lemma 10 and claim 1, replacing the quantities with the ones in our setup as derived above}
\begin{equation}\label{MBQA_entropy}
H_{\min}^{\kappa}(A^nB^nC^n|Q,h) > n \cdot \max\limits_{L_{cut}} \bigg(f_{\min}(L_b,L_{cut}) - \frac{\nu(L_{cut},\kappa,\eps_{EA})}{\sqrt{n}} \bigg){,}
\end{equation}
with $\kappa$ a smoothing parameter (which will also be optimised on) and $\eps_{EA}$ a (small) security parameter corresponding to the failure probability of this entropy accumulation step -- i.e. one can show \cite{arnon2018practical} that either the entropy accumulation step fails with probability higher than $\eps_{EA}$ or the total accumulated entropy respects \eqref{MBQA_entropy}. {Note that since the total min-entropy relates to the guessing probability according to $H_{min}=-\log_2(P_g)$, this equation \eqref{MBQA_entropy} is written in terms of $P_g$ in \eqref{EAT}.} The security parameters will be taken into account (and optimised) later when we consider the security of the full protocol in Sec.\ref{Altogether_security}. The entropy rate of the quantum device, i.e. the average amount of entropy per bit in the $3n$-bit string containing the outcomes of all rounds, is given by
\begin{equation}\label{MBQA_entropyRate}
\beta \equiv \frac{H_{\min}^{\kappa}(A^nB^nC^n|Q,h)}{3n}> \frac{1}{3} \max\limits_{L_{cut}} \bigg(f_{\min}(L_b,L_{cut}) - \frac{\nu(L_{cut},\kappa,\eps_{EA})}{\sqrt{n}} \bigg){.}
\end{equation}
Remark that because we are using a bound on the guessing probability of two of the three outcomes, we have that $\beta \leq \frac{2}{3}$, which is a limitation of our protocol. It would be interesting to be able to bound the guessing probability of two outcomes directly whilst being able to use the EAT to accumulate entropy over $n$ rounds. Note that in our case this is currently not possible, as our channels would then not satisfy the three conditions exposed above in order to use the EAT (see the interesting discussion about this limitation of the EAT in \cite{arnon2018reductions}, just after lemma 11.3).\\

Finally, the ``penalty'' term $\nu$ can be computed {(see {Lemma} 10 in \cite{arnon2019simple}, but also the proof of {Claim} 1 in \cite{Friedman17})} and gives
\begin{equation}\label{MBQA_penalty}
\nu \equiv \nu(L_{cut},\kappa,\eps_{EA}) = 2(\log_2(9)+\ceil{ \big| \frac{\textrm{d} H_{\min}}{\textrm{d} L_{\textrm{MDL}}^{obs}}\bigg|_{L_b=L_{cut}}\big|}) \cdot \sqrt{1-2\log_2(\kappa \cdot \eps_{EA})}
\end{equation}
with $\bigg|\bigg| \frac{\textrm{d} H_{\min}}{\textrm{d} L_{\textrm{MDL}}^{obs}} \bigg|\bigg|_{\infty} =  \big| \frac{\textrm{d} H_{\min}}{\textrm{d} L_{\textrm{MDL}}^{obs}}\bigg|_{L_b=L_{cut}}\big|$ the maximal value of $\big|\frac{\textrm{d} H_{\min}}{\textrm{d} L_{\textrm{MDL}}^{obs}}\big|$ (which is obtained at $L_b$).


\section{Proofs for the randomness post-processing}\label{APP:math}

\begin{figure}[!ht]\begin{center}
\scalebox{0.2}{\includegraphics{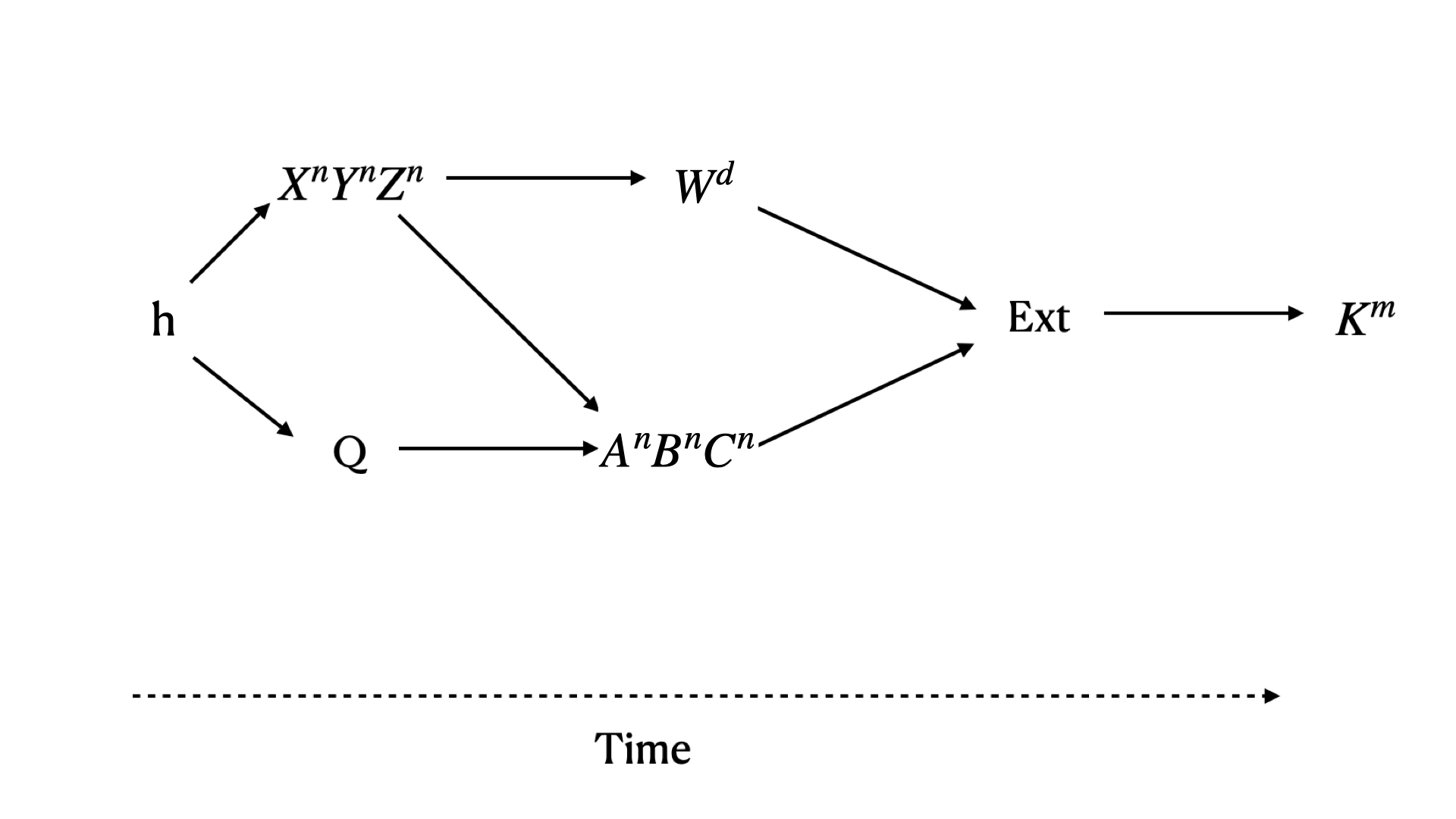}}
\end{center}
\caption{\label{Plot:Markov} (From \cite{Friedman16,Friedman17}) A diagram of the variables and their relation in our setup and assumptions. First, the adversary learns the (classical) history $h$ from the previous outputs of the imperfect RNG and possibly other information. This information gives the adversary predictive power over the outcomes of the imperfect RNG, which is first used to choose the measurement basis in the experiment ($X^nY^nZ^n$) and later as fresh input to the randomness post-processing ($W^{d}$, where for example using our implemented Dodis extractor $d=3n$). Before the string $W^{d}$ is generated, the quantum device produces its output $A^nB^nC^n$. Finally, both $W^{d}$ and $A^nB^nC^n$ are processed together by a 2-source extractor secure in the Markov model. Note, in particular, that the variables $W^{d}$ and $A^nB^nC^n$ are assumed to be independent conditionally on the other variables generated before, see \eqref{MarkovCond}. This Markov condition is necessary to obtain security in our setup and corresponds to assumption 6 in our list \ref{assumptions}.}
\end{figure}

Before giving the formal proofs and details regarding our randomness extraction routines, following \cite{Friedman16,Friedman17} we explain why the so-called \textit{Markov} model is needed in our setup (see Fig. \ref{Fig:DIRAsetup} and \ref{Fig:ImperfectRNG} for our setup). In Fig. \ref{Plot:Markov}, one can see the relations between the variables in the protocol and an explanation of why those relations are valid or correspond to one of the required assumptions. The difficulty in this setup comes from the fact that the inputs to the 2-source extractor are not independent of each other but only \textit{conditionally} independent (on the other variables generated before and that are known to the adversary), see \eqref{MarkovCond}. 2-source extractors secure in the Markov model work in the presence of this particularly strong form of possible side information. 

\subsection{Two-source extractor}\label{sec:two-source}

The following definition is standard, see for example \cite[Definition 1]{Raz05}.

\begin{definition}[Two-source extractor]
Let $n_1,n_2\in\mathbb{N}$, $k_1\in[0,n_1],k_2\in[0,n_2]$, $m\in\mathbb{N}$, and $\eps\in[0,1]$. A function $\text{Ext}:\{0,1\}^{n_1}\times\{0,1\}^{n_2}\to\{0,1\}^m$ is called a $(n_1,k_1,n_2,k_2,m,\eps)$ two-source extractor if for independent $X_1,X_2$ with $H_{\min}(X_1)\geq k_1$ and $H_{\min}(X_2)\geq k_2$, we have
\begin{align}
\frac{1}{2}\left\|\text{Ext}(X_1,X_2)-U_M\right\|_1\leq\eps\,,
\end{align}
where $U_M$ denotes the uniform random variable on $m$ bit strings. The function $\text{Ext}$ is called strong in the $i=1,2$ input if
\begin{align}
\frac{1}{2}\left\|\text{Ext}(X_1,X_2)-U_M\circ X_i\right\|_1\leq\eps\,.
\end{align}
We call $n_1,n_2$ the input length of the first and second source, resp., $m\in\mathbb{N}$ the output length, and $\eps\in[0,1]$ the security parameter.
\end{definition}

Following Arnon-Friedman {\it et al.}~\cite{Friedman16}, we extend this security criteria to adversaries holding quantum information $Q$ about $X_1$ and $X_2$. The enabling concept are quantum Markov chains.

\begin{definition}
A classical-classical-quantum state $\rho_{X_1X_2Q}$ is a Markov source if
\begin{align}\label{MarkovCond}
I(X_1:X_2|Q)_\rho=0\,,
\end{align}
where $I(X_1:X_2|Q)_\rho:=H(X_1Q)_\rho+H(X_2Q)_\rho-H(X_1X_2Q)_\rho-H(Q)_\rho$ with $H(Q)_\rho:=-\text{Tr}\left[\rho_Q\log\rho_Q\right]$ denotes the conditional mutual information.
\end{definition}

The stronger security criteria in the presence of quantum adversaries is then as follows \cite[Definition 8]{Friedman16}.

\begin{definition}[Quantum-proof two-source extractor]
Let $n_1,n_2\in\mathbb{N}$, $k_1\in[0,n_1],k_2\in[0,n_2]$, $m\in\mathbb{N}$, and $\eps\in[0,1]$. A function $\text{Ext}:\{0,1\}^{n_1}\times\{0,1\}^{n_2}\to\{0,1\}^m$ is called a quantum-proof $(n_1,k_1,n_2,k_2,m,\eps)$ two-source extractor if for Markov sources $\rho_{X_1X_2Q}$ with $H_{\min}(X_1|Q)_\rho\geq k_1$ and $H_{\min}(X_2|Q)_\rho\geq k_2$, we have
\begin{align}
\frac{1}{2}\left\|\rho_{\text{Ext}(X_1,X_2)Q}-\tau_{M}\otimes\rho_Q\right\|_1\leq\eps\,,
\end{align}
where $\rho_{\text{Ext}(X_1,X_2)Q}:=(\text{Ext}(X_1,X_2)\otimes\mathcal{I}_Q)(\rho_{X_1X_2Q})$ and $\tau_M$ denotes the fully mixed state on $\mathbb{C}^{2^m}$. The function $\text{Ext}$ is called quantum-proof strong in the $i=1,2$ input if
\begin{align}
\frac{1}{2}\left\|\rho_{\text{Ext}(X_1,X_2)X_iQ}-\tau_{M}\otimes\rho_{X_iQ}\right\|_1\leq\eps\,.
\end{align}
\end{definition}

We note that this definition is composable \cite{portmann2014cryptographic}. The following finding of Arnon-Friedman {\it et al.}~\cite[Theorem 2]{Friedman16} relates two-source extractors with its quantum-proof analogue.

\begin{prop}\label{lem:friedman-step}
Every $(n_1,k_1,n_2,k_2,m,\eps)$ two-source extractor is a quantum-proof
\begin{align}
\left(n_1,k_1+\log(1/\eps),n_2,k_2+\log(1/\eps),m,\sqrt{3\eps}\cdot2^{m/2-1}\right)
\end{align}
two-source extractor. Moreover, if the two-source is strong in one source, then it is also quantum-proof strong in the same source. 
\end{prop}

We note that it is on open question if the parameter loss $\eps\mapsto\sqrt{3\eps}\cdot2^{m/2-1}$ is needed. For our protocols, we implement the following quantum-proof version of the Dodis {\it et al.}~\cite{Dodis04} construction based on cyclic shift matrices commonly attributed to Vazirani \cite{Vazirani87}.

\begin{prop}\label{prop:Dodis}
Let $A_i$ be $n\times n$-matrices such that the $j$-th row has a 1 in the $j-i+1\;\text{mod}\;n$ column and zero elsewhere, i.e., $A_i$ implements a cyclic shift by $i-1$ bits. For $n$ prime with $2$ as primitive root,\footnote{That is, for $i=1,\ldots,n-1$ there exists $p_i\in\mathbb{N}$ such that $2^{p_i}=i\;\text{mod}\;n$.} the function $\text{Ext}_A:\{0,1\}^n\times\{0,1\}^n$ defined as
\begin{align}
(x,y)\mapsto\text{Ext}_A(x,y)=\Big((A_1x)y,\cdots,(A_mx)y\Big)
\end{align}
gives a quantum-proof $(n,k_1,n,k_2,m,\eps)$ two-source extractor with\footnote{It is implicitly claimed in \cite[Sec.6.1]{Friedman16} that the Dodis et al.~\cite{Dodis04} construction is directly quantum-proof with better parameters than stated here. However, the exact parameters of this remain to be worked out.} $\eps=\frac{\sqrt{3}}{2}\cdot2^{-\frac{1}{8}\left((k_1+k_2+1-n)-5m\right)}${,}
\begin{align}
\text{or equivalently,}\label{m2Size}\quad {m=\frac{1}{5}\left(k_1+k_2-8\log\left(1/\eps\right)+9-4\log(3)-n\right)\,,}
\end{align}
{which can be rewritten as in \eqref{DodisOut} using $-\log(p_{SV}[n]\cdot p_Q[n])=-\log(p_{SV}[n])-\log(p_Q[n])=k_1+k_2$. }\\
The construction is quantum-proof strong in either one source.
\end{prop}
A couple of comments are in order:
\begin{itemize}
\item The threshold for this construction to work is that the difference between $k_1+k_2+1$ and $n$ is positive. Then, roughly one fifth of this difference can be extracted.
\item As outlined in \cite[App.D.G]{Hayashi16}, an $n$ that is prime with $2$ as primitive root can be found efficiently for at least up to $n\leq10^{50}$. For $10^i$ with $i=1,\ldots,12$ some of these {numbers} are listed in \cite[Eq.~127]{Hayashi16}:
\begin{align*}
\begin{aligned}
\noindent & 11, 101, 1'019, 10'037, 100'003, 1'000'003, 10'000'139, 100'000'137, 1'000'000'021,10'000'000'019,\\
& 100'000'000'003.
\end{aligned}
\end{align*}
Alternatively, a complete list of the first $10^5$ entries can be found \href{https://oeis.org/A001122}{here}\,---\,the largest number in that list is $310'091$. Note that $n$ corresponds to the number of bits in the input and hence we need at most $n=10^8$ for our purposes. We comment more in Sec.\ref{sec:code-dodis} on the complexity of this one-time pre-processing step and how we implemented it.
\item We show in App.\ref{sec:code-dodis} that the function $\text{Ext}_A(x,y)$ can be implemented in complexity $O(n\log n)$.
\end{itemize}

\begin{proof}[Proof of Prop.\ref{prop:Dodis}]
This is proven as in \cite[Corollary 21]{Friedman16}, employing Prop.\ref{lem:friedman-step} together with the parameters of the Dodis {\it et al.}~construction based on cyclic shift matrices \cite[Sec.3.2]{Dodis04}.
\end{proof}

Finally, we need to operate quantum-proof two-source extractors on sources for which we only have a guarantee on the smooth min-entropy instead of the min-entropy. This is covered by the following lemma.

\begin{lemma}\label{lem:friedman}
Let $\text{Ext}:\{0,1\}^{n_1}\times\{0,1\}^{n_2}\to\{0,1\}^m$ be a quantum-proof {$(n_1,k_1,n_1,k_2,m,\eps)$} two-source extractor, $\kappa\in(0,1]$, and $\eps_2\in(0,1)$. Then, for any Markov source $\rho_{X_1X_2Q}$ with
\begin{align}
\text{$H_{\min}(X_1|Q)_\rho\geq k_1$ and $H_{\min}^{\kappa}(X_2|Q)_\rho\geq k_2+\log\left(1/\eps_2\right)$}
\end{align}
we have that
\begin{align}\label{m2Error}
\frac{1}{2}\left\|\rho_{\text{Ext}(X_1,X_2)Q}-\tau_M\otimes\rho_Q\right\|_1\leq3\kappa+\eps_2+\eps\,.
\end{align}
\end{lemma}

\begin{proof}[Proof of Prop.\ref{lem:friedman}]
This is proven as in \cite[Lemma 17]{Friedman16} that covers the variant of two smooth min-entropy sources.
\end{proof}


\subsection{Seeded extractor}\label{app:seeded}

For the special case when the second source is already perfectly random and independent of any side information $Q$, i.e., $n_2=k_2$, two-source extractors are known as seeded extractors with the seed $d:=n_2=k_2$. In particular, we need quantum-proof seeded extractors that are quantum-proof strong in the seed and call such functions strong quantum-proof seeded extractors. We reproduce the self-contained definition here.

\begin{definition}
Let $n\in\mathbb{N}$, $k\in[0,n]$, $d\in\mathbb{N}$, $m\in\mathbb{N}$, and $\eps\in[0,1]$. A function $\text{Ext}:\{0,1\}^{n}\times\{0,1\}^d\to\{0,1\}^m$ is called a strong quantum-proof $(n,k,d,m,\eps)$ seeded extractor if for sources $\rho_{XQ}$ with $H_{\min}(X|Q)_\rho\geq k$, we have
\begin{align}
\frac{1}{2}\left\|\rho_{\text{Ext}(X,U_D)DQ}-\tau_{M}\otimes\tau_D\otimes\rho_{Q}\right\|_1\leq\eps\,,
\end{align}
where $\tau_D$ denotes the fully mixed state on $\mathbb{C}^{2^d}$. We call $n\in\mathbb{N}$ the input length, $d\in\mathbb{N}$ the seed size, $m\in\mathbb{N}$ the output length, and $\eps\in[0,1]$ the security parameter.
\end{definition}

Note that the extractor $\text{Ext}$ can be seen as a family $\left\{f_i\right\}_{i\in2^d}$ of functions $f_i:\{0,1\}^n\to\{0,1\}^m$. For our purposes, we use a quantum-proof version of the Hayashi-Tsurumaru construction with seed size $d=n-m$ \cite[Sec.V.B]{Hayashi16}.

\begin{prop}\label{prop:HT}
Let $n\in\mathbb{N}$, $k\in[0,n]$, $\eps\in[0,1]$, and set $c:=1+\lceil\frac{m}{n-m}\rceil$. Regarding the input as {made of $c$ blocks $x_1,x_2, ..., x_c$ of $n-m$ bits each, i.e.} $\{0,1\}^n\cong\mathbb{F}_2^n\cong\left(\mathbb{F}_{2^{n-m}}\right)^c${, the seed r of length $d=n-m$ bits ($\{0,1\}^{n-m}\cong\mathbb{F}_2^{n-m}$),} and the output as $\{0,1\}^m\cong\mathbb{F}_2^m$, the family $\left\{f_i\right\}_{i\in2^{n-m}}$ of functions $f_i:\{0,1\}^n\to\{0,1\}^m$ with
\begin{align}\label{eq:hashing}
f_i:(x_1,\cdots,x_c)\mapsto(x_1+r\cdot x_c,\cdots,x_{c-1}+r^{l-1}\cdot x_c)
\end{align}
defines a strong quantum-proof $(n,k,d=n-m, m {= (c-1)d},\eps)$ seeded extractor $\text{Ext}_{HT}$ with output length
\begin{align}
m=k-2\log\left(1/\eps\right)-\log\left\lceil\frac{m}{n-m}\right\rceil\,.
\end{align}
The product operation $\cdot$ in \eqref{eq:hashing} stands for binary finite field multiplication in $\mathbb{F}_{2^{n-m}}$.
\end{prop}

A couple of comments are in order:
\begin{itemize}
\item Compared to standard two-universal hashing constructions with seed size $n$ or $m$ (see, e.g., \cite{tomamichel2011leftover}), the Hayashi-Tsurumaru construction $\text{Ext}_{HT}$ comes with the seed size $d=n-m$. This gives a relatively short seed size for high min-entropy sources $k\approx n$, while keeping the near optimal output size $m\approx k$.
\item Following \cite[App.D]{Hayashi16}, we discuss in App.\ref{sec:code-ht} how the Hayashi-Tsurumaru construction $\text{Ext}_{HT}$ is implemented with complexity $O(n\log n)$, when $\frac{n}{n-m}\in\mathbb{N}$ and the seed size $d=n-m$ is prime with $2$ as primitive root (cf.~the related comments in Sec.\ref{sec:two-source}).
\end{itemize}

We need to operate strong quantum-proof seeded extractors on sources for which we only have a guarantee on the smooth min-entropy instead of the min-entropy. This is covered by the following lemma.

\begin{lemma}\label{lem:smooth-seeded}
Let $\text{Ext}:\{0,1\}^{n}\times\{0,1\}^d\to\{0,1\}^m$ be strong quantum-proof $(n,k,d,m,\eps)$ seeded extractor and $\kappa\in(0,1]$. Then, for any source $\rho_{XQ}$ with $H_{\min}^{\kappa}(X|Q)\geq k$ we have that
\begin{align}
\frac{1}{2}\left\|\rho_{\text{Ext}(X,U_D)DQ}-\tau_{M}\otimes\tau_D\otimes\rho_{Q}\right\|_1\leq\eps+\kappa\,.
\end{align}
\end{lemma}

\begin{proof}
The lemma is proven straightforwardly as, e.g., \cite[Corollary 7.8]{tomamichel-book} by employing the equivalence of the trace distance and the purified distance, together with the triangle inequality for the trace distance.
\end{proof}


\section{Security analysis: putting everything together}\label{Altogether_security}

We now put together the statistical analysis of the quantum part of the protocol, as discussed in App.\ref{APP:Bell}, with the classical post-processing steps, as discussed in App.\ref{APP:math}, in order to arrive at a full security proof. This is straightforward as all cryptographic components used are composable \cite{portmann2014cryptographic}. Namely, after having performed the Bell test from App.\ref{APP:Bell}, following Fig.\ref{Fig:RandProc_rev}{,} the first step of the randomness post-processing is as follows: if one desires to make the I.I.D assumption (see Sec.\ref{app:identical}), one can then use the min-entropy as given in \eqref{IIDminE} instead. (In the I.I.D setting, we get $2n$ bits from the quantum device and so, only require $2n$ bits from the imperfect RNG. {Note: here we do the analysis in the device-independent setting (MBQA).})
\begin{itemize}
\item From the imperfect RNG, which we model as {an} SV source \eqref{SVsource}, we get $3n$ bits with min-entropy rate $\alpha$, abbreviated as $[3n,0,k_1=3\alpha n]$. Here, $\alpha$ is a function of the SV source parameter $\delta$, namely $\alpha = -\log(\frac{1}{2}+\delta)$.

\item The output of the quantum device is $3n$ bits with $\kappa$-smooth min-entropy rate $\beta$ as given in \eqref{MBQA_entropyRate} and abbreviated as $[3n,\kappa,k_2^{\kappa}=3\beta n]$. Note that in the case of MBQA (i.e. non-I.I.D), we have an additional security parameter $\eps_{EA}$ associated to the failure probability of the entropy accumulation step and a smoothing parameter $\kappa$. {The smoothing parameter comes from concentration inequalities used in the proof of the EAT, whereby for large $n$ the total min-entropy for the whole experiment is well approximated by the sum of the von Neumann entropy in the individual rounds.}. By imposing $\eps_{EA} \leq \eps_{sec}$, we can effectively ignore this parameter when computing the final security parameter $\eps_{sec}$ (note, however, that $\eps_{EA}$ still contributes indirectly in the accumulated entropy during the entropy accumulation step, i.e. \eqref{MBQA_entropy} with $\eps_{EA}=\eps_{sec}$). The smoothing parameter $\kappa$ is effectively a free parameter which will be chosen to optimise the final randomness generation rates.
\item The quantum-proof {$(3n,k_1,3n,k_2,m_2,\eps_{ext})$} two-source extractor from Prop.\ref{prop:Dodis} is run on the inputs $[3n,0,k_1]$ and $[3n,\kappa,k_2^{\kappa}]$, in the smooth min-entropy form of Lemma \ref{lem:friedman}. We note that the two-source extractor can be chosen quantum-proof strong in any of the two sources (but not both).
\item The output is $m_2$ bits of randomness with security parameter $\eps_1$, abbreviated as $[m_2,\eps_1]$. Here, $m_2$ is a function of the parameters of the min-entropy sources (via Prop.\ref{prop:Dodis}) and $\eps_1$ is a function of the parameters $\eps$ and $\kappa$ (because of working with the smooth min-entropy) and the security parameter $\eps_{ext}$ of the two-source extractor (via Lemma \ref{lem:friedman}).
\end{itemize}

The output size $|m_2|$ of the 2-source extractor is then given by using \eqref{m2Size}
\begin{equation}
m_2 = \frac{1}{5}\big(k_1+k_2-3n+1-8\log(\frac{1}{\eps_{ext}})-8\log(\frac{\sqrt{3}}{2})\big)
\end{equation}
with total error $\eps_1 = 3\kappa+\eps+\eps_{ext}$ as in \eqref{m2Error}. When applying the MBQA statistical analysis, we further compute $k_2 = k_2^{\kappa}-\log(\frac{1}{\eps})$ with $H^{\kappa}_{min}$ from \eqref{MBQA_entropy}, because we need to account for the smoothing parameter {since randomness extractors take min-entropy (not smooth min-entropy) as an initial input, so we must lower bound the min-entropy.} From the Santha-Vazirani condition \eqref{SVsource}, we get $k_1 = 3n \alpha$ with $\alpha = -\log(\frac{1}{2}+\delta)$ computed. Finally, the optimal value for $m_2$ (to get the largest output size at a given security parameter) without a seeded extractor and for a desired overall security parameter $\eps_1 = \eps_{sec}$, for a given $\delta$ and observed $L_{\textrm{MDL}}^{obs}$ can be obtained as the result of the following optimisation:
\begin{equation}\begin{aligned}\label{FinalEntropy_2}
m_2^{opt}(L_{b}, \eps_{sec}) = \max_{\substack{\kappa, \eps, \eps_{ext} \\ \text{s.t. }3 \kappa + \eps + \eps_{ext} \leq \eps_{sec}}} \frac{1}{5}\bigg( 3n\alpha+n \max\limits_{L_{cut}} & \big( f_{min}(L_{b},L_{cut})-\frac{\nu(L_{cut},\kappa,\eps_{EA}=\eps_{sec})}{\sqrt{n}}\big) \\ & -3n-\log(\frac{1}{\eps})+1-8\log(\frac{1}{\eps_{ext}})-8\log(\frac{\sqrt{3}}{2}) \bigg)
\end{aligned}\end{equation}
with $\nu(L_{cut},\delta,\eps_{EA}=\eps_{sec})$ the penalty term in \eqref{MBQA_penalty}, $L_b=\frac{L^{obs}}{4(\frac{1}{2}-\delta)^2}$ as in \eqref{amplifProof2} and $f_{min}(L_{b},L_{cut})$ in \eqref{fminEQU}.

This analysis is the basis for the plots in Fig. \ref{Plot:IIDvsEAT}, \ref{Plot:OutvsEps}, and \ref{Plot:GoodRNG}.\\

For the further randomness post-processing, i.e. adding a seeded extractor, we start by discussing {\it randomness amplification} as in Fig.  \ref{Fig:RandProc_rev}:
\begin{itemize}
\item From the previous randomness post-processing step, we get $m_2$ bits of randomness with security parameter $\eps_1$, abbreviated as $[m_2,\eps_1]$. We fix the two-source extractor to be quantum-proof strong in the imperfect RNG $[3n,0,k_1=3\alpha n]$ as opposed to the output of the quantum device.
\item From the imperfect RNG, we take $n_S$ additional bits with min-entropy rate $\alpha_S$, abbreviated as $[n_S,0,k_S=\alpha_S n_S]$.
\item The quantum-proof $(n_S,k_S,m_2,m_S,\eps_2)$ seeded extractor from Prop.\ref{prop:HT} is run on the source $[n_S,0,k_S=\alpha_S n_S]$ with the seed $m_2$ as given from $[m_2,\eps_1]$.
\item The output is $m_S$ bits of randomness with security parameter $\eps_1+\eps_2$, abbreviated as $[m_S,\eps_1+\eps_2]$. This follows as in \cite[Lemma 38]{Friedman16}, where the added errors come from a simple monotonicity and triangle inequality argument for the trace distance. Here, $m_S$ is a function of the parameters of the min-entropy source (via Prop.\ref{prop:HT}).
\end{itemize}
Importantly, this {protocol} allows to achieve $m_S\gg m_2$ for the final output $[m_S,\eps_1+\eps_2]$.\\

Next, we discuss {\it randomness amplification and privatisation} as in Fig. \ref{Fig:RandProc}:
\begin{itemize}
\item From the previous randomness post-processing step, we get $m_2$ bits of randomness with security parameter $\eps_1$, abbreviated as $[m_2,\eps_1]$. Here,  we fix the two-source extractor previously used to be quantum-proof strong in the source from the quantum device.
\item From the quantum device, we again take the $3n$ output bits with $\kappa$-smooth min-entropy rate $\beta$, abbreviated as $[3n,\kappa,k_2=3\beta n]$.
\item We run a quantum-proof $(3n,k_2,m_2,m_S,\eps^s_{ext})$ seeded extractor on the source $[3n,\kappa,k_2=3\beta n]$, in the smooth min-entropy form of Lemma \ref{lem:smooth-seeded} and with the seed $m_2$ as given from $[m_2,\eps_1]$.
\item The output is $m_S$ bits of randomness with security parameter $\eps_1+\eps^s_{ext}+\kappa$, abbreviated as $[m_S,\eps_1+\eps^s_{ext}+\kappa]$. This again follows as in \cite[Lemma 38]{Friedman16}. Here, $m_S$ is a function of the parameters of the smooth min-entropy source and given by the characteristics of the quantum-proof seeded extractor of choice.
\end{itemize}

Importantly, this protocol allows to achieve $m_S\gg m_2$ for the final output $[m_S,\eps_1+\eps^s_{ext}+\kappa]$ when using a quantum-proof seeded extractor that works with an exponentially small seed in the input size\,---\,such as Trevisan based constructions \cite{trevisan01,portmann09}. We did not implement this part ourselves, but instead refer to \cite{Mauerer12} for off-the-shelf implementations of Trevisan based seeded extractors.

When appending Trevisan's extractor, one can then roughly extract all the min-entropy of the output from the quantum device (up to logarithmic terms). The total randomness generation rate is then roughly equal to the min-entropy rate of the quantum device $3\beta$ (see \eqref{MBQA_entropyRate}), i.e. the final rate $\eta_S \approx 3\beta$. This analysis forms the basis for the plot in Fig.\ref{Plot:seeded_efficiency}. Note that as one can not assume that the imperfect RNG's output is private to the user and because the 2-source extractor needs to be made strong in the quantum input (for it to be re-used as input to the seeded one), one can not concatenate the seed together with the output of the seeded extractor as the seed therefore also is not private.


\section{Implementation of the randomness post-processing}\label{APP:code}

\subsection{Two-source extractor}\label{sec:code-dodis}

This appendix explains how the Dodis {\it et al.}~construction from Prop.\ref{prop:Dodis} is implemented with quasi-linear complexity in the input size. The first step is to note that our version of the Dodis {\it et al.}~construction only runs for input sizes $n$ prime with $2$ as primitive root. Hence, for general input bit strings of size $n'\in\mathbb{N}$, the idea is to give an $n\leq n'$ such that $n$ is prime with $2$ as primitive root\,---\,while keeping $n'-n$ as small as possible. The Dodis {\it et al.}~extractor is then run on the input size $n$, discarding $n'-n$ bits for potential later use. We view finding an appropriate $n$ as a pre-computation step and generated a sufficiently dense list of such integers up to $n'\approx10^8$. As mentioned in App.\ref{sec:two-source}, this is based on \cite[App.D.G]{Hayashi16} employing efficient algorithms for primality testing and integer factorization \cite{Shoup09}\,---\,where the latter is only needed for integers of size $o(\log n')$.

For $n$ prime with $2$ as primitive root, our Dodis {\it et al.}~extractor is then defined as taking two $n$-bit strings $x,y$ to the $m$-bit string
\begin{align}\label{eq:dodis-function}
z=\Big((A_1x)y,\cdots,(A_mx)y\Big)
\end{align}
with $A_i$ the $n\times n$-matrices such that the $j$-th row has a one in the $j-i+1\;\text{mod}\;n$ column and zero elsewhere. In the following, we give an algorithm based on the number-theoretic transform (NTT) \cite{Assche06} that provably computes $z$ with complexity $O(n\log n)$. This is inspired from \cite[App.C]{Hayashi16}.

We start with a definition.

\begin{definition}\label{def:convolution-reversal}
For $n\in\mathbb{N}$ and $x, y \in \mathbb{Z}^n$ the convolution $x * y$ is defined by
\begin{align}
(x * y)_i = \sum_{j+k \equiv i \pmod n} x_j y_k
\end{align}
and the reversal function $R : \mathbb{Z}^n \to \mathbb{Z}^n$ is defined by
\begin{align}
R(x_0, x_1, \ldots, x_{n-1}) = (x_0, x_{n-1}, x_{n-2}, \ldots, x_1)\,.
\end{align}
\end{definition}

It is then straightforward to verify that the extractor function in \eqref{eq:dodis-function} can be rewritten for $i=0,\cdots,m-1$ as
\begin{align}\label{eq:extractor-convolution}
(z)_i=\left(R(x)*y\right)_i\pmod 2\,.
\end{align}
Hence, it is sufficient to give an algorithm to compute the convolution $R(x)*y$ in complexity $O(n\log n)$. Such algorithms have been proposed in the literature based on the Fast Fourier Transform (FFT), see, e.g., \cite[App.C]{Hayashi16}. However, these algorithms are not information-theoretically secure because of potential rounding errors due to the floating point arithmetic.

In contrast, we now present an information-theoretically secure algorithm based on the NTT that is equally fast as FFT based implementations. For this, let $L\geq2n-1$, $p>n$, and consider the ring
\begin{align}
R:=\frac{\mathbb{Z}_p[X]}{\langle x^L - 1 \rangle}\,.
\end{align}
Now, assume for a moment that we can do multiplication efficiently in $R$. Given $a, b \in \{0,1\}^n$, in order to compute the convolution $a * b$ modulo $2$ as in \eqref{eq:extractor-convolution}, we may instead compute the product
\begin{align}
\left(\sum_i a_i X^i\right)\left(\sum_i b_i X^i\right) = \sum_i c_i X^i\quad\text{in $R$}
\end{align}
with $0 \leq c_i < p$. It is easily checked that the sought after convolution can then be written as
\begin{align}
(a*b)_i\pmod 2 = \sum_{j \equiv i \pmod n} c_j \pmod 2\,.
\end{align}
This shifts the problem of computing convolution as in \eqref{eq:extractor-convolution} to multiplication in $R$\,---\,or equivalently convolutions in $\mathbb{Z}_p^L$. However, we now have freedom to choose $p$ and $L$ and if we choose them appropriately we can make use of the NTT \cite{Assche06}.

\begin{definition}
For $L$ a power of $2$, $p$ a prime with $p \equiv 1 \pmod L$, and $\omega$ a primitive $L$-th root of unity in $\mathbb{Z}_p$ the number-theoretic transform $\mathcal{N} : \mathbb{Z}_p^L \to \mathbb{Z}_p^L$ is defined by
\begin{align}
\mathcal{N}(x)_j:= \sum_{0 \leq i < L} \omega^{ij} x_i
\end{align}
and its inverse is
\begin{align}
\mathcal{N}^{-1}(y)_i = L^{-1} \sum_{0 \leq j < L} \omega^{-ij} x_j\,,
\end{align}
where $L^{-1}$ is the inverse of $L$ in $\mathbb{Z}_p$.
\end{definition}

Crucially, since $L$ is a power of $2$, the NTT can be computed in complexity $O(L\log L)$ by the same divide-and-conquer technique as in the FFT \cite{Assche06}. Moreover, like the Fourier transform, the NTT has the property that \cite{Assche06}
\begin{align}
\mathcal{N}(x * y) = \mathcal{N}(x)  \mathcal{N}(y)\,,
\end{align}
allowing convolutions in $\mathbb{Z}_p^L$ to be computed by point-wise multiplication of terms. We note that in order to speed up the modular multiplications, we can choose $p$ in advance and take advantage of compiler optimization or custom code for computing remainders with fixed modulus \cite{lemire19}. For example, we can choose $p = 3 \times 2^{30} + 1$ allowing for values of $n$ up to $2^{29}> 5 \times 10^8$.

The implementation of computing the extractor function \eqref{eq:dodis-function}  via the convolution form \eqref{eq:extractor-convolution} then proceeds as follows. Formally, let us define the embedding $\eta : \mathbb{Z}_2^n \to \mathbb{Z}_p^L$ by
\begin{align}
\eta(x)_i := \begin{cases} 1 & i < n \textrm{ and } x_i = 1 \pmod 2 \\ 0 & \textrm{otherwise} \end{cases}
\end{align}
and the coercion function $\phi : \mathbb{Z}_p^L \to \mathbb{Z}_2^n$ by
\begin{align}
\phi(x)_i := \sum_{j \equiv i \pmod n} x_i \pmod 2
\end{align}
with $0 \leq x_i < p$ for all $i$. For $x,y\in\mathbb{Z}_2^n$ the convolution $x*y$ can then be computed in the form
\begin{align}
x*y=\phi\circ\mathcal{N}^{-1}\Big(\big(\mathcal{N}\circ\eta(x)\big) \big(\mathcal{N}\circ\eta(y)\big)\Big)\,,
\end{align}
with appropriately chosen $L$ and $p$, and in complexity $O(L\log L)$. Thus, the extractor function \eqref{eq:dodis-function} is computed via the convolution form \eqref{eq:extractor-convolution} in complexity $O(n\log n)$.


\subsection{Seeded extractor}\label{sec:code-ht}

This appendix explains how the Hayashi-Tsurumaru construction from Prop.\ref{prop:HT} is implemented with quasi-linear complexity in the input size. This largely follows the implementation from \cite[App.C]{Hayashi16}, but in contrast to this previous work our algorithm is information-theoretically secure as it is based on the number-theoretic transform. The way we employ Prop.\ref{prop:HT}, we are given $d'$ random bits and a linear min-entropy source with rate $k=\alpha n$ for input size $n$, with the goal of expanding the $d'$ random bits to a longer string of random bits.

First, we need to give $d\leq d'$ such that $d$ is prime with 2 as primitive root\,---\,while keeping $d'-d$ as small as possible. The Hayashi-Tsurumaru construction is then run on the seed size $d$, discarding $d'-d$ bits for potential later use. As in App.\ref{sec:code-dodis}, we view finding an appropriate $n$ as a pre-computation step (cf.~the comments in App.\ref{sec:code-dodis}). Second, for seed size $d$ prime with 2 as primitive root, the Hayashi-Tsurumaru construction proceeds by choosing $c\in\mathbb{N}$ with the requirement $c<\frac{1}{1-\alpha}$. For an input size $n=c d$, the extractor then generates $m=(c-1) d$ random bits\footnote{Since the Hayashi-Tsurumaru construction is a strong quantum-proof seeded extractor (Prop.\ref{prop:HT}), the seed can safely be outputted as well if wanted\,---\,leading to the total output size $(c-1)d+d=cd$.} with security parameter
\begin{align}\label{HT_eps}
\eps\leq{\sqrt{(c-1)}}2^{-d/2(1+c(\alpha-1))}\,.
\end{align}

{Equ.\eqref{HT_eps} is obtained starting from \cite[Sec. V, B, after theorem 5]{Hayashi16}, where they obtain\footnote{{Note that in \cite{Hayashi16} they use $t$ to denote the min-entropy whereas here we use $k$.}} a seeded quantum-proof extractor}
{\begin{align}
\big( n,k,d=n-m, m,\eps = \sqrt{\frac{m}{n-m}}2^{\frac{m-k}{2}} \big).
\end{align}}
{Note the seed length $d = n-m$ which can get very small for large $m$, e.g., from a high quality source with high min-entropy rate. For our purposes we are given the seed of size $d = n-m$ coming from two-source extractor (its output) and we want to choose $n$, $m$ such that $c:=\frac{n}{n-m} \in \mathbb{N}$ as the scheme then becomes much simpler. Moreover, the $n$-bit string comes from the SV source and typically has linear min-entropy of $k=\alpha n$ for some $\alpha \in (0,1)$. Hence, it becomes sufficient to choose c such that}
{\begin{align}\label{blablabla}
\sqrt{(c-1)}2^{-d/2(1+c(\alpha-1))}\leq \eps'
\end{align}}
{for a desired $\eps'$. One can see that \eqref{blablabla} then implies \eqref{HT_eps}, as $\eps\leq \eps'$, which is obtained using $c-1=\frac{m}{n-m}$ together with $m=(c-1)d$ and $n=d+m=d+(c+1)d$. For a given $d$ and $\eps$ we can now numerically evaluate equ.\eqref{blablabla} to find a sufficient $c$ that will then yield the needed input size $n = cd$ as well as the corresponding output size $m = (c-1)d$.}\\

Following \cite[App.D.F]{Hayashi16}, the algorithm {uses field arithmetic using circulant matrices (FACM) to compute the output bits of the Hayashi-Tsurumaru extractor. Because of this, all operations denoted $\cdot$ are now matrix-matrix or matrix-vector multiplications. The algorithm is as follows:}
\begin{itemize}
\item Let $E:\{0,1\}^c\to\{0,1\}^{c+1}$ be defined as
\begin{align}\label{eq:algo-first}
E\left(a=(a_0,\ldots,a_{c-1})\right)=\bar{a}=\left(a_0,\ldots,a_{c-1},\sum_{i=0}^{c-1}a_i\pmod2\right)
\end{align}
and $D:\{0,1\}^{c+1}\to\{0,1\}^c$ as
\begin{align}
D(\bar{a}=(a_0,\ldots,a_c))=a=(a_0,\ldots,a_{c-1})\,.
\end{align}
\item Let $C:\{0,1\}^{c+1}\to\{0,1\}^{c+1}\times\{0,1\}^{c+1}$ be defined as the circulant matrix
\begin{align}
C\left(a=(a_0,\ldots,a_c)\right)=\begin{pmatrix} a_0 & a_1 & \ldots & a_c\\ a_c & a_0 & \ldots & a_{c-1}\\\vdots & \vdots & \ldots& \vdots \\ a_1 & a_2 & \ldots & a_0\end{pmatrix}\,.
\end{align}
\item Inputs: Seed $r\in\{0,1\}^d$ and input $\left(x^{(1)},\ldots,x^{(c)}\right)\in\{0,1\}^n$ with $x^{(i)}\in\{0,1\}^d$.
\item Step I: Compute $y^{(c)}=E\left(x^{(c)}\right)$, $s=E(r)$, and $y^{(1)}=E\left(x^{(1)}\right)+C(s)\left(y^{(c)}\right)^T$.
\item Step II: For $i=2,\dots,c-1$ iterations
\begin{align}
s&=C(E(r))\cdot s^T\\
y^{(i)}&=E\left(x^{(i)}\right)+C(s)\cdot\left(y^{(c)}\right)^T\,.\label{eq:algo-last}
\end{align}
\item Output: $\left(D\left(y^{(1)}\right),\ldots,D\left(y^{(c-1)}\right)\right)\in\{0,1\}^m$.
\end{itemize}
We note that the total number of iterations is $c-1$, which is constant. Furthermore, all operations have complexity $O(d)=O(n)$ except for the bottleneck matrix-vector multiplications of the form $C(s)\cdot\left(y^{(c)}\right)^T$. However, following Definition \ref{def:convolution-reversal} it is straightforward to verify that this can be rewritten as the convolution of vectors of size $d$:
\begin{align}
\text{$R(s)*y^{(c)}$ in terms of the reversal function $R$}\,.
\end{align}
This makes the implementation amenable to the NTT methods as introduced in App.\ref{sec:code-dodis} and ultimately leads to the overall complexity $O(n\log n)$.

\begin{figure}[!ht]
\begin{center}
\includegraphics[scale=0.6]{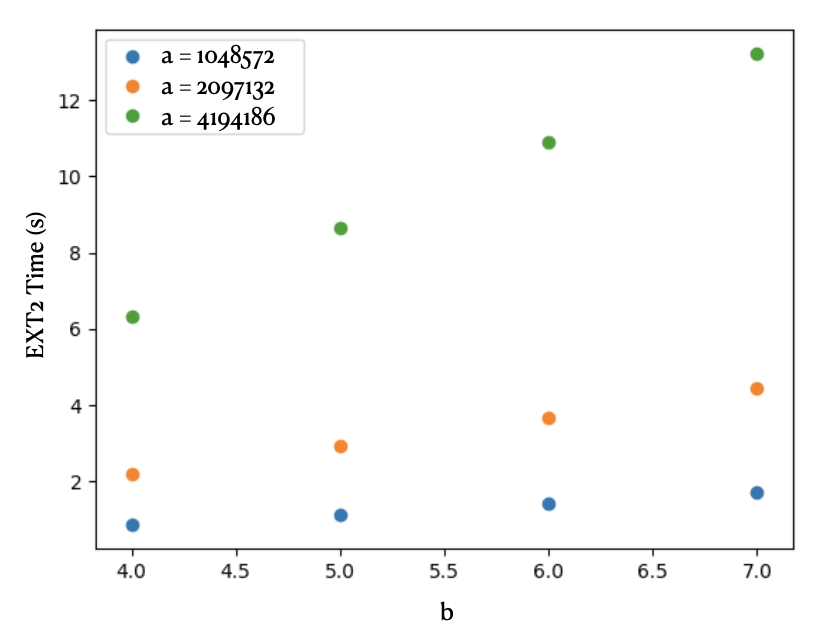}
\caption{\label{perf-ntt}Performance of the implementation using the algorithm of App.\ref{sec:code-ht}, shows execution times for various values of $a$ and $b$, run on an Intel i7-9750H 2.60GHz laptop.}
\end{center}
\end{figure}

In the following, we reuse the material on the NTT from App.\ref{sec:code-dodis}, including in particular the functions $\mathcal{N},\mathcal{N}^{-1},\eta,\phi$ with the only difference that the vector arguments are now of size $d$. Furthermore, we define $\eta^\prime$ as $\eta$ applied to the reversed sequence
\begin{align}
\eta^\prime(x_0, x_1, x_2, \ldots, x_{d-1}):=\eta(x_0, x_{d-1}, x_{d-2}, \ldots, x_1)\,,
\end{align}
as well as the consecutive functions $\mathcal{N}_0:= \mathcal{N} \circ \eta$, $\mathcal{N}_0^\prime:=\mathcal{N} \circ \eta^\prime$, and $\rho:=\phi \circ \mathcal{N}^{-1}$. We then have the convolution
\begin{align}
R(s)*y^{(c)}= \rho\Big(\mathcal{N}_0\big(R(s)\big) \cdot \mathcal{N}_0\big(y^{(c)}\big)\Big)\,.
\end{align}
This in turn leads itself to the following efficient algorithm for \eqref{eq:algo-first} -- \eqref{eq:algo-last}:
\begin{itemize}
\item Let $s_0 = E(r)$.
\item Let $\sigma = \mathcal{N}_0^\prime(s_0)$.
\item Let $\zeta = \mathcal{N}_0(E(x_{c-1}))$.
\item Output $x_0 + D(\rho(\sigma \cdot \zeta))$.
\item Let $s_1 = \rho(\sigma \cdot \mathcal{N}_0(s_0))$.
\item Output $x_1 + D(\rho(\zeta \cdot \mathcal{N}_0^\prime(s_1)))$.
\item Let $s_2 = \rho(\sigma \cdot \mathcal{N}_0(s_1))$.
\item Output $x_2 + D(\rho(\zeta \cdot \mathcal{N}_0^\prime(s_2)))$.
\item \ldots
\item Let $s_{c-2} = \rho(\sigma \cdot \mathcal{N}_0(s_{c-3}))$.
\item Output $x_{c-2} + D(\rho(\zeta \cdot \mathcal{N}_0^\prime(s_{c-2})))$.
\end{itemize}
This involves $4c-5$ forward or inverse NTTs, each of complexity $O(d\log d)=O(n\log n)$. However, the computation of $s_{i+1}$ depends only on $s_i$ and thus we can execute most of these computations (which require $2$ NTTs each) in parallel with the output computations (which also require $2$ NTTs each). The initial computations of $\sigma$ and $\zeta$ can also be parallelized. Therefore, with two concurrent threads we can reduce the algorithm to $4c-5 - 1 - 2(c-2) = 2c-2$ consecutive NTTs. The performance of this algorithm is depicted in Fig. \ref{perf-ntt} and at least as fast as the state-of-the-art FFT based implementations discussed in \cite[App.E]{Hayashi16}.

\section{Extension\,---\,Raz' two-source extractor}\label{app:Raz}

\subsection{Improved parameters}

Raz' extractor \cite{Raz05} allows to work with randomness sources of lower quality than needed for the Dodis {\it et al.}~construction. The relevant statement is \cite[Corollary 24]{Friedman16}, giving a quantum-proof $(n_1,k_1,n_2,k_2,m,\eps)$ two-source extractor\,---\,strong in either one source\,---\,for any $\delta\in(0,19/32)$ with
\begin{align}
n_1&\geq6\log n_1+2\log n_2\\
k_1&\geq\left(0.5+\delta\right)n_1+3\log n_1+\log n_2\\
k_2&\geq\frac{163}{32}\log\left(\left(1+\frac{3\delta}{19}\right)n_1-k_1\right)\label{eq:raz3}\\
m&\leq\frac{16\delta}{19}\min\left\{\frac{n_1}{8},\frac{4k_2}{163}\right\}-1\\
\eps&\leq\frac{\sqrt{3}}{2}2^{-m/4}\,.
\end{align}
For our purposes, we use the symmetric input size $n:=n_1\equiv n_2$, and linear min-entropy rates $k_1=\alpha_1 n$ and $k_2=\alpha_2 n$ for some $\alpha_1,\alpha_2\in(0,1)$. Consequently, for $n$ large enough, the only non-tirival conditions become
\begin{align}\label{eq:extractor-parameters}
m\leq\frac{64}{3097}\delta\alpha_2n-1\quad\text{and}\quad\eps\leq\frac{\sqrt{3}}{2}2^{-m/4}
\end{align}
for some $\delta\approx\alpha_1-0.5$. The main result of this appendix is to present an improvement on the undesirable pre-factor $\frac{64}{3097}\approx2\%$ while keeping the security parameter exponentially small in $m$.

\begin{prop}\label{prop:second-try}
Let $\delta\in(0,(9+\beta)/16)$ and $\beta\in(0,0.5]$ with
\begin{align}\label{eq:raz1}
\begin{aligned}
n_1\geq6\log n_1+2\log n_2,\quad k_1\geq\left(0.5+\delta\right)n_1+3\log n_1+\log n_2,\\
k_2\geq\frac{247.5+162
\beta}{26.5+17\beta}\log\left(n_1\left(1+\frac{1+\beta}{9+\beta}\delta\right)-k_1\right)\,,
\end{aligned}
\end{align}
as well as $k_2\leq2(n_1-k_1)$.
Then, Raz' construction gives a quantum-proof $(n_1,k_1,n_2,k_2,m,\eps)$ two-source extractor with output size and security parameter as
\begin{align}\label{eq:raz2}
m\leq\delta\frac{16k_2}{(9+\beta)(27.5+18\beta)}-1\quad\text{and}\quad\eps\leq\sqrt{3}2^{-(1.25+\beta/2)}2^{-m\beta/2}\,,
\end{align}
quantum-proof strong in either one source.
\end{prop}

We postpone the proof to App.\ref{sec:raz-proofs} and only note that Prop.\ref{prop:second-try} is the basis for the plots in Fig. \ref{Plot:RazVSDodis} in the main text\,---\,comparing the Dodis {\it et al.}~and Raz construction.
Compared to Raz' \cite[Theorem 1]{Raz05}, the improved pre-factor for the output condition in \eqref{eq:raz2} becomes possible because we additionally ask for the condition $k_2\leq2(n_1-k_1)$. Now, for example choosing $\beta=0.25$ leads for large enough input size $n=n_1\equiv n_2$ and linear min-entropy rates $k_1=\alpha_1 n$ and $k_2=\alpha_2 n$ with $\alpha_2\leq2(1-\alpha_1)$ to the conditions
\begin{align}
m\leq\frac{1}{18.5}\delta\alpha_2n-1\quad\text{and}\quad\eps\leq\sqrt{3}2^{-1.375}2^{-m/8}
\end{align}
for some $\delta\approx\alpha_1-0.5$. That is, we now have the improved ratio $\frac{1}{18.5}\approx5.4\%$ for the security parameter scaling as $2^{-m/8}$.
Next, we check by means of a numerical example that these parameters indeed work well in practice.

\begin{example}\label{example}
For two linear min-entropy sources
\begin{align}\label{eq:sources}
k_1=0.75 n_1,\quad k_2=0.1 n_2,\quad n=n_1\equiv n_2\,,
\end{align}
typical for our use case, we are interested in the efficiency $\eta:=\frac{m}{n}$ of Prop.\ref{prop:second-try} for sufficiently small security parameter $\eps\geq0$.\footnote{For the Dodis {\it et al.}~construction from Prop.\ref{prop:Dodis}, these min-entropy are of too low quality to be extracted as $0.75+0.1<1$. In contrast, the Raz construction as discussed here allows randomness extraction.} For that we need to check the following conditions:
\begin{itemize}
\item The condition $k_2\leq2(n_1-k_1)$ holds.
\item The first condition in \eqref{eq:raz1} is fulfilled for $n\geq64$.
\item For the second condition in \eqref{eq:raz1} we can, e.g., choose $\delta=0.2$, which then requires $n\geq1024$.
\item Choosing $\beta=0.25$, $n\geq1024$ is sufficient for the third condition in \eqref{eq:raz1}.
\item The output size condition in \eqref{eq:raz2} is then given as $m\leq 0.0001081n-1$, which becomes greater than one for $n\geq1513$.
\item The security parameter condition in \eqref{eq:raz2} then becomes $\eps=1.8888 \times 2^{-m/8}$, which requires $n\gg1850$ for a meaningfully small security parameter. For example, setting $\eps=2^{-14}\approx6.1 \times 10^{-5}$ we need $m\geq120$, which in turn requires $n\approx1.1 \times 10^6$.
\end{itemize}
Hence, for the example min-entropy sources from \eqref{eq:sources} and input size 
\begin{align}
\text{$n=1.1 \times 10^6$ we get $m=120$ random bits with security parameter $\eps=6.1\times 10^{-5}$.}
\end{align}
Alternatively, we can go to larger input sizes $n=10^7/10^8$ leading to output sizes of $m\approx10^4/10^5$ for security parameters $\eps\leq2^{-10^3}/2^{-10^4}$, respectively. To conclude, starting from an input size of at least $n=10^6$, we find the efficiency $\eta\approx10^{-3}$ for sufficiently small security parameters $\eps\ll10^{-5}$.
\end{example}

We end this appendix with a remark about the potential implementation of Raz' construction.

\begin{remark}
Besides the engaging numbers discussed in Example \ref{example}, the bottleneck of Raz' construction is the runtime of the algorithm needed for its implementation. Namely, following Raz' construction \cite[Theorem 1]{Raz05} based on the algorithmic building blocks \cite[Prop.6.5]{Alon86} and \cite[Sec.5]{Alon90}, one can verify that the complexity is indeed polynomial in $n$, but without further improvements at least $O(n^4)$. This does not allow to go to the necessary input sizes of $n\approx10^6$ and consequently we are working with the Dodis {\it et al.}~construction in the main text. Employing fast finite field arithmetics might allow to improve on the $O(n^4)$ dependence, but it seems rather unclear how to arrive at a quasi-linear complexity as required for input sizes $n\geq10^6$. To the best of our knowledge, this efficiency bottleneck is not addressed in the relevant literature.
\end{remark}


\subsection{Full proofs}\label{sec:raz-proofs}

Here, we give a proof of Prop.\ref{prop:second-try} following Raz' general proof technique \cite[Theorem 1]{Raz05}. In particular, Raz' construction is based on sequences of $0-1$ valued random variables $Y_1,\cdots,Y_N$ that are {\it $\zeta$-biased for linear tests of size $p'$}. That is, one asks for all $r\in\{1,\cdots,p'\}$ and all different $i_1,\cdots,i_r\in\{1,\cdots,N\}$ that
\begin{align}\label{eq:eps-bias}
\text{$X=Y_{i_1}\oplus\cdots\oplus Y_{i_r}$ has $\left|\text{Pr}[X=0]-\text{Pr}[X=1]\right|\leq\zeta$.}
\end{align}
Following Raz \cite[Theorem 1]{Raz05}, we employ a construction from references \cite{Naor93,Alon90} requiring
\begin{align}\label{eq:numberofbits}
\text{$n_1=2\left\lceil\log(p'-1)+\log\log(N+1)+\log\frac{1}{\zeta}-1\right\rceil\leq2\left\lceil\log(p')+\log\log(N)+\log\frac{1}{\zeta}\right\rceil$ random bits.}
\end{align}
The following lemma from Raz then gives two-source extractors constructed from sequences of $0-1$ valued random variables $Y_1,\cdots,Y_N$ that are $\zeta$-biased for linear tests of size $p'$.

\begin{lemma}\cite[Lemma 3.4]{Raz05}\label{lem:raz-step}
Let $N=m2^{n_2}$ and $Z_1,\ldots,Z_N$ be $0-1$ random variables that are $\zeta$-biased for linear tests of size $p'$ that can be constructed using $n_1$ random bits. Furthermore, define the function
\begin{align}
E:\{0,1\}^{n_1}\times\{0,1\}^{n_2}\to\{0,1\}^m\quad\text{as}\quad E_i(x,y)=Z_{(i,y)}(x)\,.
\end{align}
Then, for any even integer $p\leq\frac{p'}{m}$ and any $k_1,k_2,\gamma$ such that
\begin{align}
\gamma\geq2^{\frac{n_1-k_1}{p}}\left(\zeta^{1/p}+k 2^{-k_2/2}\right)
\end{align}
the function $E$ is a $(n_1,k_1',n_2,k_2',m,\gamma')$ two-source extractor with
\begin{align}
k_1'=k_1+\frac{m}{2}+2-\log\gamma,\quad k_2'=k_2+\frac{m}{2}+2-\log\gamma,\quad \gamma'=\gamma2^{m/2+1}\,.
\end{align}
\end{lemma}

Employing the construction of $0-1$ random variables as given in \eqref{eq:numberofbits} leads to the following two-source extractor.

\begin{lemma}\label{lem:intermediate}
For $\delta\in(0,0.5)$ and $\beta\in(0,0.5]$ with
\begin{align}
n_1\geq6\log n_1+2\log n_2,\quad k_1\geq\left(0.5+\delta\right)n_1+3\log n_1+\log n_2,\quad k_2\geq9\log(n_1-k_1),\quad k_2\leq2(n_1-k_1)\,.
\end{align}
the Raz construction gives a $(n_1,k_1,n_2,k_2,m,\gamma')$ two-source extractor with
\begin{align}
m\leq\delta\frac{k_2}{13.25+8.5\beta}-1\quad\text{and}\quad\gamma'=2^{-(0.5+\beta)}\times 2^{-m(1+\beta)}\,.
\end{align}
\end{lemma}

We note that this slightly improves on Raz' original construction \cite[Theorem 1]{Raz05} for our needed range of parameters\,---\,but is in contrast to \cite[Corollary 24]{Friedman16} not quantum-proof yet.

\begin{proof}[Proof of Lemma \ref{lem:intermediate}]
Based on Lemma \ref{lem:raz-step}, we first construct a $(n_1,k_1',n_2,k_2',m,\gamma')$ two-source extractor with
\begin{align}\label{eq:interfirst}
\begin{aligned}
k_1'=k_1+(2.5+\beta)(m+1),\quad k_2'=k_2+(2.5+\beta)(m+1),\quad m\leq\delta\frac{k_2}{6+4\beta}-1,\\
\gamma'=2^{-(0.5+\beta)}2^{-m(1+\beta)}
\end{aligned}
\end{align}
for the parameters
\begin{align}
n_1\geq6\log n_1+2\log n_2,\quad k_1\geq\left(0.5+\delta\right)n_1+3\log n_1+\log n_2,\quad k_2\geq8\log(n_1-k_1),\quad k_2\leq2(n_1-k_1) \,.\label{eq:interlast}
\end{align}
For that, let $N:=m2^{n_2}$, $p':=m\max\{n_1-k_1,2\}$, and $\zeta:=2^{-n_1/2+3\log n_1+\log n_2}$. Note that
\begin{align}
n_1\geq2\left\lceil\log(p')+\log\log(N)+\log\frac{1}{\zeta}\right\rceil
\end{align}
and hence by \eqref{eq:numberofbits}, $0-1$ random variables $Z_1,\ldots,Z_N$ that are $\zeta$-biased for linear tests of size $p'$ can be constructed using $n_1$ random bits. In Lemma \ref{lem:raz-step}, we then choose $p$ equal to the smallest even integer larger than $\frac{4(n_1-k_1)}{k_2}$, and estimate
\begin{align}
2^{\frac{n_1-k_1}{p}}\left(\zeta^{1/p}+p2^{-k_2/2}\right)\leq2 \times 2^{-\delta k_2/4}\leq2 \times 2^{-(m+1)(1.5+\beta)}=:\gamma\,,
\end{align}
where the first inequality follows by similar steps as in the proof of \cite[Lemma 3.6 Case A]{Raz05} and the second inequality is by the assumption $m\leq\delta\frac{k_2}{6+4\beta}-1$. Hence, by Lemma \ref{lem:raz-step} we get a $(n_1,k_1',n_2,k_2',m,\gamma')$ two-source extractor with parameters as set out in \eqref{eq:interfirst} -- \eqref{eq:interlast}.

Now, we restate the claim of Lemma \ref{lem:intermediate} as: For $\delta'\in(0,0.5)$ and $\beta\in(0,0.5]$ with
\begin{align}
n_1\geq6\log n_1+2\log n_2,\quad k_1'\geq\left(0.5+\delta'\right)n_1+3\log n_1+\log n_2,\quad k_2'\geq9\log(n_1-k_1'),\quad k_2'\leq2(n_1-k_1')\,.
\end{align}
the Raz construction gives a $(n_1,k_1',n_2,k_2',m,\gamma')$ two-source extractor with
\begin{align}
m\leq\delta'\frac{k_2'}{13.25+8.5\beta}-1\quad\text{and}\quad\gamma'=2^{-(0.5+\beta)}2^{-m(1+\beta)}\,.
\end{align}
By choosing $\delta=\delta'/2$ and $k_i=k_i'-(2.5+\beta)(m+1)$ for $i=1,2$ in \eqref{eq:interfirst} -- \eqref{eq:interlast} and following the arguments in \cite[Proof of Theorem 1]{Raz05}, we find that indeed the $(n_1,k_1',n_2,k_2',m,\gamma')$ two-source extractor from \eqref{eq:interfirst} -- \eqref{eq:interlast} does the job. This is the Raz construction.
\end{proof}

To finish the proof of Prop.\ref{prop:second-try}, it remains to make the construction from Lemma \ref{lem:intermediate} quantum-proof.

\begin{proof}[Proof of Prop.\ref{prop:second-try}]
We apply Lemma \ref{lem:friedman-step} to Lemma \ref{lem:intermediate}, in the same way as done for the proof of \cite[Corollary 24]{Friedman16}. 
\end{proof}


\section{Signalling effects in Bell tests}\label{APP:Signaling}

In a setup where there is possible signalling in a certain fraction of rounds $n_s \in [0,1]$ only, the input-output probability distribution decomposes as the convex mixture of rounds with signalling and rounds without
\begin{equation}\label{SNSdecompo}
{p}_{obs}(abc|xyz) = \sum\limits_{q_{ns}} q_{ns} {p}_{ns} (abc|xyz{q_{ns}}) + \sum\limits_{q_{s}} q_{s} {p}_{s}(abc|xyz{q_{s}}){,}
\end{equation}
where $\sum\limits_{q_{ns}} q_{ns} + \sum\limits_{q_{s}} q_{s} = 1$, $\sum\limits_{q_{s}} q_{s} = n_s$, $\sum\limits_{q_{ns}} q_{ns} = 1-n_s$ and we have denoted the probability terms in which signalling is possible with a subscript $s$ and the one where no signalling occurs with the subscript $ns$. Recall $z = x \oplus y$.\\

The observed value $M_{obs}$ we obtain is therefore a mixture
\begin{equation}\label{MSNSdecompo}
M_{obs} = \sum\limits_{q_{s}} q_{s} M({p}_{s}(abc|xyz{q_{s}})) + \sum\limits_{q_{ns}} q_{ns} M({p}_{ns}(abc|xyz{q_{ns}}))
\end{equation}
of a signalling contribution and a non signalling one. Randomness can, of course, only be obtained during the no-signalling rounds, as when signalling occurs there exist deterministic strategies capable of saturating the Mermin inequality $M=4$. From now on, we omit to label the inputs and outputs and use instead the notation ${p}_s^{q_{s}} \equiv {p}_{s}(abc|xyz)$ and ${p}_{ns}^{q_{ns}} \equiv 
{p}_{ns}(abc|xyz)$.\\

For clarity, we state the assumptions again:
\begin{itemize}
\item \textit{Assumption $A$: The effect of signalling (eg cross-talk) is random, in the sense that it is not tailored to the Bell test that is ran.}
\item \textit{Assumption $B$: The effect of signalling (eg cross-talk) is not random in the sense of $A$, but is fixed in the sense that its effect is the same each time.}
\item \textit{Assumption $C$: The effect of signalling (eg cross-talk) in the quantum computer is a mixture of the effects described in assumptions $A$ and $B$.}
\end{itemize}

The consequences of each of these assumptions and how to account for the signalling effects is derived in the next few paragraphs.

\begin{figure}[!ht]
\begin{center}
\scalebox{0.5}{\includegraphics{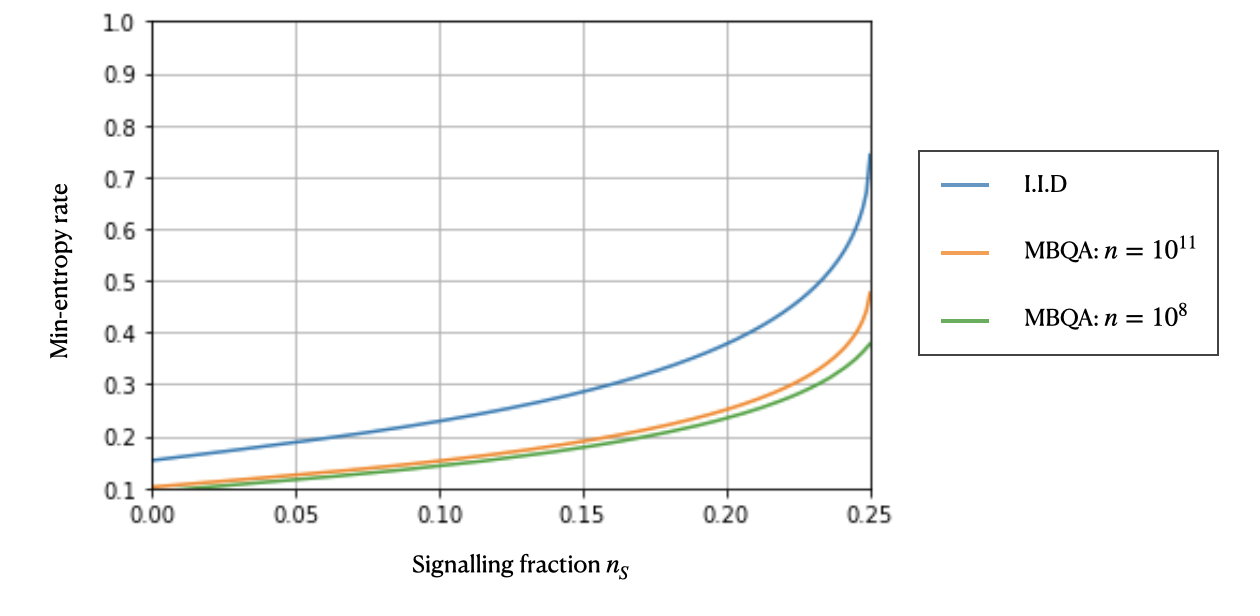}}
\caption{\label{Fig:NSA}An example of the quantum device randomness rate as a function of the signalling fraction of rounds $n_s$ for $M_{obs}=3$ and $\delta=0.05$. The rates were computed taking into account that $n(1-n_s)$ rounds only can be used for randomness generation.}
\end{center}
\end{figure}\vspace{0.5cm}

\paragraph{Assumption $A$.} In the case that one does not bias the decomposition of probability distribution towards having higher weights on contributions  with a positive contribution to the Bell inequality value, we get that
\begin{equation}\label{Ass1}
\sum\limits_{q_{s}} q_{s} M({p}_s^{q_{s}}) = 0{.}
\end{equation}
Remember that the possible values that the Mermin inequality takes over all probability distributions is contained in the interval $M \in [-4,4]$ and a random sampling of contributions over the entire set of possible quantum distributions would, for example, give $M=0$. The random probability distribution ${p}_{\unit}(abc|xyz) = \frac{1}{4} \hspace{0.2cm} \forall a,b,c,x,y,z = x \oplus y$ is a particular case giving $M=0$ and sampling randomly over the signalling probability distribution space also gives $M=0$. This assumption thus means that, when signalling occurs, it does not ``sample'' more from the signalling distribution giving a positive contribution to the Mermin inequality. We believe that this assumption is reasonable if the user believes that the quantum device was not purposely built with signalling effects tailored to the Bell inequality that is chosen and/or that there are no systematic forms of noise (which are captured in assumption B).\\

Using this assumption, we therefore obtain that
\begin{equation}\label{Ass1b}
M_{obs} = \sum\limits_{q_{ns}} q_{ns} M({p}_{ns}^{q_{ns}}){.}
\end{equation}

If we now denote $\hat{M}_{ns} = \frac{1}{1-n_s}\sum\limits_{q_{ns}} q_{ns} M({p}_{ns}^{q_{ns}})$ the average Mermin value during the rounds in which there is no signalling (i.e.~in $1-n_s$ fraction of rounds), we have that
\begin{equation}\label{Ass1c}
\hat{M}_{ns} = \frac{M_{obs}}{1-n_s}{,}
\end{equation}
which also sets a limit $n_s \leq 1-\frac{M_{obs}}{4}$ since the maximum possible value for the Mermin value is $M=4$\,---\,and in particular $\hat{M}_{ns} \leq 4$. The single-round min-entropy during a no-signalling round is therefore
\begin{equation}\label{Ass1dI}
H_{min} = -\log(P_g(\hat{M}_{ns})) =-\log(P_g(\frac{M_{obs}}{1-n_s})){,}  
\end{equation}
where the logarithm is, again, taken in base 2.

Now, when making the I.I.D. assumption, we get that the total accumulated min-entropy over $n$ rounds is
\begin{equation}\label{Ass1dII}
H^n_{min} = -n(1-n_s) \log(P_g(\frac{M_{obs}}{1-n_s})){,}  
\end{equation}
which is a strictly increasing function in $n_s$ for the guessing probability $P_g$ defined in \eqref{Pg_M}. This means that such signalling having random effect actually contributed positively to the total amount of entropy that is generated in the protocol and therefore the worst-case for us to consider it does not occur. This is not very surprising as adding such random signalling contribution actually increases the no-signalling Bell inequality value in the no-signalling contributions. The same increasing behaviour happens when considering an adversary using memory based quantum attacks (MBQA). To gain some intuition, we exemplify this increasing behaviour in Fig.~\ref{Fig:NSA}.


\paragraph{Assumption $B$.} We first recall the no-signalling condition, which without loss of generality we state for signalling occurring from qubit $A$ to $B$ (for now ignoring qubit $C$)
\begin{equation}\label{NSI}
{p}(b|y) \equiv {p}(b|y,x=0) = {p}(b|y,x=1) \hspace{1cm} \forall b,y{,}
\end{equation}
where ${p}(b|y,x) = \sum\limits_{a} {p}(ab|xy)$. In the case where there is no-signalling from qubit $A$ to $B$, the local behaviour of qubit $B$, ${p}(b|y)$, should therefore be independent of the input choice $x$ which should only affect qubit $A$.\\

The no-signalling condition \eqref{NSI} can be modified and used as a signalling quantifier
\begin{equation}\label{NSII}
s_{b,y}^{A \rightarrow B}({p}) = \bigl\vert {p}(b|y,x=0) - {p}(b|y,x=1) \bigl\vert{,}
\end{equation}
which gives $s_{b,y}^{A \rightarrow B} = 0$ for every $b$ and $y$ in the case in which there is no signalling occurring from $A$ to $B$. Remark that this quantifier can be evaluated from the observed behaviour only.\\

Now, what we call a \textit{fixed} signalling strategy $\bar{{p}}_s^{q_s}$ is such that $b = f(x,y)$ where $f: \{0,1\} \times \{0,1\} \rightarrow \{0,1\}$ is a deterministic function mapping the inputs $x,y$ to the output $b$\footnote{In a particular example, this could mean that the output $b=x \oplus y \oplus 1$ where $\oplus$ denotes the sum modulo 2.}. All extremal signalling strategies have this property and a mixture of such strategies can saturate the Mermin inequality $M=4$, which is the worst-case scenario we need to take (see below). Therefore, for a single such fixed signalling strategy (which is our assumption) one can check that there exists a pair $b,y$ such that 
\begin{equation}\label{signC}
s_{b,y}^{A \rightarrow B}(\bar{p}_s^{q_s})  = 1{.}
\end{equation}
This is what we will use in order to \textit{quantify} the amount of such signalling contribution to the observed statistics. It will have a double hit, first by reducing the actual usable Mermin inequality value (i.e.~certifying less randomness) and then also by reducing the number of rounds that can be used for generating certified randomness (since one should consider the outcomes during a signalling round as having no entropy).\\

We first discuss the impact of a given value of $s_{b,y}^{A \rightarrow B}$ on the Bell inequality value and amount of generated randomness and then explain how this quantity is measured in a worst-case scenario from the observed behaviour of the device. One can find in the main text the actual numbers that were obtained from quantum computers. The impact is minimal (although some reduction of the efficiency is observed, as expected) and quantum computers perform well in this aspect, allowing us to run our protocol efficiently and with trust. As expected, we observe that ion-trap devices have a very low signalling fraction when the one for superconducting devices is larger, but still only implies a small hit on the protocol efficiency.\\

Now, from the decomposition of our observed behaviour into signalling and no-signalling contributions \eqref{SNSdecompo}, we get that
\begin{equation}\label{Ass2a}
M_{obs} = \sum\limits_{q_{ns}} q_{ns} M(p_{ns}^{q_{ns}}) + \sum\limits_{q_{s}} q_{s} M(p_{s}^{q_{s}}){,}
\end{equation}
which in turn gives
 \begin{equation}\label{Ass2b}
\hat{M}_{ns} = \frac{M_{obs} - \sum\limits_{q_{s}} q_{s} M(p_{s}^{q_{s}})}{1-n_s}
\end{equation}
with $\hat{M}_{ns}$ defined as above $\hat{M}_{ns} = \frac{M_{obs}}{1-n_s}$. The worst-case (lowest $\hat{M}_{ns}$) is obtained when taking $M(p_{s}^{q_{s}}) = 4 \hspace{0.2cm} \forall q_{s}$, giving
 \begin{equation}\label{Ass2c}
\hat{M}_{ns} = \frac{M_{obs} - 4 n_s}{1-n_s}
\end{equation}
because $\sum\limits_{q_{s}} q_{s} = n_s$ the fraction of signalling rounds. Remember that $\hat{M}_{ns}$ is the average Mermin inequality value in the rounds in which no signalling occurs, i.e.~in a fraction $1-n_s$ of them. The single min-entropy in these no-signalling rounds only is then
 \begin{equation}\label{Ass2d}
H_{min} =-\log(P_g(\frac{M_{obs}-4n_s}{1-n_s}))  
\end{equation}
and when making the I.I.D. assumption we get that the total accumulated entropy over $n$ rounds is
\begin{equation}\label{Ass2e}
H^n_{min} = -n(1-n_s)\log(P_g(\frac{M_{obs}-4n_s}{1-n_s}))  
\end{equation}
because entropy is accumulated in $n(1-n_s)$ rounds only.\\

The same idea can be applied in the case that the I.I.D. assumption is not valid. In that case, one needs to evaluate the total (smooth) min-entropy $H^{\bar{n}}_{min}$ accumulated during the no-signalling rounds only by using $\bar{n} = n(1-n_s)$ instead of $n$ in \eqref{EAT}. One should then remember that this is only accumulated during the no-signalling rounds, so the final min-entropy rate, for example, is only $\frac{H^{\bar{n}}_{min}}{n}$ and not $\frac{H^{\bar{n}}_{min}}{\bar{n}}$.\\

We now discuss how to evaluate $n_s$, the fraction of signalling rounds in which assumption $B$ applies, from the observed statistics. Again, from the decomposition  of the observed statistics in signalling and no-signalling contributions \eqref{SNSdecompo}, we get that
\begin{equation}\label{signCb}
s_{b,y}^{A \rightarrow B}({p}_{obs})  = \sum\limits_{q_{ns}} q_{ns} s_{b,y}^{A \rightarrow B}(p_{ns}^{q_{ns}}) + \sum\limits_{q_{s}} q_{s} s_{b,y}^{A \rightarrow B}(\bar{p}_{s}^{q_{s}}) = \sum\limits_{q_{s}} q_{s} s_{b,y}^{A \rightarrow B}(\bar{p}_{s}^{q_{s}}) = \sum\limits_{q_{s}} q_{s} = n_s
\end{equation}
for a certain pair $b,y$ and where we have used that $s_{b,y}^{A \rightarrow B}(\bar{p}_{s}^{q_{s}}) = 1$ \eqref{signC} for that pair (generalising it slightly to go over all distributions we called fixed in the sense that $s_{b,y}^{A \rightarrow B}(\bar{p}_{s}^{q_{s}}) = 1$). Therefore, $n_s$ can be evaluated using $s_{b,y}^{A \rightarrow B}(p_{obs})$ on the observed behaviour.\\

Now, several additional considerations need to be made because of some simplifications we have made. The first is that we have ignored the last sub part of the device $C$ in the analysis. In order to make the considerations we have made above, we therefore always apply the worst case result when including $C$. For example, we use
\begin{equation}
s_{b,y}^{A \rightarrow B}(p) = \max\limits_{z} \hspace{0.1cm} \bigl\vert p(b|y,x=0) - p(b|y,x=1) \bigl\vert
\end{equation}
since before we used $p(b|y,x) = \sum\limits_{a} p(ab|xy)$ when it really should be ${p}(b|y,x) = \sum\limits_{a,c} p(abc|xyz)$ for some $z$.

Finally, we have considered signalling from $A$ to $B$, when in reality it might well occur in any direction between all three $A$, $B$ and $C$. Since there are 6 such possibilities, we use
\begin{equation}\label{signCc} 
n_s = 6 \max\limits_{\alpha,\beta,\Gamma,\Xi} s_{\alpha,\beta}^{\Gamma \rightarrow \Xi}(p) = \max\limits_{\gamma} \hspace{0.1cm} \bigl\vert p(\alpha|\beta,\gamma,\xi=0) - p(\alpha|\beta,\gamma,\xi=1) \bigl\vert,
\end{equation}
where $\Gamma,\Xi \in \{A,B,C\}$, $(\alpha,\beta)$ is a pair of output and input of $\Xi$, $\xi$ is the input of $\Gamma$ and $\gamma$ labels the input of the last party $\notin \{\Gamma,\Xi\}$ which is traced out. This quantity corresponds to taking the maximal value of the signalling quantifier $s(p)$ between any two sub-parts of the quantum device, maximised also on the pair of input-output that exhibits most signalling and on the input of the last sub part that is not involved in the signalling. Finally, the factor 6 comes because in the worst case, it is possible that this type of signalling occurs between any pair of sub parts and in any direction. The factor 6 is a worst case in which none of these effects overlap in a single round.\\

Table. 2 summarises the results we have observed and their typical effect on the Bell inequality value is plotted in Fig.~\ref{Fig:Signaling}. As noted in the main text, it is interesting that \textit{ibmq\textunderscore ourense} exhibits almost the double amount of signalling than \textit{ibmq\textunderscore valencia}.

\clearpage

\begin{table}\label{nsresults}
\centering
\begin{tabular}{ |p{2cm}||p{1cm}|}
 \hline
 \multicolumn{2}{|c|}{\textit{ibmq\textunderscore ourense}} \\
 \hline
 & $n_s$ \\
 \hline
 Average   & 0.0276\\ 
 max & 0.0378\\ 
 min & 0.0150\\ 
 \hline
\end{tabular} \hspace{1cm}
\begin{tabular}{ |p{2cm}||p{1cm}|}
 \hline
 \multicolumn{2}{|c|}{\textit{ibmq\textunderscore valencia}} \\
 \hline
 & $n_s$ \\
 \hline
 Average   & 0.0158\\ 
 max & 0.0252\\ 
 min & 0.0084\\ 
 \hline
\end{tabular}
\caption{Observed values obtained for $n_s$ in~\eqref{signCc} for the quantum computers \textit{ibmq\textunderscore ourense} and \textit{ibmq\textunderscore valencia}.}\label{typicalS}
\end{table}


\printbibliography

\end{document}